\documentclass[12pt, hidelinks]{article}
\usepackage[english]{babel}
\usepackage{amssymb,amsmath,amsthm,amsfonts,mathtools}
\usepackage[top=2.8cm, bottom=2.8cm, left=2.8cm, right=2.8cm]{geometry}
\usepackage{graphicx}
\usepackage{amsfonts}
\usepackage[utf8]{inputenc}
\usepackage{tikz}
\usepackage{tikz-cd}
\usetikzlibrary{trees}
\usepackage{verbatim}
\usepackage{appendix}
\usepackage[pagebackref=true, hyperindex=true, breaklinks=true]{hyperref}
\usepackage[hyperpageref]{backref}
\usepackage{natbib}
\usepackage{datetime}
\usepackage{setspace}
\usepackage{bookmark}
\usepackage{quoting} 
\usepackage{enumerate}
\usepackage{floatrow}

\usepackage{xcolor}
\usepackage{pgfplots}
\pgfplotsset{width=7cm,compat=1.8}
\usepackage{pgfplotstable}

\definecolor{bblue}{HTML}{4F81BD}
\definecolor{rred}{HTML}{C0504D}
\definecolor{ggreen}{HTML}{9BBB59}
\definecolor{ppurple}{HTML}{9F4C7C}

\theoremstyle{theorem}
\newtheorem{theorem}{Theorem}
\theoremstyle{definition}
\newtheorem{definition}{Definition}
\theoremstyle{remark}
\newtheorem{remark}{Remark}
\theoremstyle{corollary}
\newtheorem{cor}{Corollary}
\theoremstyle{lemma}

\theoremstyle{proposition}
\newtheorem{proposition}{Proposition}

\DeclareMathOperator*{\argmax}{arg\,max}

\renewcommand*{\backref}[1]{}
\renewcommand*{\backrefalt}[4]{[%
    \ifcase #1 Not cited.%
          \or Cited on p.~#2%
          \else Cited on p.~#2%
    \fi%
    ]}

\begin{document}

\title{The strategy of conflict and cooperation\footnote{{I thank Martin Antonov for an online implementation of the algorithm I introduce in this paper via GTE for three-player games, which is available at app.test.logos.bg under `BFI' and at git.io/JfQTX. I gratefully acknowledge the financial support of the DPE. I am especially grateful to Steven Brams who has helped shaped my views about cooperation. I also thank Lorenzo Bastianello, Roland B\'{e}nabou, Philippe Bich, Steven Brams, Eric van Damme, Dominik Karos, Andrew Mackenzie, Ronald Peeters, Andr\'{e}s Perea, Arkadi Predtetchinski, Martin Strobel, Peter Wakker, Xavier Venel, and participants at Maastricht University (2019), the University of Paris 1 (2019), conference in honour of Hans Peters (2020), 2020 ASSET Virtual Meeting, University of Paris 2 Panth\'eon-Assas (2021), Games, Agents, and Incentives Workshop (AAMAS) 2021, GAMES 2020 Budapest, King's College London (Department of Informatics), London School of Economics and Political Science (Department of Mathematics), DARK Seminar at the University College London Centre for Artificial Intelligence, and Imperial College London (Department of Mathematics) for their valuable comments.}}}

\author{Mehmet S. Ismail\footnote{Department of Political Economy, King's College London, London, UK. e-mail: mehmet.s.ismail@gmail.com}}

\date{Revised: 24 September 2023 \\ First version: 19 August 2018}

%\date{\today}

\maketitle

\begin{abstract}
This paper introduces a unified framework called \textit{cooperative extensive form games}, which (i) generalizes standard non-cooperative games, and (ii) allows for more complex coalition formation dynamics than previous concepts like coalition-proof Nash equilibrium. Central to this framework is a novel solution concept called \textit{cooperative equilibrium system} (CES). CES differs from Nash equilibrium in two important respects. First, a CES is immune to both unilateral and multilateral `credible' deviations. Second, unlike Nash equilibrium, whose stability relies on the assumption that the strategies of non-deviating players are held fixed, CES allows for the possibility that players may regroup and adjust their strategies in response to a deviation. The main result establishes that every cooperative extensive form game, possibly with imperfect information, possesses a CES. For games with perfect information, the proof is constructive. This framework is broadly applicable in contexts such as oligopolistic markets and dynamic political bargaining.
\end{abstract}

\newpage
\onehalfspacing

\section{Introduction}

\begin{quoting}
\noindent I hope we can go back to von Neumann and Morgenstern's vision of cooperative games and resurrect it. What is missing from their cooperative game theory was that it ignored externalities created by coalition forming which affect other players in strategic situations, e.g., competition among big firms. (Eric Maskin's paraphrased response to a question on the open problems in game theory in Nobel Symposium ``One Hundred Years of Game Theory: Future Applications and Challenges.'')
\end{quoting}

Eric Maskin's observation highlights a significant gap in economic theory: the absence of a comprehensive framework to accommodate the complex interplay of coalition formation, externalities, and strategic behavior. This is not a recent concern. As early as in his seminal work, \citet{neumann1928} also acknowledged this problem. He specifically pointed out that the strategic cooperation between any two players in a three-person zero-sum game could destabilize the maximin solution, summarizing this issue by stating, ``a satisfactory general theory is as yet lacking'' (see section~\ref{sec:literature}).

This long-standing open problem serves as the motivation for the current paper. The aim here is twofold: (i) to present a unified framework, called \textit{cooperative extensive form games}, which accommodates both externalities and synergies arising from strategic coalition formation, and (ii) to introduce a novel solution concept called the \textit{cooperative equilibrium system} (CES). I show that non-cooperative extensive form games are a special case of cooperative extensive form games, wherein  players may strategically cooperate---e.g., by writing a possibly costly contract---or act non-cooperatively. The main result of this paper establishes that every cooperative extensive form game with possibly imperfect information possesses a CES (Theorem~\ref{thm:imperfect}). In particular, in perfect information games, a CES consisting solely of pure strategies exists (Theorem~\ref{thm:perfect}), in which case the proof is constructive. To the best of my knowledge, this paper is the first to offer a comprehensive solution to this problem.

I situate these contributions within an extensive literature on both cooperative and non-cooperative games dating back to the ground-breaking book by \citet{neumann1944}.  Prior literature includes, but is not limited to, \citet{aumann1959}, \citet{harsanyi1974}, \citet{rubinstein1980}, \citet*{bernheim1987}, \citet{aumann1988}, \citet{bloch1996}, \citet{ray1997}, \citet*{brams2005}, and \citet{chander2020}; the authors defined key concepts and discussed main issues. The list is by no means comprehensive; for further discussion, see section~\ref{sec:literature}.

It is well known that the Nash equilibrium is susceptible to `multilateral' deviations by coalitions comprising two or more players. Key solution concepts, such as the coalition-proof Nash equilibrium \citep{bernheim1987} and equilibrium binding agreements \citep{ray1997}, partially address this issue. However, these concepts encounter limitations when faced with complex coalition formation dynamics. \citet[p. 33]{ray1997} articulate this challenge clearly:
\begin{quote}
    ``We must state at the outset that our treatment is limited by the assumption that agreements can be written only between members of an existing coalition... This is also the assumption in the definition of a coalition proof Nash equilibrium. It must be emphasized that an extension of these notions to the case of arbitrary blocking is far from trivial.''
\end{quote}
Cooperative extensive form games address this limitation by permitting deviating players to form coalitions not only within their subcoalitions but also with other players. Prior to making a decision, each player rationally evaluates this dynamic to determine both the coalition to join and the strategy to choose.

It is helpful to distinguish between a Nash equilibrium point and a CES. Recall that a Nash equilibrium point is characterized by the absence of unilateral profitable deviations, given that the strategies of other players are held \textit{fixed}. In contrast, a CES is a solution where neither unilateral nor multilateral `credible' deviations exist. This leads to two key distinctions. First, unlike a CES, an equilibrium point is immune only to unilateral deviations. Second, the stability of an equilibrium point relies on the assumption that the strategies of non-deviating players remain fixed---an assumption that is not generally rational in cooperative extensive form games. Players might, for instance, adapt by forming new coalitions (i.e., regrouping) and choosing new strategies in response to a deviation. Consequently, a CES aims for robustness against deviations that  are not only profitable but also credible. Credibility, a recursive notion formally defined in section~\ref{sec:CES}, ensures that a deviating player is better off when compared to a recursively defined `counterfactual.' This counterfactual considers the potential chain reactions triggered by the initial deviation.

Finally, in section~\ref{sec:applications}, I apply the CES to various theoretical domains. These include oligopolistic competition with mergers, dynamic logrolling as conceptualized by \citet{casella2019}, and `corruption,' where one player buys another's cooperation to influence a particular action. Specifically, I introduce a general logrolling model and show that it always possesses a CES. This existence result extends \citeauthor{casella2019}'s (2019) result to logrolling games with (i) `dynamically rational' voters, (ii) any type of voting rule, (iii) any type of preferences, and (iv) imperfect information. 

\section{Informal discussion with an illustrative example}
\label{sec:informal_discussion}
\vspace{0.2cm}
\noindent\textbf{Short definition of the framework}: A cooperative extensive form game or a cooperative strategic game is denoted by $\Gamma=(P, T , I, u, S, H)$, which extends an extensive form game by incorporating a coalitional utility function $u_C$ for each feasible coalition $C$. $P$ denotes the set of players, $X$ the game tree, $I$ the player function, $H$ the set of all information sets, and $S$ the set of all mixed strategy profiles. For every feasible (possibly singleton) coalition $C$, $u_{C|P'}$ denotes the von Neumann-Morgenstern utility of player $C$ given a partition of players $P'$. This formalization accommodates the potential for externalities and synergies, either positive or negative, that arise when a player joins a coalition.

In the context of this framework, the term `cooperation' is used as defined by the Oxford Advanced Learner's Dictionary: ``the fact of doing something together or of working together towards a shared aim.''\footnote{This is distinct from labeling a strategy `cooperate' in non-cooperative games. For more details, see section~\ref{subsec:repeated_games}.} The objective or aim of any coalition within this setting is to maximize the coalition's utility function, paralleling an individual player's aim to maximize their individual utility function.\footnote{It is worth noting that neither coalitions nor individual players `receive' utility in a material sense. Utility functions are merely a conceptual mechanism for representing preferences. Furthermore, the distinction between individual and coalitional utility arises because an individual's preferences may not fully align with those of the coalition to which they belong.}

\vspace{0.2cm}
\noindent\textbf{Interpretation of the framework}:  First, it should be noted that cooperative extensive form games generalize non-cooperative extensive form games in the sense that every non-cooperative extensive form game is a cooperative extensive form game but not vice versa. Put differently, a cooperative extensive form game $\Gamma=(P, T , I, (u_i)_{i\in P}, S, H)$ would reduce to a non-cooperative extensive form game when the only feasible coalitions are the singleton coalitions.

Second, in contrast to non-cooperative games, the set of players is endogenous---that is, it may evolve over the course of the game according to the following rule: if two or more players form a coalition, they each become an \textit{agent} of that coalition. For example, if $i$ and $j$ form a coalition $C=\{i,j\}$, then each of them becomes an agent of player $C$---i.e., $i$ and $j$ are no longer independent players.\footnote{A related but somewhat opposite approach, known as `player splitting,' is studied by \citet{monsuwe2000} and \citet{mertens1989}, and was used in the refinement of Nash equilibrium.} The expectation is that, once a coalition $C$ is formed, its agents will select strategies based on player $C$'s utility function, $u_C$. At each information set, any coalition may form, and any available action may be chosen.

Third, the formation of coalitions may entail monetary and/or psychological costs. Once a coalition is formed, agents retain the freedom to exit the coalition, regardless of whether they have entered into a potentially costly and binding contract. However, exiting a coalition may also incur monetary and/or psychological costs. The idea that a contract is `binding' merely implies that the cost of breaching it exceeds the cost of compliance. Nevertheless, players retain full agency and may make any choices, even those considered `irrational.'\footnote{To clarify, players are not physically constrained to adhere to the terms of a contract.} Neither the formation of coalitions nor the strategies selected by players are necessarily disclosed at each information set. While rational players may infer potential coalition formations and strategy selections, these considerations pertain to the solution concept rather than the framework itself.

Fourth, in cooperative strategic games, players possess the freedom to either act independently or form coalitions, which could be via formal or informal mechanisms/institutions. However, the ability to form a coalition may be subject to restrictions, and coordinating actions could be unfeasible under certain plausible circumstances. If, for whatever reason, some players are unable to form a coalition, this constraint is incorporated into the cooperative strategic game framework, allowing all players to rationally account for this constraint.\footnote{For a comprehensive discussion on enforcement issues in both cooperative and non-cooperative games, see \citet{serrano2004}.}

\vspace{0.2cm}
\noindent\textbf{Informal definition of the solution concept}: In cooperative extensive form games, the term \textit{system} refers to a solution concept that comprises a family of collections of strategy profiles and a family of collections of coalitions. In contrast, in non-cooperative games, a solution concept comprises a single strategy profile. The term \textit{parallel game} is used to describe a clone of an original game, altered such that the player acting at the root joins a coalition with one or more players. (These terms are formally defined in section~\ref{sec:setup}.) A parallel game is intended to model counterfactual scenarios of coalition formation.

A \textit{cooperative equilibrium system} (CES) is a system from which no unilateral or multilateral credible deviation exists, compared to the corresponding parallel game. To elaborate, the following conditions must hold true in every subgame and every parallel game: Given an appropriate parallel game that is determined recursively, singleton players have no individual incentive to deviate; coalitional players have no joint incentives to deviate, as determined by the coalitional utility function; and each coalition is stable, meaning that its agents prefer their current coalition over any alternative coalition or becoming a singleton player.

\subsection{Illustrative example}

\begin{figure}
	\centering
	\begin{tikzpicture}[font=\footnotesize,edge from parent/.style={draw,thick}]
	% Two node styles: solid and hollow
	\tikzstyle{solid node}=[circle,draw,inner sep=1.2,fill=black];
	\tikzstyle{hollow node}=[circle,draw,inner sep=1.2];
	% Specify spacing for each level of the tree
	\tikzstyle{level 1}=[level distance=15mm,sibling distance=50mm]
	\tikzstyle{level 2}=[level distance=15mm,sibling distance=25mm]
	\tikzstyle{level 3}=[level distance=15mm,sibling distance=15mm]
	% The Tree
	\node(0)[hollow node]{}
	child{node[solid node]{}
		child{node[solid node]{}
			child{node[below]{$\begin{pmatrix}0\\60\\40\\0\end{pmatrix}$} edge from parent node[left]{$F$}}
			child{node[below]{$\begin{pmatrix}10\\30\\60\\0\end{pmatrix}$} edge from parent node[right]{$A$}}
			edge from parent node[above left]{$F$}
		}
		child{node[below]{$\begin{pmatrix}30\\40\\30\\0\end{pmatrix}$}
			edge from parent node[above right]{$A$}
		}
		edge from parent node[above left]{Enter Small Market}
	}
	child{node[solid node]{}
		child{node[below]{$\begin{pmatrix}20\\0\\0\\100\end{pmatrix}$}
			edge from parent node[above left]{$F$}
		}
		child{node[below]{$\begin{pmatrix}25\\0\\0\\95\end{pmatrix}$}
			edge from parent node[above right]{$A$}
		}
		edge from parent node[above right]{Enter Large Market}
	};
	% information sets
	%\draw[loosely dotted,very thick](0-1-1)to[out=-15,in=195](0-2-1);
	%\draw[loosely dotted,very thick](0-1-2)to[out=-15,in=195](0-2-2);
	% movers
	\node[above,yshift=2]at(0){Firm 1};
	%\foreach \i in {1,2} \node[above,yshift=2]at(0-\i){Firm 2};
	\node[left,yshift=2]at(0-1){Firm 2};
	\node[right,yshift=2]at(0-2){Firm 4};
	\node[left,yshift=2]at(0-1-1){Firm 3};
	\end{tikzpicture}	
	\caption{An international market entry game}
	\label{fig:market}	
\end{figure}

Figure~\ref{fig:market} depicts a stylized international market entry game. In this game, Firm 1 has the option to enter either a small (S) or a large (L) country's market. In the small country, the market structure consists of a leader (Firm 2) and a follower (Firm 3); in the large country, a monopolist (Firm 4) operates. Firms in both markets S and L have the options to either fight (\textit{F}) or accommodate (\textit{A}). The total market value is 100 units for market $S$ and 120 units for market $L$. The distribution of this value among the firms is as shown in the figure. Pre-entry profits for all firms are normalized to zero. If Firm 1 opts to enter a market in either country, then the incumbent firm in that market reciprocally gains access to Firm 1's own market. Consequently, all else being equal, Firm 1's entry would be advantageous for the local firms. However, the distribution of these additional gains is subject to the strategic choices of the firms involved.

%Fig 2
\begin{figure}
	\centering
		\begin{tikzpicture}[font=\footnotesize,edge from parent/.style={draw,thick}]
		% Two node styles: solid and hollow
		\tikzstyle{solid node}=[circle,draw,inner sep=1.2,fill=black];
		\tikzstyle{hollow node}=[circle,draw,inner sep=1.2];
		% Specify spacing for each level of the tree
		\tikzstyle{level 1}=[level distance=15mm,sibling distance=50mm]
		\tikzstyle{level 2}=[level distance=15mm,sibling distance=25mm]
		\tikzstyle{level 3}=[level distance=15mm,sibling distance=15mm]
		% The Tree
		\node[draw] at (-5,0.1) {A};
		\node(0)[hollow node]{}
		child{node[solid node]{}
			child{node[solid node]{}
				child{node[below]{$\begin{pmatrix}0\\60\\40\\0\end{pmatrix}$} edge from parent node[left]{$F$}}
				child{node[below]{$\begin{pmatrix}10\\30\\60\\0\end{pmatrix}$} edge from parent[->, thick] node[right]{$A$}}
				edge from parent node[above left]{$F$}
			}
			child{node[below]{$\begin{pmatrix}30\\40\\30\\0\end{pmatrix}$}
				edge from parent[->, thick] node[above right]{$A$}
			}
			edge from parent[->, thick] node[above left]{S}
		}
		child{node[solid node]{}
			child{node[below]{$\begin{pmatrix}20\\0\\0\\100\end{pmatrix}$}
				edge from parent[->, thick] node[above left]{$F$}
			}
			child{node[below]{$\begin{pmatrix}25\\0\\0\\95\end{pmatrix}$}
				edge from parent node[above right]{$A$}
			}
			edge from parent node[above right]{L}
		};
		% information sets
		%\draw[loosely dotted,very thick](0-1-1)to[out=-15,in=195](0-2-1);
		%\draw[loosely dotted,very thick](0-1-2)to[out=-15,in=195](0-2-2);
		% movers
		\node[above,yshift=2]at(0){1};
		%\foreach \i in {1,2} \node[above,yshift=2]at(0-\i){2};
		\node[left,yshift=2]at(0-1){2};
		\node[right,yshift=2]at(0-2){4};
		\node[left,yshift=2]at(0-1-1){3};
		\end{tikzpicture}		
		
		\begin{tikzpicture}[font=\footnotesize,edge from parent/.style={draw,thick}]
		% Two node styles: solid and hollow
		\tikzstyle{solid node}=[circle,draw,inner sep=1.2,fill=black];
		\tikzstyle{hollow node}=[circle,draw,inner sep=1.2];
		% Specify spacing for each level of the tree
		\tikzstyle{level 1}=[level distance=15mm,sibling distance=50mm]
		\tikzstyle{level 2}=[level distance=15mm,sibling distance=25mm]
		\tikzstyle{level 3}=[level distance=15mm,sibling distance=15mm]
		% The Tree
		\node[draw] at (-5,0.1) {B};
		\node(0)[hollow node]{}
		child{node[solid node]{}
			child{node[solid node]{}
				child{node[below]{$\begin{pmatrix}0\\60\\40\\0\end{pmatrix}$} edge from parent[->, thick] node[left]{$F$}}
				child{node[below]{$\begin{pmatrix}10\\30\\60\\0\end{pmatrix}$} edge from parent node[right]{$A$}}
				edge from parent[->, thick] node[above left]{$F$}
			}
			child{node[below]{$\begin{pmatrix}30\\40\\30\\0\end{pmatrix}$}
				edge from parent node[above right]{$A$}
			}
			edge from parent node[above left]{S}
		}
		child{node[solid node]{}
			child{node[below]{$\begin{pmatrix}20\\0\\0\\100\end{pmatrix}$}
				edge from parent[->, thick] node[above left]{$F$}
			}
			child{node[below]{$\begin{pmatrix}25\\0\\0\\95\end{pmatrix}$}
				edge from parent node[above right]{$A$}
			}
			edge from parent[->, thick] node[above right]{L}
		};
		% information sets
		%\draw[loosely dotted,very thick](0-1-1)to[out=-15,in=195](0-2-1);
		%\draw[loosely dotted,very thick](0-1-2)to[out=-15,in=195](0-2-2);
		% movers
		\node[above,yshift=2]at(0){1};
		%\foreach \i in {1,2} \node[above,yshift=2]at(0-\i){2};
		\node[left,yshift=2]at(0-1){2,3};
		\node[right,yshift=2]at(0-2){4};
		\node[left,yshift=2]at(0-1-1){2,3};
		\end{tikzpicture}	

		\begin{tikzpicture}[font=\footnotesize,edge from parent/.style={draw,thick}]
		% Two node styles: solid and hollow
		\tikzstyle{solid node}=[circle,draw,inner sep=1.2,fill=black];
		\tikzstyle{hollow node}=[circle,draw,inner sep=1.2];
		% Specify spacing for each level of the tree
		\tikzstyle{level 1}=[level distance=15mm,sibling distance=50mm]
		\tikzstyle{level 2}=[level distance=15mm,sibling distance=25mm]
		\tikzstyle{level 3}=[level distance=15mm,sibling distance=15mm]
		% The Tree
		\node[draw] at (-5,0.1) {C};
		\node(0)[hollow node]{}
		child{node[solid node]{}
			child{node[solid node]{}
				child{node[below]{$\begin{pmatrix}0\\60\\40\\0\end{pmatrix}$} edge from parent node[left]{$F$}}
				child{node[below]{$\begin{pmatrix}10\\30\\60\\0\end{pmatrix}$} edge from parent[->, thick] node[right]{$A$}}
				edge from parent node[above left]{$F$}
			}
			child{node[below]{$\begin{pmatrix}30\\40\\30\\0\end{pmatrix}$}
				edge from parent[->, thick] node[above right]{$A$}
			}
			edge from parent[->, thick] node[above left]{S}
		}
		child{node[solid node]{}
			child{node[below]{$\begin{pmatrix}20\\0\\0\\100\end{pmatrix}$}
				edge from parent[->, thick] node[above left]{$F$}
			}
			child{node[below]{$\begin{pmatrix}25\\0\\0\\95\end{pmatrix}$}
				edge from parent node[above right]{$A$}
			}
			edge from parent node[above right]{L}
		};
		% information sets
		%\draw[loosely dotted,very thick](0-1-1)to[out=-15,in=195](0-2-1);
		%\draw[loosely dotted,very thick](0-1-2)to[out=-15,in=195](0-2-2);
		% movers
		\node[above,yshift=2]at(0){1,2};
		%\foreach \i in {1,2} \node[above,yshift=2]at(0-\i){2};
		\node[left,yshift=2]at(0-1){1,2};
		\node[right,yshift=2]at(0-2){4};
		\node[left,yshift=2]at(0-1-1){3};
		\end{tikzpicture}	
	\caption{Illustration of a CES in three steps A--C (parallel games).}
	\label{fig:market_solution1}	
\end{figure}

The CES is derived through an algorithm, formally defined in section~\ref{sec:CES}. To explain the steps involved in arriving at a CES in a simplified context, I make two assumptions. Firstly, I assume the absence of any cost associated with coalition formation for the sake of this example. It should be noted, however, that even if forming a coalition incurs costs, firms will still engage in such alliances as long as the mutual benefits outweigh these costs. Secondly, for each non-empty coalition $C\subseteq \{1,2,3,4\}$, the coalitional utility function is defined as $u_C(\,\cdot\,)\coloneqq \sum_{i\in C}u_i(\,\cdot\,)$.\footnote{It is worth noting that the framework of cooperative strategic games, as introduced in section~\ref{sec:setup}, accommodates any form of von Neumann-Morgenstern coalitional preferences.}

Figure~\ref{fig:market_solution1} illustrates the CES for this market entry game in a three-step process. Step A starts with the standard subgame perfect Nash equilibrium (SPNE), in which each player acts independently and non-cooperatively. The outcome of this solution is (30, 40, 30, 0). Step B shows that Firm 2 and Firm 3 would both benefit from a collusion in which they both opt for action F, yielding the outcome (0, 60, 40, 0). In this case, the coalitional utility $u_{2,3}(0, 60, 40, 0)=100$ is maximized compared to $u_{2,3}(30, 40, 30, 0)=70$ and $u_{2,3}(10, 30, 60, 0)=90$.\footnote{The notation ``$2,3$'' signifies a coalition between Firm 2 and Firm 3.} Anticipating this collusion, Firm 1 opts to enter market L, leading to the outcome (20, 0, 0, 100). 

Step C shows that both Firm 1 and Firm 2 collude by best responding to the outcome in Step B. Specifically, Firm 1 chooses to enter market S and Firm 2 accommodates, leading to the outcome (30, 40, 30, 0), which is strictly better for both compared to the outcome in Step B. In response to the actions of Firm 1 and Firm 2, Firm 3 and Firm 4 choose the best-response actions A and F, respectively. Importantly, there are no additional opportunities for beneficial cooperation between any players.

As a result, the `on-path' CES of this game can be summarized as 
\[ [\{S\},\{A\},\{A\},\{F\};\{1,2\},3,4],\]
where 1 and 2 form a coalition, agent 1 chooses S, agent 2 chooses A, player 3 chooses A, and player 4 chooses F. The outcome of this CES is (30, 40, 30, 0), as depicted in Figure~\ref{fig:market_solution1} (C).

\subsubsection{The significance of coalitional credible threats}
\label{subsec:credible_threats}

At the outset, the collusion between Firm 1 and Firm 2 to obtain (30, 40, 30, 0) may appear self-evident, particularly because it coincides with the SPNE outcome. However, the following two modifications of the game illustrate that the collusion heavily relies on `off-path' credible threats.

First, suppose that, all else being equal, the outcome (20, 0, 0, 100) is changed to (20, 50, 0, 100). Recall that the reason Firm 2 colluded with Firm 1 in Step C was the credible threat of Firm 1 opting for market L. In this modified game, however, such a threat would no longer be credible. Firm 2 would have no incentive to collude with Firm 1 in Step C, given the new outcome specified in Step B, because $50>40$. Consequently, the CES in the modified game can be summarized by 
\[
[\{L\},\{F\},\{F\},\{F\};1,\{2,3\},4],
\]
whose outcome is (20, 50, 0, 100), even though the SPNE remains unchanged.

Second, suppose that the outcome (25, 0, 0, 95) is changed to (60, 0, 0, 70), holding everything else constant in the original game (Figure~\ref{fig:market}). In this modified game, the CES outcome would be (60, 0, 0, 70). This is because Firm 4, foreseeing the potential collusion between Firm 1 and Firm 2 in Step C, would extend an offer of collusion to Firm 1 in the new `Step D.' Firm 1 and Firm 4 would form a coalition, denoted $\{1,4\}$, because (i) both Firm 1 and Firm 4 are strictly better off compared to the outcome in Step C, and (ii) choosing L and A maximizes player $\{1,4\}$'s utility in Step D.\footnote{Importantly, the incentive for Firm 4 to collude with Firm 1 in Step D arises from the credible threat of collusion between Firm 1 and Firm 2 in Step C, which would result in an unfavorable outcome for Firm 4.} Therefore, the CES in this modified game can be summarized by 
\[
[\{L\},\{F\},\{F\},\{A\};\{1,4\},\{2,3\}],
\]
whose outcome is (60, 0, 0, 70).

These modifications underscore the importance of `off-path' credible coalitional threats in finding and sustaining the CES. This is akin to the role of the credibility of threats in non-cooperative games  \citep{selten1965,schelling1980,brams1994}.

\section{Related literature}
\label{sec:literature}

The study of connections between non-cooperative games and cooperative games, which abstract away from strategic interaction, has its origins in the seminal work of \citet{neumann1928}. In a lesser-known portion of this influential paper, von Neumann introduced the maximin solution in a three-person zero-sum game (see, p. 311). Notably, von Neumann observed that any two players within this three-person game could form a strategic coalition for mutual benefit, thereby destabilizing the maximin solution. Furthermore, he pointed out that the opportunities for strategic cooperation grow as the number of players increases. He concludes by stating, ``aber eine befriedigende allgemeine Theorie fehlt zur Zeit noch,'' which has been translated by Sonya Bargmann as
\begin{quote}
\centering
``\textit{a satisfactory general theory is as yet lacking}.''
\end{quote}
Von Neumann (1928) appears to be the first to articulate the issue of strategic cooperation as an open problem. His sustained interest in the subject came to light through his subsequent discussions with mathematician Kurt G{\"o}del. Evidence of these interactions can be found in G{\"o}del's personal notes, which have recently been published by \citet{plato2021}. In these notes, G{\"o}del observes:
\begin{quote}
``\textit{with games of more than two players, there does not always exist any reasonable statistical system of play}.''
\end{quote}

Following this early work, numerous studies---specifically in the subfields of repeated games and the farsighted approach to cooperative games---have provided valuable insights into the nature of cooperation. In the following subsections, I examine how the contributions of the current paper align with this extensive body of existing research. It is pertinent to note that the literature is vast and fragmented; I do not attempt to provide a comprehensive survey of it in this paper.

\subsection{Cooperation in repeated games and the Nash program}
\label{subsec:repeated_games}

The approach in this paper diverges from that of repeated games and the Nash program in several key aspects. The theory of repeated games primarily investigates whether cooperative outcomes in a one-shot game can be sustained as either Nash equilibria or subgame perfect equilibria when the game is repeated a sufficient number of times. Similarly, under the framework of the Nash program, the central question is whether it is feasible to construct a non-cooperative game in which the equilibrium outcome coincides with a cooperative solution. Thus, these frameworks are not primarily concerned with the cooperative behavior of players \textit{per se} but rather focus on non-cooperative behavior that yields cooperative \textit{outcomes}. Seminal works in this area include those of \citet{nash1953} and \citet{rubinstein1982}.
 
To elaborate, in repeated games, mutually beneficial outcomes are sustained as subgame perfect equilibria through the use of credible threats. However, this is accomplished in a completely non-cooperative manner---that is, players select their actions individually and independently at each stage. In contrast, in a cooperative strategic game, credible threats are used not only by individual players but also by coalitions of players who strategically collaborate and coordinate their actions.

\subsection{Cooperative approaches in non-cooperative games}
\label{subsec:cooperative}
A closely related strand of literature is the study of coalition-proofness in non-cooperative games, exemplified by concepts such as strong Nash equilibrium by \citet{aumann1959}, strong perfect equilibrium by \citet{rubinstein1980}, and coalition-proof Nash equilibrium by \citet{bernheim1987}. Roughly speaking, a strong Nash equilibrium is a Nash equilibrium in which no coalition can deviate in a manner that would benefit all of its members.  Coalition-proof Nash equilibrium is a weaker notion than strong Nash equilibrium; it requires that coalitional deviations---while holding the strategies of other players fixed---must be internally consistent, meaning that subcoalitions should not have an incentive to further deviate.

The most salient difference between these solutions and the CES presented in this paper lies in their relationship to Nash equilibrium. Specifically, these established equilibrium concepts refine the set of Nash equilibria to the extent that they may not always exist. In contrast, the CES is neither a refinement nor a coarsening of Nash equilibrium, but its existence is guaranteed, as shown in Theorem~\ref{thm:imperfect}.

Of note, as \citet[p. 7]{bernheim1987} themselves point out, coalitional deviations in coalition-proof Nash equilibrium are restrictive in the sense that deviating subcoalitions do not consider forming a coalition with non-deviating players:
\begin{quote}
``This rules out the possibility that some member of the deviating coalition might form a pact to deviate further with someone not included in this coalition. Such arrangements are clearly much more complex than those made entirely by members of the coalition itself... Clearly, further is required to resolve these issues in a fully satisfactory way.''
\end{quote}
These complexities regarding the strategic formation of coalitions are addressed in cooperative strategic games: every coalition structure is considered and may potentially emerge as part of the CES solution in such games.

Finally, another major difference lies in the nature of the predictions made by each solution concept. The concepts in the coalition-proofness literature predict a certain \textit{strategy profile} that is coalition-proof according to a specific notion. In contrast, the CES is formally a family of collections of strategy profiles and coalitions, where the prediction includes not only strategy profiles but also a set of `stable' coalitions. 

\subsection{Coalition formation and farsighted (non-cooperative) approaches to cooperative games}
\label{subsec:farsighted}

\citeauthor{harsanyi1974}'s (1974) seminal work in cooperative games has led to a recently burgeoning body of literature incorporating elements of non-cooperative games into cooperative games, such as farsightedness and backward induction. This integration has significantly enhanced our understanding of both frameworks and their interrelations. 

A vast literature exists which examines coalition formation, networks, team reasoning, and farsighted cooperative approaches in various contexts; see, for example, \citet{ichiishi1981}, \citet{moulin1982}, \citet{greenberg1990}, \citet{zhao1992}, \citet{sugden1993,sugden2003}, \citet{chwe1994}, \citet{perry1994}, \citet{moldovanu1995}, \citet{bloch1996}, \citet{ray1999}, \citet{bacharach1999}, \citet{maskin2003}, \citet*{brams2005}, \citet{bachrach2009}, \citet*{dutta2001}, \citet*{dutta2005}, \citet*{herings2006}, \citet{brandenburger2007}, \citet*{acemoglu2008}, \citet*{herings2009}, \citet{grabisch2010}, \citet{jackson2008}, \citet*{jackson1996}, \citet{karos2021}, and \citet{kimya2020}. For an informative and extensive review of the literature in economics and computer science, see, e.g., \citet{ray2007}, \citet{ray2015}, \citet*{chalkiadakis2011}, \citet{dafoe2020}, \citet*{elkind2013} and \citet{aziz2016}, along with the references therein.

As mentioned in the introduction, similar to Bernheim et al. (1987), \citet[p. 33]{ray1997} note that their study is limited to `internally consistent' deviations: 
\begin{quote}
    ``We must state at the outset that our treatment is limited by the assumption that agreements can be written only between members of an existing coalition; once a coalition breaks away from a larger coalition it cannot forge an agreement with any member of its complement. Thus, deviations can only serve to make an existing coalition  structure finer---never coarser.''
\end{quote}
\citet{diamantoudi2007} extend Ray and Vohra's (1997) equilibrium binding agreement notion by characterizing it with von Neumann-Morgenstern (vNM) stable sets in an appropriate cooperative game, and then allowing for any internal or external deviations. As is well-known, vNM stable sets may not exist in general, and this also applies to the extended equilibrium binding agreement notion of \citet{diamantoudi2007}. In addition, both Ray and Vohra (1997) and Diamantoudi and Xue (2007) study a consistency notion that is not based on extensive form representation, and they assume that contracts are costless in their models. They also show that inefficiency issues are not resolved by coalition formation, nor are they resolved in cooperative strategic games. For further details and an extension of these models, I refer the interested reader to `Equilibrium Process of Coalition Formation' in \citet{ray2015}, which provides an extensive survey of the recent literature. 

Starting from a standard extensive form game with perfect information, \citet{chander2020} construct a cooperative game and introduce the notion of the subgame perfect core; like the standard core, it may be empty. Their main finding contributes to the Nash program: If the subgame perfect core of a perfect information game is non-empty, then it can be implemented as an SPNE payoff of a relevant non-cooperative game. As discussed with respect to \citeauthor{bernheim1987}'s (1987) and \citeauthor{ray1997}'s (1997) concepts, \citeauthor{chander2020}'s (2020) concept assumes that once a coalition forms, the remaining players stay as independent players and do not `regroup'; though, \citeauthor{chander2020} do briefly discuss a special case in which the remaining players form a single coalition, as discussed in \citet{maskin2003}. In addition, \citeauthor{chander2020} discuss extending their analysis to more general extensive form games. They define the notion of a subgame perfect strong Nash equilibrium in general extensive form games, which is a strong Nash equilibrium in every subgame of the game and is akin to the strong perfect equilibrium of \citet{rubinstein1980}.

\citet{aumann1988} were the first to introduce endogenous network formation games in a strategic setting. Given a cooperative game, they construct an auxiliary link formation game, which is a non-cooperative extensive form game with perfect information, in which pairs of players are offered the opportunity to form links in a sequential order.\footnote{For more details on this model, see e.g. \citet{nouweland2005}} Subgame perfect equilibrium links in this non-cooperative extensive form game with perfect information give rise to a `cooperation structure' \`a la \citet{myerson1977}, and players receive their Myerson value in the cooperative game restricted to this cooperation structure. Because such link formation games with perfect information are a subset of non-cooperative extensive form games, they are also a subset of cooperative extensive form games. To see this, note that any non-cooperative link formation game is non-cooperative in nature---i.e., players choose their actions independently. The labels of the actions---whether they are called `links' between two players or named alphabetically---do not matter in terms of the predictions. An interesting direction for future research would be to extend non-cooperative link formation games to more general cooperative strategic games and explore whether the CES predictions coincide with cooperative solutions such as the Shapley and Myerson values. The network formation literature is not restricted to non-cooperative link-formation games; there is a vast literature on general network formation games, which is distinct from but closely related to the coalition formation literature.\footnote{For more details, see the references provided earlier in this subsection; e.g., \citet{jackson2008}.}

The current paper mainly differs from the above literature in two respects: the framework and the solution concept. First, various setups used in this literature are not directly comparable to either standard extensive form games or cooperative strategic games. This is in part due to the `cyclic' behavior in coalition formation frameworks and the fact that more than one player or coalition can choose an `action' at a given decision node, which cannot occur in extensive form games.\footnote{Note that this is not unusual because the frameworks in this literature emerged from cooperative games.} Second, in cases where a coalition formation setup such as \citeauthor{kimya2020}'s (2020) is comparable to an extensive form game with perfect information, the solution concept in question generally coincides with standard non-cooperative concepts such as the backward induction. This is not surprising because the main idea is to incorporate non-cooperative notions into cooperative games, as first proposed by \citet{harsanyi1974}.

\subsection{The differences between cooperative strategic games and non-cooperative games}
\label{subsec:coalitional_vs_noncoop}

Cooperative strategic games differ from non-cooperative games in three main respects: (i) their philosophical and conceptual underpinnings, (ii) the employed solution concept, and (iii) the mathematical framework that structures them.

Firstly, in cooperative strategic games, the fundamental assumption is that players can form coalitions to choose and coordinate their actions. This contrasts with the stance in non-cooperative games, as outlined by \citet[p. 286]{nash1951}: ``Our theory, in contradistinction, is based on the absence of coalitions in that it is assumed that each participant acts independently, without collaboration or communication with any of the others.'' It is worth noting that, in cooperative strategic games, should conditions permit the enforcement of non-cooperation among certain players, this would be formally integrated into the framework. Secondly, while the CES concept builds on the equilibrium ideas of \citet{cournot1838}, \citet{neumann1928}, \citet{neumann1944}, \citet{nash1951}, and \citet{selten1965}, it is neither a refinement nor a coarsening of Nash equilibrium. This is unsurprising because it is well known that Nash equilibrium is generally not immune to coalitional deviations. Thirdly, the framework of cooperative strategic games includes non-cooperative extensive form games as a special case (see Remark~\ref{rem:generalize}).

\subsection{Computational aspects}
\label{subsec:computation}

In $n$-person cooperative extensive form games, the computation of a CES becomes increasingly complex as the number of players, $n$, increases. Specifically, the number of subgames and parallel games requiring solution grows exponentially with $n$. This complexity arises because one must solve not only parallel games but also every conceivable subgame within those parallel games, as well as every parallel game of those subgames, and so forth.\footnote{For formal definitions of these concepts, refer to section~\ref{sec:setup}.} In the most generalized case where every coalition is feasible and all utilities depend on the partition of players, it is clear that computing a CES is at least as computationally challenging as computing a non-cooperative SPNE. Furthermore, even the task of verifying whether a given system qualifies as a CES presents its own set of challenges. Unlike a Nash equilibrium, which comprises a single strategy profile, a CES is characterized by a family of collections of strategy profiles and a family of collections of coalitions.

In the special case where no coalitions (except the singletons) are feasible, standard hardness results in non-cooperative games apply; see, e.g., \citet{gilboa1989}, \citet{papadimitriou1994}, \citet*{daskalakis2009}, and \citet{etessami2010}. This is because cooperative extensive form games reduce to standard non-cooperative games in this case.

\section{The setup and the solution concept}
\label{sec:setup}

\begin{table}[h!]
\[
\arraycolsep=1.1pt\def\arraystretch{1.4}
\begin{array}{ r|c|c|c|}
\multicolumn{1}{r}{}
&  \multicolumn{1}{c}{\text{Name}}
& \multicolumn{1}{c}{\text{Notation}}
& \multicolumn{1}{c}{\text{Element}} \\
\cline{2-4}
& \text{Agents} & N=\{1,2,\ldots,n\} & i \\
& \text{Players or coalitions} & P\subseteq 2^N & i \text{ or } C \\
& ~ \text{Players, including a non-singleton coalition } C ~ ~& P_C &  \\
& \text{Information sets} & H & h \\
& \text{Non-terminal nodes} & X & x \\
& \text{Terminal nodes} & Z & z \\
& \text{Game tree} & T = X \cup Z &  \\
& \text{Player function} & I:X\to N &  \\
& \text{Player}~i\text{'s actions at } h & A_i(h) & a_i(h) \\
& \text{Mixed strategy profiles} & S & s \\
& \text{Behavior strategies} & \bigtimes_{h_i\in H_i}\Delta(A(h_i)) & b_i \\
& \text{Utility of player } C \text{ given } P & u_C(\,\cdot\,|\,P) &  \\
& \text{Cooperative extensive form game} & ~~ \Gamma=(P, T , I, u, S, H) ~~&  \\

& \text{Subgame at } x & \Gamma(x) &  \\
& \text{Subtree at } h & T(h) &   \\
& \text{Root of subgame containing } h & root(h) &  \\
& \text{Set of all subgames at } root(h) & G(h) & g_h \\
& \text{Immediate successors of } x & f(x) &  \\
& \text{Information set following } h & f' (h) &  \\
& \text{All successors of } x & F(x) &  \\
& \text{Parallel game of } \Gamma(x) & \Gamma_{P_C} &  \\
& \text{Parallel tree of } T(h) & T_{P_C} &  \\
& \text{Strategy system} & \sigma=(s_{h,g})_{h\in H,g\in G(h)} &  \\
& \text{Strategy profile in subgame } g\in G(h) &  \sigma(h, g) & \\
& \text{Partition system} & \pi =(\pi_{h,g})_{h\in H,g\in G(h)} &  \\
& \text{Reference system at } h \text{ indexed by } j & r_j(h) &  \\
& \text{System} & (\sigma, \pi) &  \\
\cline{2-4}
\end{array}
\]
\caption{Cooperative extensive form game notation}
\label{table:terminology}
\end{table}

\subsection{The setup}
Let $\Gamma=(P, T , I, u, S, H)$ denote a \textit{cooperative extensive form game} with perfect recall and with possibly imperfect information, which is a standard extensive form game with an addition of coalitional utility function for each feasible coalition, as explained below. $\Gamma$ will also be referred to as a \textit{cooperative strategic game} or coalitional extensive form game.\footnote{I refer to standard textbooks such as \citet{fudenberg1991}, whose notation I mostly adapt, for details about (non-cooperative) extensive form games.} The notation and terminology used in this paper are summarized in Table~\ref{table:terminology}.

\vspace{0.2cm}
\noindent\textbf{Players:}
Let $P\subseteq 2^N$ be a non-empty finite set of players, where $N=\{1,2,...,n\}$. Each element of $P$ is either called a \textit{coalition}, denoted by $C\in P$, or a \textit{player}, denoted by $i\in P$. With a slight abuse of notation, the singleton coalition $\{i\}$ is represented as $i$.  Each $i\in N$ has a finite set of pure actions $A_i$.

\begin{remark}
The terms, player and coalition, are used interchangeably throughout the paper. 
\end{remark}

During the game, two or more players in $P$ may cooperate to create a new player, as follows.

\vspace{0.2cm}
\noindent\textbf{Forming a coalition:} 
If, for some $k$, players $i_1, i_2,...,i_k$ form a coalition $C=\cup_{j=1}^{k} i_j$, then each singleton in $C$ becomes an \textit{agent} of coalition $C$, which is also called player $C$.\footnote{If a player does not want to join a coalition, then the coalition cannot form---i.e., any member can veto the formation of the coalition they belong. For different assumptions on coalition formation, see section~\ref{sec:modification}.} If each $i_j$ is a singleton, then $C=\{i_1, i_2,...,i_k\}$.

\begin{remark}
As previously emphasized in section~\ref{sec:informal_discussion}, the formation of coalitions may entail monetary and/or psychological costs. Agents maintain the option to dissolve their coalitions even in the presence of a binding contract. However, the termination of such contracts could also entail both monetary and psychological costs. Importantly, players have the freedom to opt for actions that may be deemed `irrational.' While the framework itself imposes no explicit constraints on player behavior, the solution concept---set to be defined in section~\ref{sec:CES}---does introduce such limitations.
\end{remark}

\begin{remark}[Incomplete information games] 
To circumvent additional complications in notation, I do not separately introduce the notation for (Bayesian) cooperative strategic games with incomplete information. These can be considered as cooperative strategic games with complete but imperfect information \'a la \citet{harsanyi1967}, where nature chooses the types at the beginning of the game.
\end{remark}

\vspace{0.2cm}
\noindent\textbf{Utility functions:} 
Let $T=X\cup Z$ denote a game tree, $x_0$ the root of the game tree, $z\in Z$ a terminal node, defined as a node that is not a predecessor of any other node, $x\in X$ a non-terminal node in the tree, and $\bar{x}=|X|$ the cardinality of $X$ (for an illustration, see Figure~\ref{fig:3pl_standard_game} in the Appendix). Let $\overline{C}\subseteq 2^N$ be the set of \textit{feasible} coalitions, which includes every $C\in P$, and let $\overline{P}$ be the set of \textit{feasible partitions} where each (player) partition $P'\in\overline{P}$ contains only feasible coalitions. The interpretation of a feasible coalition $C\in\overline{C}$ is that its members have the potential to form $C$, but this  does not necessarily imply that its members have a preference for forming $C$.\footnote{It is possible that a feasible partition $P'\in\overline{P}$ is equal to any partition of e.g. $\{1,\{1,2\}, 3\}$ in a subgame of $\Gamma$ where $\{1,2\}$ is treated as a distinct player. The underlying intuition is that a player (e.g., player 1) may prefer to stay independent at some information sets, while preferring to form a coalition at other information sets within a given subgame.}

Let $u$ denote a profile of utility functions for each feasible coalition and partition. Let $u_C(\,\cdot\,|P'):Z\to\mathbb{R}$ denote the payoff function of feasible coalition $C\in\overline{C}$ given feasible partition $P'\in\overline{P}$. For each terminal node $z$, $u_C(z|P')$ denotes the von Neumann-Morgenstern utility of player $C$ if $z$ is reached, given the partition of players $P'$. With a slight abuse of notation, $u_C(\,\cdot\,|P')$ is abbreviated as $u_{C|P'}(\,\cdot\,)$.

\begin{remark}
As previously noted, if a coalition is not feasible, then there is no utility function for that coalition.\footnote{Alternatively, the individual utility from an infeasible coalition could be defined as minus infinity.}
\end{remark}

\begin{remark}
Although a coalitional utility function may generally be any vNM utility function, in some situations one could imagine that the coalitional utility may originate from a cooperative game.
\end{remark}

For a player $i\in P$, the utility given a player partition $P'$ is denoted as $u_{i}(\,\cdot\,|P')$. Note that $u_{i}(\,\cdot\,|P)$ is not necessarily equivalent to $u_i(\,\cdot\,|P')$, where $P'\neq P$, due to the potential for externalities---either positive or negative---and synergies that may arise when a player joins a coalition. For example, a player $i$ may derive different utility from an outcome depending on whether $i$ forms a coalition with player $j$ as opposed to joining a coalition with $j'\neq j$. (More generally, the formation of a coalition could also impact the utility of players external to the coalition.) This distinction will be useful in determining which coalitions are `individually rational.' Upon joining a coalition $C$, the strategic decisions within that coalition are guided by its utility function $u_C$ of the coalition $C$, whereas player $i$ makes the decision to join a coalition based on the individual utility function $u_{i}$. 

\begin{remark}[The interpretation of forming a coalition]
When forming a coalition $C$, the agents within player $C$ select their strategy based on the utility function of player $C$. For an illustration of the formation of a coalition, refer to Figure~\ref{fig:coalition-forming} in the Appendix.
\end{remark}

\begin{remark}
\label{rem:cost}
Any psychological or monetary costs incurred due to the formation of a coalition are incorporated into the utility functions, given the player partition. More specifically, there could be a fixed cost of forming a coalition.
\end{remark}

\vspace{0.2cm}
\noindent\textbf{Strategies:} Consider a player function $I:X\to\mathbb{N}$, where $I(x)$ specifies the `active' player at node $x$. Define $A(x)$ as the set of pure actions available at this node. 

Let $h\in H$ and $h(x)$ denote an information set and the information set at node $x$, respectively. It is conceivable that another node $x'\neq x$ may exist such that $x'\in h(x)$. In this scenario, the active player at $h(x)$ does not know whether she is at $x$ or at $x'$. If $h(x)$ is a singleton, then, with a slight abuse of notation, $h(x)=x$. Moreover, if $x'\in h(x)$, then $A(x)=A(x')$.  For each player $i$, $H_i$ specifies player $i$'s set of all information sets.  Let $A(h)$ denote the set of pure actions at $h$ and $A_i=\bigcup_{h_i\in H_i}A(h_i)$ denote player $i$'s set of all pure actions.

A pure strategy, $s'_i\in S'_i$, of player $i$ is a function $s'_i:H_i\rightarrow A_i$ satisfying $s'_i(h_i)\in A(h_i)$ for all $h_i\in H_i$. The set of all pure strategies of $i$ is denoted by $S'_i=\bigtimes_{h_i\in H_i} A(h_i)$. A pure strategy profile is represented by $s'\in S'$.

Let $\Delta (A(h_i))$ denote the set of probability distributions over $A(h_i)$, $b_i\in \bigtimes_{h_i\in H_i}\Delta (A(h_i))$ a behavior strategy of $i$, and $b$ a profile of behavior strategies. The profile $s\in S$ denotes the mixed strategy profile that is equivalent to behavior strategy profile $b$ in the sense of \citet{kuhn1953}. I assume that $\Gamma$ is common knowledge.

Let $u_{C|P'}(s)$ denote the von Neumann-Morgenstern expected utility (or payoff) for player $C$ given $P'$. The \textit{outcome} of a mixed strategy profile $s$ given some partition $P'$ is a profile of expected utilities, $(u_C(s|P'))_{C\in P'}$, one for each player in $P'$. 

As standard in the literature, a deviation from a given strategy profile by an individual player is called a \textit{unilateral} deviation. A deviation by a non-singleton coalition is called a \textit{multilateral} deviation.

\begin{remark}
When two or more agents form a coalition, $C$, their information sets merge. In incomplete information games, their types are also revealed to each other.
\end{remark}
 
\vspace{0.2cm}
\noindent\textbf{Subgames and subtrees:} 
A \textit{subgame} $G$ of a game $\Gamma$ is the game $\Gamma$ restricted to an information set with a singleton node and all of its successors in $\Gamma$. The subgame at a node $x$ is denoted by $\Gamma(x)$. Analogously, the set of players in subgame $\Gamma(x)$ is denoted by $P(x)$. At the root of a game, the subgame is the game itself. Let $G$ be a subgame and $H'_i$ be the set of $i$'s information sets in the subgame. Then, for every $h_i\in H'_i$ function $s_i(\,\cdot\,|h_i)$ denotes the restriction of strategy $s_{i}$ to $h_i$ in the subgame $G$. If all information sets are singletons, then $\Gamma$ is called a game of perfect information.

A \textit{subtree} $T$ of a game $\Gamma$ is defined as  the game tree of $\Gamma$ restricted to a specific  information set and all of its successors in $\Gamma$. It is important to note that the root of such a subtree may be a non-singleton information set.\footnote{Note also that the root of an information set $h$ is in general different from the root of a subtree $T(h)$.} For a given  information set $h$, $T(h)$ denotes the subtree whose root is $h$. A subgame is a special case of a subtree whose root is singleton.

Let $f(x)$ and $F(x)$ be correspondences that denote the set of immediate successors and all successors of a node $x$ (excluding $x$ in both definitions), respectively. Let $root(h)$ denote the node that is the root of the subgame containing information set $h$, such that there is no other subgame starting at an information set between $root(h)$ and $h$---i.e., $root(h)$ is the closest singleton `ancestor' of $h$. Note that in perfect information games, $x=root(x)$ for all nodes $x\in X$. $F(h)$ and $f(h)$ denote the set of all successor information sets and the set of immediate successor information sets of information set $h$ (excluding $h$ in both definitions), respectively. 

In certain games with imperfect information, the notion of a `preceding' set of information sets for a given $h$ within a single subtree can be ambiguous. In imperfect information games, define the correspondences $F'$ and $f'$ as extensions of $F$ and $f$, respectively. These extensions satisfy the following conditions. If $h=x$, i.e., $h$ is a singleton, then $F'(h)=F(h)$. This condition ensures that the extended correspondence $F'$ coincides with the successor correspondence $F$ in singleton information sets. For any $h'\in F(h)$,  $h\notin F'(h')$. In words, if $h'$ is a successor of $h$, then $h$ cannot be a successor of $h'$ according to $F'$. Similarly, if $h=x$, then $f'(h)=f(h)$. And, if $h$ is not singleton, then $f'(h) = F'(h) \setminus F'(\hat{h})$, where $\hat{h}\in f'(h)$. In words, in non-singleton information sets, an immediate successor of $h$ according to $f'$ cannot be in the set of successors of the immediate successor of $h$. These conditions are natural extensions of the original correspondences. It is worth noting that in games with perfect information, $F'$ and $f'$ reduce to $F$ and $f$, respectively.

\vspace{0.2cm}
\noindent\textbf{Parallel games and parallel trees:} 
Let $\Gamma$ be a game, $x\in X$ a non-terminal node, $i=I(x)$ the active player at $x$, $\Gamma(x)$ the subgame starting at $x$, and $P'$ the set of players in $\Gamma(x)$, which is a subset of $2^N$, where $N=\{1,2,...,n\}$. Let $h\in H$ be a non-terminal history, which is equal to $x$ when $\Gamma(x)$ is a game of perfect information.

For a non-singleton coalition $C$ of players in $P'$, define $P_C=\{C'\in P'|C'\notin C\}\cup C$, i.e., the player partition in which only the agents in $C$ form a coalition. For example, if $P'=\{1,2,3,4,5,6\}$ and $C=\{1,2,4\}$, then $P_C=\{\{1,2,4\},3,5,6\}$. For a player $i$, define $FC_i(h)=\{C'\in\overline{C}~|~i\in C', C'\subseteq \bigcup_{h\in F(h)}I(h)\}$, i.e., the set of all feasible coalitions $i$ can form with players who act after $i$.\footnote{In section~\ref{sec:modification}, I discuss the extension of this model to the case in which coalitions can form more generally. For example, it is possible that an `outside' player who is not part of the game, i.e., who has no action to choose from, may form a coalition with some of the players in the game. While this may be a reasonable assumption in some strategic situations, I do not include this in the baseline model, for the sake of simplicity.} As an example, suppose that $P'=\{1,2,3\}$, $h=x_0$, $I(x_0)=i$, and $i=1$. Then, $FC_1=\{\{1,2\},\{1,3\},\{1,2,3\}\}$.

A \textit{parallel game} of $\Gamma(x)$, denoted by $\Gamma_{P_C}$, is the game in which the set of players $P'$ is replaced with the set of players given by the partition $P_C$ where $C\in FC_i(x)$. The set of all parallel games is defined as follows:
\[
par(\Gamma(x)) = \{\Gamma_{P_C} | C\in FC_i(x)\}.
\]
For an illustration of parallel games of a game, see Figure~\ref{fig:parallelgames} in the Appendix. Note that if the original game is an $n$-player game, then its parallel game $\Gamma_{P_C}$ is an $(n-|C|+1)$-player game. Note also that each $j\in C$ acts as an agent of a single player $C$ in $\Gamma_{P_C}$. For example, if $\Gamma$ is a six-player game, $1=I(x_0)$ (player 1 acts at the root), and $P'=\{1,2,3,4,5,6\}$, then $\Gamma_{P_C}$, where $C=\{1,2,4\}$, is the game in which the players are $\{1,2,4\},3,5$, and $6$. $1,2,$ and $4$ are the agents of coalitional player $\{1,2,4\}$.

\begin{remark}
A subgame or parallel game of a cooperative extensive form game is a cooperative extensive form game (not a non-cooperative game).
\end{remark}

Let $\Gamma(h)$ be the subgame starting at $root(h)$, $i\in I(h)$, and $T(h)$ be a subtree of game $\Gamma(h)$.  A \textit{parallel tree} of $T(h)$, denoted by $T_{P_C}$, is a subtree of $\Gamma(h)$ in which the set of players $P'$ is replaced with the set of players given by the partition $P_C$ where $C\in FC_i(h)$. The set of all parallel trees is defined as follows:
\[
par(T(h)) = \{T_{P_C} | C\in FC_i(h)\}.
\]

\vspace{0.2cm}
\noindent\textbf{Systems:} 
A \textit{system} or a `solution profile' is a pair $(\sigma,\pi)$ which is defined as follows. Let $G(h)$ be the set of all subgames of the subgame starting at $root(h)$ for some information set $h$. For example, if $h$ is the root of game $\Gamma$, then $G(h)$ is the set of all subgames of $\Gamma$, including itself. For a given $h$, let $S_{|g_h}$ denote the set of all strategy profiles in subgame $g_h\in G(h)$.

Let $\sigma$ be a \textit{strategy system} defined as a mapping that maps the pair $(h,g_h)$, where $h\in H$ and $g_h\in G(h)$, to a strategy profile $s_{h,g}\in S_{|g_h}$:
\[(h,g_h)\mapsto s_{h,g_h}.\] 
Note that $g$ depends on $h$. For simplicity, subscripted notation $g_h$ will not be used from now on. 

For a given $h$, $\sigma(h,g)$ denotes a strategy profile in some subgame $g\in G(h)$. Note that function $\sigma(h,\,\cdot\,)$ gives for each information set a collection of (possibly different) strategy profiles, one for each successor subgame:  $\sigma(h,\,\cdot\,)=(s_{g})_{g\in G(h)}$. More compactly, $\sigma\coloneqq(s_{h,g})_{h\in H,g\in G(h)}$. In summary, $\sigma$ is a family of collections of strategy profiles.

\begin{remark}
Let $h\in H$ be an information set and $g\in G(h)$. Suppose that $g'$ is a subgame of $g$. In general, $\sigma(h,g|g')\neq \sigma(h,g') $. In other words, $\sigma(h,g')$ is not necessarily equal to the restriction of the strategy profile $\sigma(h,g)$ to the subgame starting at the root of $g'$. This is because a set of agents may prefer not to form a coalition in $g'$ but prefer to form a coalition in $g$, which would affect the choices in its subgame $g'$. Thus, the strategy profile chosen in $g'$ can be different than the strategy profile chosen in $g$ restricted to subgame $g'$.
\end{remark}

Similarly, let $\pi$ be a \textit{partition system} defined as a mapping that maps the pair $(h,g)$, where $h\in H$ and $g\in G(h)$, to a feasible partition $\pi_{h,g}$ in $\overline{P}$ restricted to subgame $g$:
\[(h,g)\mapsto \pi_{h,g}.\] 

In other words, given some $h$, $\pi(h,g)$ assigns a feasible coalition to each information set in subgame $g\in G(h)$. Let $\pi(h,\,\cdot\,)\coloneqq(\pi_{g})_{g\in G(h)}$ and $\pi\coloneqq(\pi_{h,g})_{h\in H,g\in G(h)}$. Function $\pi(h,\,\cdot\,)$ gives for each information set a collection of (possibly different) partition of players, one for each successor subgame. In summary, $\pi$ is a family of collections of player partitions.

The interpretation of a pair $(\sigma(h,g),\pi(h,g))$ is as follows. Let $C$ be some coalition in $\pi(h,g)$. Each agent $i$ of player $C$ chooses their actions in $g$ based on the strategy profile $\sigma(h,g)$. Their actions are guided by the coalitional utility function $u_C(\,\cdot\,|\pi(h,g))$ in the subgame $g$. In summary, a system is a complete description of what actions will be played by which coalitions at every subtree and every parallel tree in the original game, every subtree of every parallel tree, and every parallel tree of every subtree, and so on. For an illustration of a system in a three-player game, refer to Figure~\ref{fig:systems} in the Appendix.

For a given $h$, let $[\sigma(h), \pi(h)]$ denote the restriction of $(\sigma,\pi)$ to the subgame starting at $h$, $\Gamma(h)$. Let $[\sigma, \pi]$ be a system defined in $\Gamma(x)$ for some $x\in X$. For each $x'\in f(x)$, let $[\sigma'(x'), \pi'(x')]$ be a system. Then, $[\sigma, \pi]$ is called an \textit{extension} of $[\sigma', \pi']$ to $x$ if for every $x'\in f(x)$,  $[\sigma(x'), \pi(x')]=[\sigma'(x'), \pi'(x')]$. Note that this implies that $[\sigma, \pi]$ and $[\sigma', \pi']$ coincide for every $x''\in F(x)$.

The \textit{outcome} of a system $(\sigma,\pi)$ in a game $\Gamma$ is defined as the outcome of the (possibly mixed) strategy profile $\sigma(x_0,g_{x_0})$ with respect to the players in the partition $\pi(x_0,g_{x_0})$ where $g_{x_0}=\Gamma$.

\vspace{0.2cm}
\noindent\textbf{Non-cooperative solution concepts:} 
Let $\Gamma=(P, T , I, u, S, H)$ be a cooperative strategic game. A mixed strategy profile $s$ is called a Nash equilibrium if for every player $j\in P$, $s_j\in \argmax_{s'_j\in S'_j} u_j(s'_j,s_{-j})$. A mixed strategy profile $s$ is a called a subgame perfect Nash equilibrium (SPNE) if for every subgame $G$ of $\Gamma$ the restriction of $s$ to subgame $G$ is a Nash equilibrium in $G$.

\begin{remark}[Generalization of non-cooperative games]
\label{rem:generalize}
Cooperative extensive form games generalize non-cooperative extensive form games. Specifically, every non-cooperative extensive form game can be considered a special case of a cooperative extensive form game. If no coalition involving more than one player is feasible, then a cooperative  extensive form game would reduce to a non-cooperative extensive form game.
\end{remark}

\section{Cooperative equilibrium system}
\label{sec:CES}

\subsection{Perfect information games}
\label{sec:CES_perfect_info}

\subsubsection{Definition}
\label{sec:CES_def}

Let $\Gamma$ be a cooperative strategic game with perfect information and without chance moves. Let $(\sigma,\pi)$ be a system of $\Gamma$, $x\in X$ be a non-terminal node, and $i=I(x)$. At $x$, consider parallel game $\Gamma_{P_C}\in par(\Gamma(x))$ where $C\ni i$ is a feasible non-singleton coalition. For the rest of this subsection, associate the system $[\tau^{C}(x),\pi^{C}(x)]$ with each parallel game $\Gamma_{P_C}$.

\begin{definition}[Non-cooperative reference system]
\label{def:autarky}
Let $[\sigma(x'), \pi(x')]$ be a system where $x'\in f(x)$. The \textit{non-cooperative reference system} at $x$, denoted as $r_0(x)$, with respect to $[\sigma(x'), \pi(x')]$ is defined as follows. Let $b^*_i(x)$  be a utility-maximizing behavior strategy of $i$ at $x$ given $[\sigma(x'),\pi(x')]$ where $x'\in f(x)$:
\[
b_i(x)\in \argmax_{b'_i(x)\in \Delta A_i(x)} u_{i}(b'_i(x)|\sigma(x'),\pi(x')).
\]
Define $r_0(x)\coloneqq[\tau^{i}(x),\pi^{i}(x)]$, which is the extension of $[\sigma(x'), \pi(x')]$ to $\Gamma(x)$, the subgame starting at node $x$ where $i$ chooses the behavior strategy $b^*_i(x)$, and $\pi^{i}(x,\Gamma(x))$ is defined as the player partition in which $i$ acts non-cooperatively at $x$, and the other players in the partition are given by $\pi(x',\Gamma(x'))$ for every $x'\in f(x)$.
\end{definition}

In simple terms, $\tau^{i}(x,\Gamma(x))$ is defined as the strategy profile in which player $i$ non-cooperatively chooses behavior action $b^*_i(x)$. Furthermore, the non-cooperative reference system at node $x$ extends a given system to the node $x$, incorporating $\tau^{i}(x,\Gamma(x))$ and player $i$. The intuition behind the definition of the non-cooperative reference system is that if no player forms a coalition with the active player $i$ at $x$, then player $i$ would choose a utility-maximizing strategy non-cooperatively. 

Next, I define `reference systems' to facilitate a comparison between the systems of parallel games at $x$ and the non-cooperative choice made by the active player at $x$.

\begin{definition}[Reference systems]
\label{def:reference}
For each parallel game $\Gamma_{P_C}$ at $x$, fix system $[\tau^{C}(x),\pi^{C}(x)]$. Let $(r_j(x))^{\bar{j}}_{j=1}$ be a sequence for player $i$ where $r_j(x)\coloneqq[\tau^{C_j}(x),$ $\pi^{C_j}(x)]$ satisfying the following two conditions.
\begin{enumerate}
	\item For every two indices $j$ and $j'$ with $j<j'$, $u_{i}(\tau^{C_j}(x)|\pi^{C_{j}}(x))\leq $ $u_{i}(\tau^{C_{j'}}(x)|\pi^{C_{j'}}(x))$; i.e., the sequence of payoffs is nondecreasing for $i$. 
	\item If $u_{i}(\tau^{C_j}(x)|\pi^{C_{j}}(x)) $ $= u_{i}(\tau^{C_{j'}}(x)|\pi^{C_{j'}}(x))$ and $C_j\subsetneq C_{j'}$, then $r_j$ precedes $r_{j'}$ for parallel game coalitions $C_j$ and $C_{j'}$ containing $i$.
\end{enumerate}
Each element of sequence $(r_j(x))^{\bar{j}}_{j=0}$, which includes $r_0(x)$, is called a \textit{reference system}.
\end{definition}
For an illustration of reference systems in a three-player game, see Figure~\ref{fig:reference_points} in the Appendix. Notice that each reference system, as an object, is a system in the relevant parallel game, where player $i$ forms a coalition with some of the players. For a coalition to form, however, every member of the coalition must agree to it, hence the next definition.

\begin{definition}[Individually rational reference systems]
\label{def:IR}
Fix the sequence $(r_j(x))^{\bar{j}}_{j=1}$ of reference systems, including the non-cooperative reference system, for player $i$.
\begin{enumerate}
		\item\textbf{Base case}: 
		Define the non-cooperative  reference system $r_0(x)$ to be individually rational and denote it by $r^*_0(x)$. The following sequence of \textit{individually rational} (IR) reference systems (with respect to the system $[\tau^{C}(x),\pi^{C}(x)]$ for  every $\Gamma_{P_C}$) are defined inductively as follows.
		
		\item\textbf{Induction step}: Assume that $r_j(x)$ is IR, denoted as $r^*_k(x)$, for some $k\geq 0$. Next, $r^*_{k+1}(x)$ is defined as follows. Let $j'>j$ be the smallest number such that $r_{j'}(x)$ is IR with respect to $r^*_k(x)$; that is, for every agent $i'\in \bar{C}_{j'}$ where $C_{j'} \subseteq \bar{C}_{j'}$ and $\bar{C}_{j'}\in \pi^{C_{j'}}(x)$, $u_{i'}(\tau^{C_{j'}}(x)|\pi^{C_{j'}}(x)) > u_{i'}(\tau^{C_{j}}(x)|\pi^{C_{j}}(x))$.\footnote{In other words, $\bar{C}_{j'}$ contains coalition $C_{j'}$ such that $\bar{C}_{j'}$ is part of the solution of the parallel game $\Gamma_{P_{C_{j'}}}$.} Then, define $r^*_{k+1}(x)\coloneqq r_{j'}(x)$.
	\end{enumerate}
	The sequence of IR reference systems at node $x$ is denoted by $(r^*_j(x))^{\bar{l}}_{k=0}$ where $\bar{l}\geq 0$.
\end{definition}

In simple words, an IR reference system is a system of a parallel game $\Gamma_{P_C}$ for some coalition $C$, including agent $i$, where every member of $C$ agrees to form $C$, i.e., every member strictly prefers forming $C$ compared to the appropriate counterfactual, the next best IR reference system. Note that the sequence $(r^*_j(x))^{\bar{l}}_{k=0}$ of IR reference systems is a subsequence of the sequence $(r_j(x))^{\bar{j}}_{j=0}$ of reference systems.

\begin{definition}[Credibility]
\label{def:credibility}
Let $x\in X$ be a non-terminal node and $(r^*_j(x))^{\bar{l}}_{k=0}$ be a
sequence of IR reference systems at $x$. The IR reference system, $r^*_{\bar{l}}(x)$, that maximizes $i$'s utility is defined as the \textit{credible} reference system at $x$, with respect to systems $(r^*_j(x))^{\bar{l}}_{k=0}$.
\end{definition}

In simple words, the credible reference system is the IR reference system that maximizes player $i$'s utility at $x$ with respect to IR reference systems. For this reason, if a reference system is not credible, then it is said to possess either a unilateral or a multilateral \textit{credible deviation}. There are a few further points to note regarding this definition. First, given the sequence of IR reference systems, the credible reference system is unique. However, at node $x$, there may be two different sequences of IR reference systems, in each of which the credible reference systems may differ yet give the same utility to player $i$. This is not surprising as there can be games in which all payoffs are the same.  Second, note that $r^*_{\bar{l}}(x)=[\tau^{C_{\bar{l}}}(x),\pi^{C_{\bar{l}}}(x)]$, i.e., the credible reference system is a system of the parallel game with coalition $C_{\bar{l}}$. It may be that the non-cooperative reference system $r^*_0(x)$ where $\bar{l}=0$ is the only IR reference system (i.e., $C_{\bar{l}}$ is singleton) and hence the credible reference system. Third, also note that IR reference systems---and by extension, credibility---are defined with respect to some arbitrary systems of parallel games, which are themselves reference systems. There is no requirement whether these systems themselves satisfy the credibility requirement. CES, as defined next, incorporates a recursive notion of credibility.

\begin{definition}[Cooperative equilibrium system, CES]
	\label{def:CES}
	Let $\Gamma=(P, T , I, u, S, H)$ be a cooperative strategic game with perfect information and without chance moves.
	\begin{enumerate}
	\item Let $|P|=1$. A CES of $\Gamma$ is defined as a system $(\sigma^*,\pi^*)$ where $\pi^*=P$ and $\sigma^*\in \argmax_{s'_i\in S_i} u_i(s'_i)$ where $i\in P$.
	\item Let $|P|\geq 2$. A CES, $(\sigma^*,\pi^*)$, of $\Gamma$ is defined as follows. Let $x\in X$ be any non-terminal node. Every reference system except the non-cooperative reference system, $[\sigma^{*C}(x),\pi^{*C}(x)]$, is a CES of the parallel game $\Gamma_{P_C}$. The non-cooperative reference system at $x$ is defined with respect to the CES $[\sigma^*(x'), \pi^*(x')]$ where $x'\in f(x)$. Moreover, the IR reference system $r^*_{\bar{l}}(x)$ is the credible reference system with respect to $(r^*_j(x))^{\bar{l}}_{k=0}$ such that, for every $j$, $r^*_j(x)=[\sigma^{*C_j}(x),\pi^{*C_j}(x)]$.
	\end{enumerate}
\end{definition}

The recursive structure of the CES warrants special attention. While credibility is defined with respect to arbitrary systems, a CES prescribes a credible reference system to every non-terminal node $x$ with respect to cooperative equilibrium systems. In the context of one-person games, the notion of credibility simplifies to straightforward utility maximization. In $n$-person games, the CES is based on a notion of credibility with respect to systems of parallel games that are themselves credible. In turn, these systems are credible with respect to other systems that are also credible, and so on. While it is not immediately evident  whether a CES always exists in $n$-person games, a constructive algorithm for its computation is provided in the subsequent subsection.

I now turn to a simple observation about the relationship between CES and SPNE in non-cooperative extensive form games.

\begin{proposition}[CES in non-cooperative games]
\label{prop:CES_SPNE}
If a cooperative extensive form game, $\Gamma$, is also a non-cooperative extensive form game, then the strategies that constitute a CES coincide with those that form a subgame perfect equilibrium in $\Gamma$.
\end{proposition}

For the proof, see Appendix, section~\ref{subsec:proof3}.

\subsubsection{A constructive algorithm for finding cooperative equilibrium systems}
\label{subsec:RBI_perfect}

Let $\Gamma$ be a cooperative strategic game with perfect information and no chance moves. I use strong mathematical induction to define a constructive algorithm for finding a CES in $\Gamma$, which I will refer to as the CES algorithm. This algorithm outputs a system on each subgame $\Gamma(x)$, starting at some node $x$, by inducting on the number of players ($n$) and number of nodes (denoted as $\bar{x}$) in the successor nodes of $x$---including $x$ itself and excluding the terminal nodes.\footnote{The termination of the algorithm is indicated by the symbol $\blacksquare$.} For instance, if $x$ is a penultimate node, then $(n,\bar{x}) = (1,1)$. Let $i=I(x)$ be the active player at $x$.

\begin{enumerate}
	\item \textbf{Base case}: Let $(n,\bar{x}) = (1,1)$. A CES at $\Gamma(x)$ is defined by the pair $(\sigma^*,\pi^*)$, where $\sigma^*(x)\in \argmax_{s\in S} u_i(s)$ and $\pi^*=i$.
	
	\item \textbf{Induction step}: Assume that a CES is defined for all subgames with parameters $(m,\bar{y})$ where $1\leq m\leq n$, $1\leq\bar{y}\leq \bar{x}$, and such that $(n,\bar{x}) \neq (1,1)$ and $(m,\bar{y}) \neq (n,\bar{x})$. 
	
	Given this assumption, $[\sigma^*(x'),\pi^*(x')]$ is defined for all $x'\in F(x)$, where $x$ is the root of subgame $\Gamma(x)$. The solution can be extended to subgame $\Gamma(x)$ with parameters $(n,\bar{x})$ as follows.\footnote{I present steps (2a), (2b), and (2c) below to make the algorithm self-sufficient. For more details, see the definitions in section~\ref{sec:CES_def}.} Subsequently, the solutions of all parallel games are compared with the non-cooperative choice of $i$ at $x$.
	\begin{enumerate}
		\item The non-cooperative reference system at $x$, denoted $r_0(x)$: Let $b^*_i(x)$  be a utility-maximizing behavior strategy of $i$ at $x$ given $[\sigma^*(x'),\pi^*(x')]$:
		\[
		b^*_i(x)\in \argmax_{b'_i(x)\in \Delta A_i(x)} u_{i}(b'_i(x)|\sigma^*(x'),\pi^*(x')).
		\]
		Define $r_0(x)\coloneqq[\tau^{i}(x),\pi^{i}(x)]$, which is the extension of $[\sigma^*(x'), \pi^*(x')]$ where $x'\in f(x)$ to the subgame starting at node $x$. Here, $i$ chooses the behavior strategy $b^*_i(x)$, and $\pi^{i}(x)$ is the partition in which $i$ and other players in the subgame at $x$ act non-cooperatively.\footnote{Put differently, $\tau^{i}(x,x)$ is defined as the strategy profile in which $i$ chooses behavior action $b^*_i(x)$, and $\pi^{i}(x)$ is defined as the partition in which $i$ and other players in the subgame at $x$ act non-cooperatively. The rest follows the strategy profile and partitions given in the system $[\sigma^*(x'), \pi^*(x')]$ for all successors $x'\in F(x)$. Note that $\tau^{i}(x)=\tau^{i}(x,x')$  for all $x'\in F(x)$, meaning it provides a strategy profile for every subgame.}
				
		\item Reference systems: For any $(n,\bar{x})$, consider parallel game $\Gamma_{P_C}$ where $C\ni i$ is a feasible non-singleton coalition.  Since $\Gamma_{P_C}$ is an $(n-|C|+1)$-person subgame with $\bar{x}$ nodes, its CES, denoted by $[\tau^{C}(x),\pi^{C}(x)]$, is defined by the induction hypothesis.

		Let $(r_j(x))^{\bar{j}}_{j=0}$ be a sequence for $i$ where $r_j(x)\coloneqq[\tau^{C_j}(x),\pi^{C_j}(x)]$. The sequence satisfies: 
  
    \noindent (i) for every two indices $j$ and $j'$ with $j<j'$, we have $u_{i}(\tau^{C_j}(x)|\pi^{C_{j}}(x))\leq u_{i}(\tau^{C_{j'}}(x)|\pi^{C_{j'}}(x))$ (i.e., it is a nondecreasing sequence for $i$), and 
  
		\noindent (ii) if $u_{i}(\tau^{C_j}(x)|\pi^{C_{j}}(x)) $ $= u_{i}(\tau^{C_{j'}}(x)|\pi^{C_{j'}}(x))$ and $C_j\subsetneq C_{j'}$, then $r_j$ precedes $r_{j'}$, where $C_j$ and $C_{j'}$ are parallel game coalitions  containing $i$.\footnote{Ties are broken arbitrarily.}
		
		\item Individually rational (IR) reference systems: Given the non-cooperative reference system $r^*_0(x)$, the following IR reference systems are defined inductively.
		
		Assume that $r_j(x)$ is IR, denoted as $r^*_j(x)$, for some $j\geq 0$. Let $j'>j$ be the smallest index such that $r_{j'}(x)$ is IR with respect to  $r^*_j(x)$, that is, $u_{i'}(\tau^{C_{j'}}(x)|\pi^{C_{j'}}(x)) > u_{i'}(\tau^{C_{j}}(x)|\pi^{C_{j}}(x))$ for every agent $i'\in \bar{C}_{j'}$ where $C_{j'} \subseteq \bar{C}_{j'}$ and $\bar{C}_{j'}\in \pi^{C_{j'}}(x)$.\footnote{More explicitly, $\bar{C}_{j'}$ contains coalition $C_{j'}$ such that $\bar{C}_{j'}$ is part of the solution of the parallel game $\Gamma_{P_{C_{j'}}}$.}
		
		Let $r^*_{\bar{l}}(x)$ be the credible reference system that maximizes $i$'s utility at $x$. Note that $r^*_{\bar{l}}(x)=[\tau^{C_{\bar{l}}}(x),\pi^{C_{\bar{l}}}(x)]$. It is possible that $r^*_0(x)$ where $\bar{l}=0$ is the only IR reference system.
	\end{enumerate}
	
	A CES of $\Gamma(x)$ is then given by system $[\sigma^*, \pi^*]$ where $[\sigma^*(x), \pi^*(x)]=r^*_{\bar{l}}(x)$, which extends $[\sigma^*(x'), \pi^*(x')]$ where $x'\in f(x)$. When $x$ is the root of the game tree of $\Gamma$, $[\sigma^*, \pi^*]$ is a CES of $\Gamma$. $\blacksquare$
\end{enumerate}

I call coalitions \textit{stable} if they survive the CES algorithm and players and agents \textit{dynamically rational} if they utilize the CES algorithm in cooperative strategic games. For a fully worked-out example, see section~\ref{sec:worked-out_ex}. The following theorem asserts the existence of a CES.

\begin{theorem}[Existence: perfect information games]
	\label{thm:perfect}
A CES, denoted as $(\sigma^*, \pi^*)$, exists in every finite $n$-person cooperative strategic game with perfect information, where $\sigma^*$ consists solely of pure strategy profiles.
\end{theorem}

The proof of this theorem can be found in the Appendix, section~\ref{subsec:proof1}. 

\subsubsection{Informal description of the algorithm}
\label{subsec:informal_RBI_perfect}

I proceed to explain each step in the CES algorithm in as informal a manner as possible. As previously defined in section~\ref{subsec:RBI_perfect}, let $\Gamma$ be a cooperative strategic game with perfect information, $\Gamma(x)$ be a subgame of $\Gamma$ with a root $x$, $n$ the number of players, and $\bar{x}$ the successor nodes of $x$ including $x$ itself (and excluding the terminal nodes). Recall that $\Gamma(x)$ is the  subgame starting at $x$. For example, if $x$ is the root of game $\Gamma$, then $\Gamma(x)$ is simply the game $\Gamma$. The CES algorithm is defined via  (strong) induction on $n$ and $\bar{x}$.

\begin{enumerate}
	\item \textbf{Base case}: Let $(n,\bar{x}) = (1,1)$. CES at $\Gamma(x)$ is simply given by the pair $(\sigma^*,\pi^*)$ where player $i$, who is active at $x$, maximizes their utility at $x$, and the only viable coalition is the singleton coalition containing $i$.
	
	\item \textbf{Induction step}: The induction hypothesis assumes that CES is defined for all `smaller' subgames (and parallel games) as given in section~\ref{subsec:RBI_perfect}. Based on this assumption, a CES is defined for all $\Gamma(x')$ where $x'$ is an immediate successor of $x$. The next step is then to extend CES to $\Gamma(x)$. This is achieved by comparing the cooperative equilibrium systems of all parallel games at $x$ with the non-cooperative choice made by $i$ at $x$.
	\begin{enumerate}
		\item The non-cooperative reference system at $x$ is defined as follows. Player $i$ non-cooperatively chooses a behavior strategy that maximizes their utility at $x$, given the CES at every $\Gamma(x')$, where $x'$ is an immediate successor of $x$. The associated coalitions with this reference system are the singleton coalitions. This is called the non-cooperative reference system because if at node $x$ player $i$ does not form a coalition, then $i$ will have to make a non-cooperative choice at $x$, given the CES at the immediate successor nodes of $x$.

		\item Reference systems: Consider a parallel game $\Gamma_{P_C}$ where some coalition $C$, including agent $i$, forms. Note that the parallel game $\Gamma_{P_C}$ includes strictly fewer players than $\Gamma(x)$, hence its CES is defined by the induction hypothesis. A CES of each such parallel game is called a reference system at $x$.
		
  	A nondecreasing sequence is constructed based on player $i$'s utility in the cooperative equilibrium systems of each parallel game at $x$. In other words, we have a nondecreasing sequence of reference systems.

		\item Individually rational reference systems: I first define the non-cooperative reference system to be individually rational (IR). Other IR reference systems are then defined inductively as follows.
		
		Assume that for some $j\geq 0$ the reference system $r_j(x)$ is IR, and denote it as $r^*_j(x)$. Find the next reference system, denoted by $r^*_{j'}(x)$, in the sequence defined in Step (b) such that (i) $j'$ is the smallest index satisfying $j'>j$, and (ii) $r^*_{j'}(x)$ is IR with respect to $r^*_j(x)$. In other words, every agent of the coalition that includes player $i$ in the new reference system strictly prefers their CES to the CES of the previous reference system.
		
		The credible reference system is defined as the IR reference system that maximizes player $i$'s utility among all IR reference systems. Note that the credible reference system might be the non-cooperative reference system at $x$.
	\end{enumerate}
	
	A CES of $\Gamma(x)$ is then given by the system that extends---according to the above steps---cooperative equilibrium systems at every node $x'$, where $x'$ is an immediate successor of $x$. Accordingly, a CES of game $\Gamma$ is a CES of $\Gamma(x)$ where $x$ is the root of the game tree of $\Gamma$.
\end{enumerate}

\subsection{Imperfect information games}
\label{sec:imperfect}

%\subsubsection{Definition}
%\label{sec:CES_def_imperfect}

Let $\Gamma$ be a cooperative strategic game with imperfect information. Let $(\sigma,\pi)$ be a system of $\Gamma$, $h$ be a non-terminal information set, and $i=I(h)$. At $h$, consider parallel tree $T_{P_C}\in par(T(h))$ where $C\ni i$ is a feasible non-singleton coalition. For the rest of this subsection, associate the system $[\tau^{C}(h),\pi^{C}(h)]$ with each parallel tree $T_{P_C}$.

\begin{definition}[Non-cooperative reference system]
\label{def:autarky}
Let $[\sigma(h'), \pi(h')]$  be a system where $h'\in f'(h)$. The \textit{non-cooperative reference system} at some $h\in H$, denoted as $r_0(h)$, with respect to $[\sigma(h'), \pi(h')]$ is defined as follows. Let $r_0(h)\coloneqq[\tau^{i}(h),\pi^{i}(h)]$ which is defined as the extension of $[\sigma(h'), \pi(h')]$ with $h'\in f'(h)$ to, $T(h)$, the  parallel tree starting at $h$, such that $[\tau^{i}(h,\Gamma(h)),\pi^{i}(h,\Gamma(h))]$ is an SPNE in subgame $\Gamma(h)$  where each player in subgame $\Gamma(h)$ excluding those who act at an information set $h'\in F'(h)$ choose their strategies non-cooperatively.\footnote{Note that SPNE is defined with respect to players, which may be coalitions, which are given by player partition $\pi^{i}(h,\Gamma(h))$.}
\end{definition}

Put differently, $[\tau^{i}(h,\Gamma(h)),\pi^{i}(h,\Gamma(h))]$ is an SPNE of $\Gamma(h)$ in which all players in subgame $\Gamma(h)$ before player $I(h)$ choose their strategies non-cooperatively, and each player who acts after player $I(h)$ is given by player partition $\pi(h')$ where $h'\in f'(h)$.

\begin{definition}[Reference systems]
\label{def:reference}
For each parallel tree $T_{P_C}$, fix the system $[\tau^{C}(h),\pi^{C}(h)]$. Let $(r_j(h))^{\bar{j}}_{j=0}$ be a sequence for player $i$ where $r_j(h)\coloneqq[\tau^{C_j}(h),$ $\pi^{C_j}(h)]$, satisfying the following two conditions.
\begin{enumerate}
	\item For every two indices $j$ and $j'$ with $j<j'$, $u_{i}(\tau^{C_j}(h)|\pi^{C_{j}}(h))\leq $ $u_{i}(\tau^{C_{j'}}(h)|\pi^{C_{j'}}(h))$; i.e., the sequence is nondecreasing for $i$. 
	\item If $u_{i}(\tau^{C_j}(h)|\pi^{C_{j}}(h)) $ $= u_{i}(\tau^{C_{j'}}(h)|\pi^{C_{j'}}(h))$ and $C_j\subsetneq C_{j'}$, then $r_j$ precedes $r_{j'}$ for parallel tree coalitions $C_j$ and $C_{j'}$ containing $i$.
\end{enumerate}
Each element of sequence $(r_j(h))^{\bar{j}}_{j=0}$ is called a \textit{reference system}.
\end{definition}

\begin{definition}[Individually rational reference systems]
\label{def:IR}
Fix the sequence $(r_j(h))^{\bar{j}}_{j=0}$ of reference systems for player $i$.
\begin{enumerate}
		\item\textbf{Base case}: 
		Define the non-cooperative  reference system $r_0(h)$ to be individually rational and denote it by $r^*_0(h)$. The following \textit{individually rational} (IR) reference systems, with respect to the system $[\tau^{C}(h),\pi^{C}(h)]$ for  every $T_{P_C}$, are defined inductively as follows.
		
		\item\textbf{Induction step}: Assume that $r_j(h)$ is IR for some $j\geq 0$. Denote $r_j(h)$ as $r^*_k(h)$ for some $k\geq 0$ such that $j=0$ if and only if $k=0$. Next, $r^*_{k+1}(h)$ is defined as follows. Let $j'>j$ be the smallest number such that $r_{j'}(h)$ is IR with respect to $r^*_k(h)$; in other words, $u_{i'}(\tau^{C_{j'}}(h)|\pi^{C_{j'}}(h)) > u_{i'}(\tau^{C_{j}}(h)|\pi^{C_{j}}(h))$ for every agent $i'\in \bar{C}_{j'}$ where $C_{j'} \subseteq \bar{C}_{j'}$ and $\bar{C}_{j'}\in \pi^{C_{j'}}(h)$.\footnote{In other words, $\bar{C}_{j'}$ contains coalition $C_{j'}$ such that $\bar{C}_{j'}$ is part of the solution of the parallel tree $\Gamma_{P_{C_{j'}}}$.} Then, define $r^*_{k+1}(h)\coloneqq r_{j'}(h)$.
	\end{enumerate}
\end{definition}

Note that the sequence $(r^*_j(h))^{\bar{l}}_{k=0}$ of IR reference systems  is a subsequence of the sequence $(r_j(h))^{\bar{j}}_{j=0}$ of reference systems.

\begin{definition}[Credibility]
\label{def:credibility2}
Let $h\in H$ be a non-terminal information set and $(r^*_j(h))^{\bar{l}}_{k=0}$ be a
sequence of IR reference systems at $h$. The IR reference system, $r^*_{\bar{l}}(h)$, that maximizes $i$’s utility is defined as the credible reference system at $h$, with respect to systems $(r^*_j(h))^{\bar{l}}_{k=0}$.
\end{definition}

Note that $r^*_{\bar{l}}(h)=[\tau^{C_{\bar{l}}}(h),\pi^{C_{\bar{l}}}(h)]$. As in the perfect information case, it may be that $r^*_0(h)$ where $\bar{l}=0$ is the only and the credible reference system. Additionally, it is worth emphasizing that credibility is defined with respect to given systems; there is no requirement whether these systems themselves are credible. This observation leads to the following definition.

\begin{definition}[Cooperative equilibrium system, CES]
	\label{def:CES_imperfect}
	Let $\Gamma=(P, T , I, u, S, H)$ be a cooperative strategic game with imperfect information without a chance move. 
	\begin{enumerate}
	\item Let $|P|=1$. A CES of $\Gamma$ is defined as a system $(\sigma^*,\pi^*)$ where $\pi^*=P$ and $\sigma^*\in \argmax_{s'_i\in S_i} u_i(s'_i)$ where $i\in P$.
	\item Let $|P|\geq 2$. A CES, $(\sigma^*,\pi^*)$, of $\Gamma$ is defined as follows. Let $h\in H$ be any non-terminal information set. Every reference system, $[\sigma^{*C}(h),\pi^{*C}(h)]$, except the non-cooperative reference system, is a CES of the parallel tree $T_{P_C}$. The non-cooperative reference system at $h$ is defined with respect to the CES $[\sigma^*(h'), \pi^*(h')]$ where $h'\in f'(h)$. Moreover, the IR reference system $r^*_{\bar{l}}(h)$ is the credible reference system with respect to $(r^*_j(h))^{\bar{l}}_{k=0}$ such that, for every $j\geq 0$, $r^*_j(h)=[\sigma^{*C_j}(h),\pi^{*C_j}(h)]$.
	\end{enumerate}
\end{definition}

As in perfect information games, this definition is recursive: a CES prescribes the credible reference system to every non-terminal information set $h$ with respect to cooperative equilibrium systems, not just arbitrary systems. In one-person games, credibility essentially means utility maximization. In $n$-person games, every reference system, except for the non-cooperative reference system, is a CES of a parallel tree at $h$. Both the non-cooperative reference system at $h$ and every credible reference system are defined with respect to cooperative equilibrium systems. 

For the extension of the CES algorithm to imperfect information games, refer to subsection~\ref{subsec:CES_algo_imperfect} in the Appendix. Analogous to the perfect information games, coalitions are called \textit{stable} if they survive the CES algorithm and players and agents are called \textit{dynamically rational} if they utilize the CES algorithm in cooperative strategic games. The following theorem asserts the existence of the CES in imperfect information games.

\begin{theorem}[Existence: imperfect information games]
	\label{thm:imperfect}
	There exists a CES in possibly mixed strategies in every finite $n$-person cooperative extensive form game.
\end{theorem}

The proof is given in the Appendix (see section~\ref{subsec:proof2}). I next give a definition of a CES in a game with chance moves.

\begin{definition}[CES with a chance move]
\label{def:CES_chance}
Let $\Gamma=(P, T , I, u, S, H)$ be a cooperative strategic game with perfect information with a chance move. Nature is denoted as player $0\in P$ which chooses a possibly mixed behavior strategy $b_0(x_0)$ at root $x_0$. But there is no utility function for player $0$, so it cannot form any feasible coalition. A CES of $\Gamma$ is defined as system $[\sigma^*, \pi^*]$ which for every $h'\in f(h)$ extends CES $[\sigma^*(h'), \pi^*(h')]$ to $x_0$ such that Nature chooses a possibly mixed behavior strategy $b_0(x_0)$ and $\pi^*(x_0)=P$.
\end{definition}

In simple words, a CES of the game in which Nature moves at the root of the game is simply the profile of the CESs of the subgames starting at each immediate successor of the root.

If the cooperative extensive form structure of a normal form game is not provided, then a CES of the game is defined as a CES of a cooperative extensive form game whose reduced normal form coincides with the given normal form game.\footnote{In the context of incomplete information games, the notion of subgame perfection used within the CES algorithm can be changed to perfect Bayesian equilibrium. Additionally, the concept of CES can be extended from finite to infinite horizon games in a manner analogous to the extension of the backward induction concept to infinite horizon games via subgame perfect equilibrium.}

Next, I discuss some of the possibilities of modifying the framework of cooperative strategic games, and how this might affect the definition and the existence of a CES.

\subsection{Modifications and extensions}
\label{sec:modification}

\vspace{0.2cm}
\noindent\textbf{Cooperative strategic games without coalitional utility}: One may construct cooperative strategic games without relying on a coalitional utility function for the players. Specifically, these games can be formulated using a set of preference relations---denoted as $\succeq_{C|P}$---rather than utility functions $u_{C|P}$. These preference relations do not have to conform to (expected) utility assumptions. The only requirement is the existence of maximal elements for coalitional players in certain parallel games. When a coalition does not have a maximal element, the coalition would not be feasible in the relevant parallel game. Importantly, the definition of CES, as well as the existence proofs found in Theorem~\ref{thm:perfect} and Theorem~\ref{thm:imperfect}, remain valid. This is because these constructs rely on the presence of maximal elements, and not on a full preference ordering that comes with a utility function.

\vspace{0.2cm}
\noindent\textbf{Partial, fixed, and mixed cooperation}:  Consider a strategic situation in which coalitions are `subgame-dependent' in parallel games associated to a given game. In this extended framework, the parallel game may consist of various subgames wherein coalition compositions differ based on specific information sets. For example, at the root  $x$ of the parallel game, player $i$ may form a coalition with player $j$, and at another information set $h'$, player $i$ may cooperate with player $j'\neq j$. Therefore, the active players at $x$ and $h'$ would be $\{i,j\}$ and $\{i,j'\}$, respectively. Given that utilities for both $\{i,j\}$ and $\{i,j'\}$ are well-defined, the CES algorithm would also be well-defined. Specifically, a CES in this context would assign a credible reference system to each information set, enabling the same player to form different coalitions even along the path of play of a CES.

In `fixed cooperation' variant of cooperative strategic games, it is feasible to model situations where two or more players cooperate at specific information sets, while acting independently at others.  To formalize this, let $P$ be the finite set of players, where each $C\in P$ is a non-empty subset of $N=\{1,2,...,n\}$. Importantly, $P$ is not constrained to be a partition of $N$; for instance, it is plausible that  $P=\{1,2,3,\{1,2\},\{1,3\}\}$. Consider the case where two players, $i$ and $j$, cooperate at an information set $h$ but act independently at other information sets. In this situation, any parallel tree coalitions at $h$ must necessarily include both $i$ and $j$. Furthermore, if a set of possibly non-singleton players, $i_1, i_2,...,i_k$, forms a coalition, then $C=\{i_1, i_2,...,i_k\}$. For instance, if $\{1,2\}$ and $\{1,3\}$ form a coalition $C$, then it may be that $C=\{\{1,2\},\{1,3\}\}$ rather than $\{1,2,3\}$. This method of modeling may be particularly useful for studying specific scenarios, such as the merger of two independent decision-making organizations, where the decision to merge is made at the organizational level.

In an alternative variant, one can model a situation where player $i$ at an information set $h$ `mixes' between two or more parallel trees, for example, between coalitions $\{i,j\}$ and $\{i,j'\}$, with the objective of attaining a more favorable individually rational reference system at $h$. Any of these variants can be incorporated into the cooperative strategic games. Mutatis mutandis, the CES would remain well-defined, and its existence would be guaranteed, in line with Theorem~\ref{thm:imperfect}.

\vspace{0.2cm}
\noindent\textbf{Cooperative strategic games with more complex contracts}: Players can write contracts to form a coalition in cooperative extensive form games. In some sense, such contracts are `simple,' but one could consider situations where the agents can write more complex contracts. For example, consider a parallel game where a coalition does not have a maximal element and the preferences of its members cycle among three outcomes, $s\succ s'\succ s'' \succ s$. Further, suppose that they prefer these three outcomes to all others.\footnote{In the context of social choice theory, this immediately brings to mind the Condorcet Paradox and Arrow's \citeauthor{arrow1950} impossibility theorem.} In such a scenario, the coalition could write a contract that prescribes a specific strategy, such as uniformly mixing over these three outcomes. This could be individually rational for the members when compared to the corresponding counterfactual. Another possibility is to define parallel games in terms of contracts rather than coalitions, allowing a player the flexibility to write multiple contracts with the same set of players.

To formalize this, let $\Gamma=(P, T , I, u, \mathcal{C}, \Sigma, H)$ be a cooperative extensive form game with complex contracts, in which $\mathcal{C}$ denotes a correspondence such that for every information set $h$, $\mathcal{C}(h)$ is a set of feasible (abstract) contracts. For example, at each $h$ where $i=I(h)$, $\mathcal{C}(h)$ could be a subset of $2^N$ that includes $i$. This extended framework accommodates a broader range of coalition forming possibilities than those introduced in section~\ref{sec:setup}. To give an example, there might exist players in $\Gamma$ who do not act at any information set but are nonetheless permitted to write contracts with those who do.

Provided that the contracts, regardless of their complexity, yield a well-defined outcome and that credible individually rational reference systems exist, the CES would remain well-defined, and its existence would be guaranteed.

\vspace{0.2cm}
\noindent\textbf{General non-cooperative/disagreement points}: The starting point for the CES algorithm is the non-cooperative reference system, wherein the active player acts independently. This serves as a de facto disagreement point in case where no one forms a coalition. However, the concept of a disagreement point can be defined more broadly. Players could have the flexibility to either exogenously establish the non-cooperative reference system for all histories, or to implement a rule-based mechanism for determining the disagreement point.

To formalize this extended framework, let $\Gamma=(P, T , I, u, \mathcal{C}, D, \Sigma, H)$ be a  cooperative extensive form game with complex contracts and with general disagreement points. Here, $D$ is a correspondence defined as follows: for every non-terminal $x\in X$, $D(x)$ represents an abstract set that prescribes a method for choosing the disagreement point in subgame $\Gamma(x)$ and all of its subtrees. 

As an example, $D(x)$ could specify an explicit sequence of players in subgame $\Gamma(x)$, wherein player $i=I(x)$ offers to write contracts (or form coalitions) with other players. Provided that this ordering is consistent with the implied order of play according to the structure of the game tree, the CES would remain well-defined, and its existence would be guaranteed.\footnote{It should be noted that, unlike in games with perfect information, the order of play in games with imperfect information may be ambiguous within specific subgames. This occurs because, at a non-singleton information set $h$ with player $i=I(h)$, choosing an action $a_i$ does not unambiguously dictate the subsequent player. Rather, the subsequent player might depend on the particular nodes following action $a_i$ at information set $h$. In such scenarios, players can write more specific contracts.}

\vspace{0.2cm}
\noindent\textbf{Cost of cooperation}: As elaborated in Remark~\ref{rem:cost}, both psychological and monetary costs, if any, associated with forming coalitions are incorporated into the utility functions, which are dependent on player partitions. One could extend this to accommodate more general cost structures. For instance, the cost of forming a coalition could be contingent on the specific outcomes that the coalition aims to achieve. Provided that the utility functions are well-defined and accurately account for these costs, the CES will remain well-defined.

\vspace{0.2cm}
\noindent\textbf{Refinements, coarsenings, and other solutions}: The CES, in its current formulation, precludes the use of non-credible threats. However, this solution concept could be extended to accommodate such threats, akin to the discourse contrasting Nash equilibrium with SPNE in non-cooperative games. Furthermore, a more generalized version of the CES could be defined via the CES algorithm. This extended definition may incorporate either refinements or coarsenings of SPNE, such as the sequential equilibrium as proposed by \citet{kreps1982}.

Let \textit{cooperative system} denote a system wherein each individual player, each coalitional player, and each agent within a coalition seeks to maximize their utility relative to specific counterfactuals. Importantly, these counterfactuals are not required to be credible---i.e., they do not necessarily satisfy the steps described in the CES algorithm. Given this, it follows that while every CES qualifies as a cooperative system, the converse is not necessarily true.

An alternative modification to the current framework could involve relaxing the strict incentive condition for joining a coalition. This adjustment would likely lead to an expanded set of solutions. Further extensions may incorporate notions like maximin strategy and maximin equilibrium (optimin criterion) \citep{ismail2014}. In `optimin cooperative system,' players would aim to simultaneously maximize their minimum utility under profitable deviations by other coalitions.

\section{A fully worked-out example}
\label{sec:worked-out_ex}

Figure~\ref{fig:3pl-1} (A) illustrates a three-player cooperative strategic game in which player 1 (P1) starts by choosing either $L$ or $R$.   To streamline the analysis, I impose two simplifying assumptions. First, the cost of coalition formation is considered negligible. It is important to note, however, that the analysis remains valid even if forming coalitions incurs some costs, as long as these costs are not `too high.' Second, for each non-empty coalition $C\subseteq \{1,2,3\}$, I define the utility function of the coalition $u_C(\,\cdot\,) \coloneqq \min_{i\in C}u_i(\,\cdot\,)$.  In other words, one outcome is preferred to another if the minimum utility a member of the coalition receives from the former is greater than the minimum utility derived from the latter. To give an example, consider the coalition $C=\{1,3\}$. In this case,  $u_C(R, d, l)>u_C(R, d, k)$ because $u_C(R, d, l)=5$ while $u_C(R, d, k)=1$. (Note that in general a coalitional utility can be any von Neumann-Morgenstern utility.)

\begin{figure}
\centering
	\begin{tabular}{c}
		\begin{tikzpicture}[font=\footnotesize,edge from parent/.style={draw,thick}]
		% Two node styles: solid and hollow
		\tikzstyle{solid node}=[circle,draw,inner sep=1.2,fill=black];
		\tikzstyle{hollow node}=[circle,draw,inner sep=1.2];
		% Specify spacing for each level of the tree
		\tikzstyle{level 1}=[level distance=15mm,sibling distance=50mm]
		\tikzstyle{level 2}=[level distance=15mm,sibling distance=25mm]
		\tikzstyle{level 3}=[level distance=15mm,sibling distance=15mm]
		% The Tree
		\node[draw] at (-4.5,0) {A};
		\node(0)[hollow node]{}
		child{node[solid node]{}
			child{node[solid node]{}
				child{node[below]{$\begin{pmatrix}5\\5\\3\end{pmatrix}$} edge from parent node[left]{$e$}}
				child{node[below]{$\begin{pmatrix}2\\2\\1\end{pmatrix}$} edge from parent node[right]{$f$}}
				edge from parent node[above left]{$a$}
			}
			child{node[solid node]{}
				child{node[below]{$\begin{pmatrix}4\\4\\5\end{pmatrix}$} edge from parent node(s)[left]{$g$}}
				child{node[below]{$\begin{pmatrix}1\\6\\4\end{pmatrix}$} edge from parent node(t)[right]{$h$}}
				edge from parent node[above right]{$b$}
			}
			edge from parent node[above left]{$L$}
		}
		child{node[solid node]{}
			child{node[solid node]{}
				child{node[below]{$\begin{pmatrix}3\\1\\2\end{pmatrix}$} edge from parent node(m)[left]{$i$}}
				child{node[below]{$\begin{pmatrix}2\\2\\6\end{pmatrix}$} edge from parent node(n)[right]{$j$}}
				edge from parent node[above left]{$c$}
			}
			child{node[solid node]{}
				child{node[below]{$\begin{pmatrix}1\\1\\6\end{pmatrix}$} edge from parent node[left]{$k$}}
				child{node[below]{$\begin{pmatrix}6\\3\\5\end{pmatrix}$} edge from parent node[right]{$l$}}
				edge from parent node[above right]{$d$}
			}
			edge from parent node[above right]{$R$}
		};
		% information sets
		%\draw[loosely dotted,very thick](0-1-1)to[out=-15,in=195](0-2-1);
		%\draw[loosely dotted,very thick](0-1-2)to[out=-15,in=195](0-2-2);
		% movers
		\node[above,yshift=2]at(0){1};
		\node[left,yshift=2]at(0){$x_0$};
		\foreach \i in {1,2} \node[above,yshift=2]at(0-\i){2};
		\node[left,yshift=2]at(0-1){$x_1$};
		\node[right,yshift=2]at(0-2){$x_2$};
		\node[above,yshift=2]at(0-1-1){3};
		\node[left,yshift=2]at(0-1-1){$x_3$};
		\node[above,yshift=2]at(0-1-2){3};
		\node[right,yshift=2]at(0-1-2){$x_4$};
		\node[above,yshift=2]at(0-2-1){3};
		\node[left,yshift=2]at(0-2-1){$x_5$};
		\node[above,yshift=2]at(0-2-2){3};
		\node[right,yshift=2]at(0-2-2){$x_6$};
		%\foreach \i in {1,2,3,4} \node[above,yshift=2]at(0-1-\i){3};
		%	\node at($.5*(s)+.5*(t)$){1};
		%	\node at($.5*(m)+.5*(n)$){1};
		\end{tikzpicture}
		\\
		\begin{tikzpicture}[font=\footnotesize,edge from parent/.style={draw,thick}]
		% Two node styles: solid and hollow
		\tikzstyle{solid node}=[circle,draw,inner sep=1.2,fill=black];
		\tikzstyle{hollow node}=[circle,draw,inner sep=1.2];
		% Specify spacing for each level of the tree
		\tikzstyle{level 1}=[level distance=15mm,sibling distance=50mm]
		\tikzstyle{level 2}=[level distance=15mm,sibling distance=25mm]
		\tikzstyle{level 3}=[level distance=15mm,sibling distance=15mm]
		% The Tree
		\node[draw] at (-4.5,0) {B};
		\node(0)[hollow node]{}
		child{node[solid node]{}
			child{node[solid node]{}
				child{node[below]{$\begin{pmatrix}5\\5\\3\end{pmatrix}$} edge from parent[->, solid, thick] node[left]{$e$}}
				child{node[below]{$\begin{pmatrix}2\\2\\1\end{pmatrix}$} edge from parent[black, solid, thick] node[right]{$f$}}
				edge from parent[ thick] node[above left]{$a$}
			}
			child{node[solid node]{}
				child{node[below]{$\begin{pmatrix}4\\4\\5\end{pmatrix}$} edge from parent[->, solid, thick] node(s)[left]{$g$}}
				child{node[below]{$\begin{pmatrix}1\\6\\4\end{pmatrix}$} edge from parent[black, solid, thick] node(t)[right]{$h$}}
				edge from parent[black, solid, thick] node[above right]{$b$}
			}
			edge from parent[thick] node[above left]{$L$}
		}
		child{node[solid node]{}
			child{node[solid node]{}
				child{node[below]{$\begin{pmatrix}3\\1\\2\end{pmatrix}$} edge from parent[black, solid, thick] node(m)[left]{$i$}}
				child{node[below]{$\begin{pmatrix}2\\2\\6\end{pmatrix}$} edge from parent[->,solid, thick] node(n)[right]{$j$}}
				edge from parent[thick] node[above left]{$c$}
			}
			child{node[solid node]{}
				child{node[below]{$\begin{pmatrix}1\\1\\6\end{pmatrix}$} edge from parent[->,black, solid, thick] node[left]{$k$}}
				child{node[below]{$\begin{pmatrix}6\\3\\5\end{pmatrix}$} edge from parent[black, solid, thick] node[right]{$l$}}
				edge from parent[black, solid, thick] node[above right]{$d$}
			}
			edge from parent[black, solid, thick] node[above right]{$R$}
		};
		% information sets
		%\draw[loosely dotted,very thick](0-1-1)to[out=-15,in=195](0-2-1);
		%\draw[loosely dotted,very thick](0-1-2)to[out=-15,in=195](0-2-2);
		% movers
		\node[above,yshift=2]at(0){1};
		\foreach \i in {1,2} \node[above,yshift=2]at(0-\i){2};
		\node[above,yshift=2]at(0-1-1){3};
		\node[above,yshift=2]at(0-1-2){3};
		\node[above,yshift=2]at(0-2-1){3};
		\node[above,yshift=2]at(0-2-2){3};
		%\foreach \i in {1,2,3,4} \node[above,yshift=2]at(0-1-\i){3};
		%	\node at($.5*(s)+.5*(t)$){1};
		%	\node at($.5*(m)+.5*(n)$){1};
		\end{tikzpicture}
	\end{tabular}
	\caption{(A) A three-player cooperative strategic game. (B) The best-responses of P3 at the penultimate nodes are shown by the lines with arrows.}
\label{fig:3pl-1}
\end{figure}

\subsection{Applying the CES algorithm}

I proceed to solve the CES of this game by following the main steps outlined in the CES algorithm. Specifically, Figure~\ref{fig:3pl-1} (B) displays the best-response actions of player 3 (P3) at the penultimate nodes $x_3, x_4, x_5$, and $x_6$, conforming to Step 1 of the CES algorithm. These best-responses are indicated by lines accompanied by arrows.

\begin{figure}
	\centering
	\begin{tabular}{c}
		\begin{tikzpicture}[font=\footnotesize,edge from parent/.style={draw,thick}]
		% Two node styles: solid and hollow
		\tikzstyle{solid node}=[circle,draw,inner sep=1.2,fill=black];
		\tikzstyle{hollow node}=[circle,draw,inner sep=1.2];
		% Specify spacing for each level of the tree
		\tikzstyle{level 1}=[level distance=15mm,sibling distance=50mm]
		\tikzstyle{level 2}=[level distance=15mm,sibling distance=25mm]
		\tikzstyle{level 3}=[level distance=15mm,sibling distance=15mm]
		% The Tree
		\node[draw] at (-4.9,0) {C};
		\node[draw] at (4.9,0) {$r^*_1(x_1)$};
		\node(0)[solid node]{}
		child{node[solid node]{}
			child{node[below]{$\begin{pmatrix}5\\5\\3\end{pmatrix}$}
				edge from parent[->, black, solid, very thick] node[above left]{$e$}
			}
			child{node[below]{$\begin{pmatrix}2\\2\\1\end{pmatrix}$}
				edge from parent[black, solid, thick] node[above right]{$f$}
			}
			edge from parent[solid] node[above left]{$a$}
		}
		child{node[solid node]{}
			child{node[below]{$\begin{pmatrix}4\\4\\5\end{pmatrix}$}
				edge from parent[solid] node[above left]{$g$}
			}
			child{node[below]{$\begin{pmatrix}1\\6\\4\end{pmatrix}$}
				edge from parent[->, black, solid, very thick] node[above right]{$h$}
			}
			edge from parent[->,black, solid, very thick] node[above right]{$b$}
		};
		% information sets
		%\draw[loosely dotted,very thick](0-1-1)to[out=-15,in=195](0-2-1);
		%\draw[loosely dotted,very thick](0-1-2)to[out=-15,in=195](0-2-2);
		% movers
		\node[above,yshift=2]at(0){2,3};
		\foreach \i in {1,2} \node[above,yshift=2]at(0-\i){2,3};
		\end{tikzpicture}
		\\
		\begin{tikzpicture}[font=\footnotesize,edge from parent/.style={draw,thick}]
		% Two node styles: solid and hollow
		\tikzstyle{solid node}=[circle,draw,inner sep=1.2,fill=black];
		\tikzstyle{hollow node}=[circle,draw,inner sep=1.2];
		% Specify spacing for each level of the tree
		\tikzstyle{level 1}=[level distance=15mm,sibling distance=50mm]
		\tikzstyle{level 2}=[level distance=15mm,sibling distance=25mm]
		\tikzstyle{level 3}=[level distance=15mm,sibling distance=15mm]
		% The Tree
		\node[draw] at (-4.9,0) {D};
		\node[draw] at (4.9,0) {$r^*_0(x_2)$};
		\node(0)[solid node]{}
		child{node[solid node]{}
			child{node[below]{$\begin{pmatrix}3\\1\\2\end{pmatrix}$}
				edge from parent[solid] node[above left]{$i$}
			}
			child{node[below]{$\begin{pmatrix}2\\2\\6\end{pmatrix}$}
				edge from parent[->,black, solid, thick] node[above right]{$j$}
			}
			edge from parent[->,black, solid, thick] node[above left]{$c$}
		}
		child{node[solid node]{}
			child{node[below]{$\begin{pmatrix}1\\1\\6\end{pmatrix}$}
				edge from parent[->,black, solid, thick] node[above left]{$k$}
			}
			child{node[below]{$\begin{pmatrix}6\\3\\5\end{pmatrix}$}
				edge from parent[solid] node[above right]{$l$}
			}
			edge from parent[solid] node[above right]{$d$}
		};
		% information sets
		%\draw[loosely dotted,very thick](0-1-1)to[out=-15,in=195](0-2-1);
		%\draw[loosely dotted,very thick](0-1-2)to[out=-15,in=195](0-2-2);
		% movers
		\node[above,yshift=2]at(0){2};
		\foreach \i in {1,2} \node[above,yshift=2]at(0-\i){3};
		\end{tikzpicture}
	\end{tabular}
	\caption{Solution steps C and D. The lines with arrows represent best-responses, which could be non-cooperative or coalitional. (C) The coalition between P2 and P3 form to play $b$ and $h$. The outcome of this reference system $r_1(x_1)$ is (1, 6, 4), which is IR with respect to the non-cooperative reference system $r^*_0(x_1)$ whose outcome is (5, 5, 3). The credible reference system at $x_1$ is $r^*_1(x_2)$. (D) No coalition forms because P3 receives their highest payoff when there is no coalition. Thus, the non-cooperative reference system, $r^*_0(x_2)$ is the credible reference system at $x_2$.}
	\label{fig:3pl-21}
\end{figure}

\begin{figure}
	\centering
	\begin{tabular}{c}
		\begin{tikzpicture}[font=\footnotesize,edge from parent/.style={draw,thick}]
		% Two node styles: solid and hollow
		\tikzstyle{solid node}=[circle,draw,inner sep=1.2,fill=black];
		\tikzstyle{hollow node}=[circle,draw,inner sep=1.2];
		% Specify spacing for each level of the tree
		\tikzstyle{level 1}=[level distance=15mm,sibling distance=50mm]
		\tikzstyle{level 2}=[level distance=15mm,sibling distance=25mm]
		\tikzstyle{level 3}=[level distance=15mm,sibling distance=15mm]
		% The Tree
		\node[draw] at (-4.5,0) {E};
		\node[draw] at (4.5,0) {$r^*_0(x_0)$};
		\node(0)[hollow node]{}
		child{node[solid node]{}
			child{node[solid node]{}
				child{node[below]{$\begin{pmatrix}5\\5\\3\end{pmatrix}$} edge from parent[->, black, solid, very thick] node[left]{$e$}}
				child{node[below]{$\begin{pmatrix}2\\2\\1\end{pmatrix}$} edge from parent[black, solid, thick] node[right]{$f$}}
				edge from parent[thick] node[above left]{$a$}
			}
			child{node[solid node]{}
				child{node[below]{$\begin{pmatrix}4\\4\\5\end{pmatrix}$} edge from parent[solid, thick] node(s)[left]{$g$}}
				child{node[below]{$\begin{pmatrix}1\\6\\4\end{pmatrix}$} edge from parent[->, black, solid, very thick] node(t)[right]{$h$}}
				edge from parent[->, black, solid, very thick] node[above right]{$b$}
			}
			edge from parent[solid] node[above left]{$L$}
		}
		child{node[solid node]{}
			child{node[solid node]{}
				child{node[below]{$\begin{pmatrix}3\\1\\2\end{pmatrix}$} edge from parent[black, solid, thick] node(m)[left]{$i$}}
				child{node[below]{$\begin{pmatrix}2\\2\\6\end{pmatrix}$} edge from parent[->,solid, thick] node(n)[right]{$j$}}
				edge from parent[->,thick] node[above left]{$c$}
			}
			child{node[solid node]{}
				child{node[below]{$\begin{pmatrix}1\\1\\6\end{pmatrix}$} edge from parent[->,black, solid, thick] node[left]{$k$}}
				child{node[below]{$\begin{pmatrix}6\\3\\5\end{pmatrix}$} edge from parent[black, solid, thick] node[right]{$l$}}
				edge from parent[black, solid, thick] node[above right]{$d$}
			}
			edge from parent[->, solid, thick] node[above right]{$R$}
		};
		% information sets
		%\draw[loosely dotted,very thick](0-1-1)to[out=-15,in=195](0-2-1);
		%\draw[loosely dotted,very thick](0-1-2)to[out=-15,in=195](0-2-2);
		% movers
		\node[above,yshift=2]at(0){1};
		\foreach \i in {1} \node[above,yshift=2]at(0-\i){2,3};
		\foreach \i in {2} \node[above,yshift=2]at(0-\i){2};
		\node[above,yshift=2]at(0-1-1){2,3};
		\node[above,yshift=2]at(0-1-2){2,3};
		\node[above,yshift=2]at(0-2-1){3};
		\node[above,yshift=2]at(0-2-2){3};
		%\foreach \i in {1,2,3,4} \node[above,yshift=2]at(0-1-\i){3};
		%	\node at($.5*(s)+.5*(t)$){1};
		%	\node at($.5*(m)+.5*(n)$){1};
		\end{tikzpicture}
		\\
	\begin{tikzpicture}[font=\footnotesize,edge from parent/.style={draw,thick}]
	% Two node styles: solid and hollow
	\tikzstyle{solid node}=[circle,draw,inner sep=1.2,fill=black];
	\tikzstyle{hollow node}=[circle,draw,inner sep=1.2];
	% Specify spacing for each level of the tree
	\tikzstyle{level 1}=[level distance=15mm,sibling distance=50mm]
	\tikzstyle{level 2}=[level distance=15mm,sibling distance=25mm]
	\tikzstyle{level 3}=[level distance=15mm,sibling distance=15mm]
	% The Tree
	\node[draw] at (-4.5,0) {F};
	\node[draw] at (4.5,0) {$r_1(x_0)$};
	\node(0)[hollow node]{}
	child{node[solid node]{}
		child{node[solid node]{}
			child{node[below]{$\begin{pmatrix}5\\5\\3\end{pmatrix}$} edge from parent[->,solid,very thick] node[left]{$e$}}
			child{node[below]{$\begin{pmatrix}2\\2\\1\end{pmatrix}$} edge from parent[black, solid, thick] node[right]{$f$}}
			edge from parent[solid] node[above left]{$a$}
		}
		child{node[solid node]{}
			child{node[below]{$\begin{pmatrix}4\\4\\5\end{pmatrix}$} edge from parent[->,solid,very thick] node(s)[left]{$g$}}
			child{node[below]{$\begin{pmatrix}1\\6\\4\end{pmatrix}$} edge from parent[solid] node(t)[right]{$h$}}
			edge from parent[->,solid,very thick] node[above right]{$b$}
		}
		edge from parent[->,solid,very thick] node[above left]{$L$}
	}
	child{node[solid node]{}
		child{node[solid node]{}
			child{node[below]{$\begin{pmatrix}3\\1\\2\end{pmatrix}$} edge from parent[solid] node(m)[left]{$i$}}
			child{node[below]{$\begin{pmatrix}2\\2\\6\end{pmatrix}$} edge from parent[->,solid,very thick] node(n)[right]{$j$}}
			edge from parent[solid] node[above left]{$c$}
		}
		child{node[solid node]{}
			child{node[below]{$\begin{pmatrix}1\\1\\6\end{pmatrix}$} edge from parent[solid] node[left]{$k$}}
			child{node[below]{$\begin{pmatrix}6\\3\\5\end{pmatrix}$} edge from parent[->,solid,very thick] node[right]{$l$}}
			edge from parent[->, solid, very thick] node[above right]{$d$}
		}
		edge from parent[solid] node[above right]{$R$}
	};
	% information sets
	%\draw[loosely dotted,very thick](0-1-1)to[out=-15,in=195](0-2-1);
	%\draw[loosely dotted,very thick](0-1-2)to[out=-15,in=195](0-2-2);
	% movers
	\node[above,yshift=2]at(0){1,2,3};
	\foreach \i in {1,2} \node[above,yshift=2]at(0-\i){1,2,3};
	\node[above left]at(0-1-1){1,2,3};
	\node[above right]at(0-1-2){1,2,3};
	\node[above left]at(0-2-1){1,2,3};
	\node[above right]at(0-2-2){1,2,3};
	%\foreach \i in {1,2,3,4} \node[above,yshift=2]at(0-1-\i){3};
	%	\node at($.5*(s)+.5*(t)$){1};
	%	\node at($.5*(m)+.5*(n)$){1};
	\end{tikzpicture}
	\end{tabular}
	\caption{(E) The non-cooperative reference system at $x_0$ whose outcome is (2, 2, 6). Given the cooperative equilibrium systems in steps C and D, P1 chooses their non-cooperative best-response. (F) The reference system associated with the parallel game of the grand coalition $\{1,2,3\}$. The outcome of this reference system is (4, 4, 5), which is not IR with respect to  (2, 2, 6).}	
	\label{fig:3pl-2}
\end{figure}

Consider the left subgame originating at node $x_1$, where player 2 (P2) is active. The non-cooperative best-response for P2 at this decision node is to choose $a$, leading to the outcome (5, 5, 3) given P3's subsequent choice of $e$. Thus, (5, 5, 3) is the outcome of the non-cooperative reference system $r^*_0(x_1)$, which is individually rational (IR) by definition (see Step 2.a in the CES algorithm). At node $x_1$, there is only one parallel game in which P2 and P3 form a coalition, as shown in Figure~\ref{fig:3pl-21} (C). Observe that coalition $\{2,3\}$'s utility is maximized at the outcome (1, 6, 4) (Step 2.b in the CES algorithm). This outcome is IR with respect to the non-cooperative reference system because both P2 and P3 prefer (1, 6, 4) to (5, 5, 3) (Step 2.c in the CES algorithm).\footnote{It is worth noting that although the outcome (4, 4, 5) also maximizes coalition $\{2,3\}$'s utility, it is not IR for P2 when compared to (5, 5, 3).} Because there is no other parallel game at $x_1$, $r^*_1(x_1)$ is the credible reference system. Figure~\ref{fig:3pl-21} (C) illustrates (1, 6, 4) as the outcome determined by the CES algorithm in this step.

Next, consider the right subgame starting at node $x_2$, where P2 is active. The non-cooperative best-response of P2 at this decision node is to choose $c$. Thus, the outcome of the non-cooperative reference system at this node is (2, 2, 6), as is illustrated in Figure~\ref{fig:3pl-21} (D). In the parallel game where the coalition $\{2,3\}$ forms, the utility $u_{2,3}$ is maximized at the outcome (6, 3, 5). However, this outcome is clearly not IR for P3 with respect to the outcome of the non-cooperative reference system (2, 2, 6). As a result, the credible reference system at this subgame is the non-cooperative reference system, $r^*_0(x_2)$.

\begin{figure}
	\centering
	\begin{tabular}{c}
	\begin{tikzpicture}[font=\footnotesize,edge from parent/.style={draw,thick}]
	% Two node styles: solid and hollow
	\tikzstyle{solid node}=[circle,draw,inner sep=1.2,fill=black];
	\tikzstyle{hollow node}=[circle,draw,inner sep=1.2];
	% Specify spacing for each level of the tree
	\tikzstyle{level 1}=[level distance=15mm,sibling distance=50mm]
	\tikzstyle{level 2}=[level distance=15mm,sibling distance=25mm]
	\tikzstyle{level 3}=[level distance=15mm,sibling distance=15mm]
	% The Tree
	\node[draw] at (-4.5,0) {G};
	\node[draw] at (4.5,0) {$r^*_2(x_0)$};
	\node(0)[hollow node]{}
	child{node[solid node]{}
		child{node[solid node]{}
			child{node[below]{$\begin{pmatrix}5\\5\\3\end{pmatrix}$} edge from parent[->,solid, thick] node[left]{$e$}}
			child{node[below]{$\begin{pmatrix}2\\2\\1\end{pmatrix}$} edge from parent[black, solid, thick] node[right]{$f$}}
			edge from parent[->,solid, very thick] node[above left]{$a$}
		}
		child{node[solid node]{}
			child{node[below]{$\begin{pmatrix}4\\4\\5\end{pmatrix}$} edge from parent[->,solid, thick] node(s)[left]{$g$}}
			child{node[below]{$\begin{pmatrix}1\\6\\4\end{pmatrix}$} edge from parent[solid] node(t)[right]{$h$}}
			edge from parent[solid] node[above right]{$b$}
		}
		edge from parent[->,solid,very thick] node[above left]{$L$}
	}
	child{node[solid node]{}
		child{node[solid node]{}
			child{node[below]{$\begin{pmatrix}3\\1\\2\end{pmatrix}$} edge from parent[black, solid, thick] node(m)[left]{$i$}}
			child{node[below]{$\begin{pmatrix}2\\2\\6\end{pmatrix}$} edge from parent[->,solid, thick] node(n)[right]{$j$}}
			edge from parent[->, solid, very thick] node[above left]{$c$}
		}
		child{node[solid node]{}
			child{node[below]{$\begin{pmatrix}1\\1\\6\end{pmatrix}$} edge from parent[->,black, solid, thick] node[left]{$k$}}
			child{node[below]{$\begin{pmatrix}6\\3\\5\end{pmatrix}$} edge from parent[black, solid, thick] node[right]{$l$}}
			edge from parent[black, solid, thick] node[above right]{$d$}
		}
		edge from parent[solid, thick] node[above right]{$R$}
	};
	% information sets
	%\draw[loosely dotted,very thick](0-1-1)to[out=-15,in=195](0-2-1);
	%\draw[loosely dotted,very thick](0-1-2)to[out=-15,in=195](0-2-2);
	% movers
	\node[above,yshift=2]at(0){1,2};
	\foreach \i in {1,2} \node[above,yshift=2]at(0-\i){1,2};
	\node[above,yshift=2]at(0-1-1){3};
	\node[above,yshift=2]at(0-1-2){3};
	\node[above,yshift=2]at(0-2-1){3};
	\node[above,yshift=2]at(0-2-2){3};
	%\foreach \i in {1,2,3,4} \node[above,yshift=2]at(0-1-\i){3};
	%	\node at($.5*(s)+.5*(t)$){1};
	%	\node at($.5*(m)+.5*(n)$){1};
	\end{tikzpicture}
	\\
	\begin{tikzpicture}[font=\footnotesize,edge from parent/.style={draw,thick}]
	% Two node styles: solid and hollow
	\tikzstyle{solid node}=[circle,draw,inner sep=1.2,fill=black];
	\tikzstyle{hollow node}=[circle,draw,inner sep=1.2];
	% Specify spacing for each level of the tree
	\tikzstyle{level 1}=[level distance=15mm,sibling distance=50mm]
	\tikzstyle{level 2}=[level distance=15mm,sibling distance=25mm]
	\tikzstyle{level 3}=[level distance=15mm,sibling distance=15mm]
	% The Tree
	\node[draw] at (-4.5,0) {H};
	\node[draw] at (4.5,0) {$r^*_3(x_0)$};
	\node(0)[hollow node]{}
	child{node[solid node]{}
		child{node[solid node]{}
			child{node[below]{$\begin{pmatrix}5\\5\\3\end{pmatrix}$} edge from parent[->,solid,very thick] node[left]{$e$}}
			child{node[below]{$\begin{pmatrix}2\\2\\1\end{pmatrix}$} edge from parent[black, solid, thick] node[right]{$f$}}
			edge from parent[->,solid,very thick] node[above left]{$a$}
		}
		child{node[solid node]{}
			child{node[below]{$\begin{pmatrix}4\\4\\5\end{pmatrix}$} edge from parent[->,solid,very thick] node(s)[left]{$g$}}
			child{node[below]{$\begin{pmatrix}1\\6\\4\end{pmatrix}$} edge from parent[solid] node(t)[right]{$h$}}
			edge from parent[solid] node[above right]{$b$}
		}
		edge from parent[solid] node[above left]{$L$}
	}
	child{node[solid node]{}
		child{node[solid node]{}
			child{node[below]{$\begin{pmatrix}3\\1\\2\end{pmatrix}$} edge from parent[->,solid,very thick] node(m)[left]{$i$}}
			child{node[below]{$\begin{pmatrix}2\\2\\6\end{pmatrix}$} edge from parent[solid] node(n)[right]{$j$}}
			edge from parent[solid] node[above left]{$c$}
		}
		child{node[solid node]{}
			child{node[below]{$\begin{pmatrix}1\\1\\6\end{pmatrix}$} edge from parent[solid] node[left]{$k$}}
			child{node[below]{$\begin{pmatrix}6\\3\\5\end{pmatrix}$} edge from parent[->,solid,very thick] node[right]{$l$}}
			edge from parent[->, solid, very thick] node[above right]{$d$}
		}
		edge from parent[->,solid,very thick] node[above right]{$R$}
	};
	% information sets
	%\draw[loosely dotted,very thick](0-1-1)to[out=-15,in=195](0-2-1);
	%\draw[loosely dotted,very thick](0-1-2)to[out=-15,in=195](0-2-2);
	% movers
	\node[above,yshift=2]at(0){1,3};
	\foreach \i in {1,2} \node[above,yshift=2]at(0-\i){2};
	\node[above,yshift=2]at(0-1-1){1,3};
	\node[above,yshift=2]at(0-1-2){1,3};
	\node[above,yshift=2]at(0-2-1){1,3};
	\node[above,yshift=2]at(0-2-2){1,3};
	%\foreach \i in {1,2,3,4} \node[above,yshift=2]at(0-1-\i){3};
	%	\node at($.5*(s)+.5*(t)$){1};
	%	\node at($.5*(m)+.5*(n)$){1};
	\end{tikzpicture}
\end{tabular}
	\caption{G: The parallel game of coalition $\{1,2\}$: P1 and P2 cooperate to play $L$ and $a$ to receive 5 each. The outcome of this reference system is (5, 5, 3), which is IR with respect to the previous IR reference system, $r^*_0(x_0)$. H: P1's forming a coalition with P2 is a credible threat to P3. Thus, P3 forms a coalition with P1: Agent P1 plays $R$ and agent P3 plays $l$ in response to P2's $d$. The outcome of this reference system is (6, 3, 5), which is IR with respect to $r^*_2(x_0)$. As a result, (6, 3, 5) is the CES outcome of this game because this is the outcome of the credible reference system at the root of the game.}
\label{fig:3pl-3}
\end{figure}

Finally, consider the root of the game at $x_0$, where P1 is active. Figure~\ref{fig:3pl-2} (E) illustrates P1's non-cooperative best-response, which is to choose $R$, taking into account the solutions determined in the preceding steps of the CES algorithm. Thus, the outcome of the non-cooperative reference system, $r^*_0(x_0)$, at the root is (2, 2, 6). To find the other reference systems at the root, I analyze the parallel games at node $x_0$, generated by the coalitions $\{1,2\}, \{1,3\}$, and $\{1,2,3\}$. 

\subsection{Parallel games at the root of the game}

Firstly, Figure~\ref{fig:3pl-2} (F) depicts the parallel game of the grand coalition $\{1,2,3\}$. This coalition maximizes its utility with the outcome (4, 4, 5) by choosing $L, b$ and $g$. Thus, the CES outcome for this parallel game is (4, 4, 5). 

Secondly, in parallel game $\Gamma_{\{1,2\}}$, the CES outcome is (5, 5, 3), as illustrated in Figure~\ref{fig:3pl-3} (G). This is because coalition $\{1,2\}$ maximizes its coalitional utility by choosing $L$ and $a$, given the best-responses of P3. 

Thirdly, Figure~\ref{fig:3pl-3} (H) shows that in parallel game $\Gamma_{\{1,3\}}$, the CES outcome is (6, 3, 5). Here, coalition $\{1,3\}$ maximizes its coalitional utility by choosing $R$ and $l$, given the best-responses of P2.

Subsequently, in accordance with Step 2.b of the CES algorithm, the reference systems arising from the parallel games $\{1,2,3\}$, $\{1,2\}$, and $\{1,3\}$ are denoted by $r_1(x_0)$, $r_2(x_0)$, and $r_3(x_0)$, respectively. Their corresponding outcomes are (4, 4, 5), (5, 5, 3), and (6, 3, 5). Importantly, these outcomes form an increasing sequence with respect to P1's utilities, who is the active player at $x_0$.
 
Proceeding to Step 2.c, the non-cooperative reference system $r^*_0(x_0)$ is IR by definition, and its outcome is (2, 2, 6). It is clear that the reference system, $r_1(x_0)$, from the parallel game $\Gamma_{\{1,2,3\}}$ is \textit{not} IR because P3 prefers the outcome of the non-cooperative reference system (2, 2, 6) to (4, 4, 5). In contrast, the reference system,  $r^*_2(x_0)$, from the parallel game $\Gamma_{\{1,2\}}$ is IR, as both P1 and P2 prefer (5, 5, 3) to (2, 2, 6). Given this IR reference system, the reference system, $r^*_3(x_0)$, from the parallel game $\Gamma_{\{1,3\}}$ is also IR, since both P1 and P3 prefer (6, 3, 5) to (5, 5, 3). As a result, (6, 3, 5) emerges as the outcome of the credible reference system for P1 at the root of the game. 

Figure~\ref{fig:solution} presents a CES solution for the game as represented in Figure~\ref{fig:3pl-1} (A). A short, `on-path,' description of this CES can be summarized by the following list of players, stable coalitions, and their corresponding strategies:
\[ [\{R\},\{a,d\},\{e,g,j,l\}; \{1,3\},2 ]. \]
In this CES, P1 and P3 form a coalition and hence act as a single player, denoted by $\{1,3\}$. Agent P1 chooses action $R$, P2 chooses $a$ and $d$, and agent P3 chooses $e$, $g$, $j$, and $l$ at their respective decision nodes. The outcome of this CES is (6, 3, 5). 

It is important to highlight that each of the potential coalitions---$\{1 ,2\}$, $\{2, 3\}$, and $\{1, 3\}$---except for the grand coalition, emerges as IR. Nevertheless, the coalition between P1 and P3 is the only stable coalition.\footnote{One interpretation of the CES is that the calculations underpinning the CES algorithm are initially conducted mentally by the players, who subsequently form coalitions based on these calculations.} 

\begin{figure}
	\begin{enumerate}
		\item Node $x_0$: The solution at the root of the game:
		\begin{enumerate}
			\item Node $x_0$: \[ [\{R\},\{a,d\},\{e,g,j,l\}; \{1,3\},2 ]. \]
			
			\item Node $x_2$ (the subgame after action $R$):
			\begin{enumerate}
				\item Node $x_2$: \[ [\{d\},\{i,l\}; 2,\{1,3\}]. \]
				\item Nodes $x_5$ to $x_6$: \[ [\{i,l\}; \{1,3\}]. \]
			\end{enumerate}
			
			\item Node $x_1$ (the subgame after action $L$):
			\begin{enumerate}
				\item Node $x_1$: \[ [\{a\},\{e,g\}; 2,\{1,3\}]. \]
				\item Nodes $x_3$ to $x_4$: \[ [\{e,g\}; \{1,3\}]. \]
			\end{enumerate}
			
			\item Nodes $x_3$ to $x_6$ (the penultimate nodes): 
			\[ [\{e,g,i,l\}; \{1,3\}]. \]
		\end{enumerate}
		
		\item Node $x_2$: The solution at the subgame after action $R$:
		\begin{enumerate}
			\item Node $x_2$: \[ [\{c\},\{j,k\}; 2,3]. \]
			
			\item Nodes $x_5$ to $x_6$:\[ [\{j,k\}; 3]. \]
		\end{enumerate}
		
		\item Node $x_1$: The solution at the subgame after action $L$:
		\begin{enumerate}
			\item Node $x_1$ (the subgame after action $L$):
			\[ [\{b\},\{h\}; \{2,3\}]. \]
			
			\item Nodes $x_3$ to $x_4$:	\[ [\{e,g\}; \{3\}]. \]
		\end{enumerate}
		
		\item Nodes $x_3$ to $x_6$: The solution at the penultimate nodes: 
		\begin{enumerate}
			\item Nodes $x_3$ to $x_6$: \[ [\{e,g,j,k\}; 3]. \]
		\end{enumerate}
	\end{enumerate}
		\caption{A CES of the game in Figure~\ref{fig:3pl-1} (A).}
	\label{fig:solution}
\end{figure}

\subsection{Coalitional credible threats}

At the outset, it might be tempting to conclude---without running the CES algorithm---that P1 and P3 would obviously form a coalition to secure the outcome (6, 3, 5).  However, such a  conclusion would be false. As an illustrative example, consider a modification to the game in Figure~\ref{fig:3pl-1} (A), wherein, all else being equal, the outcome (3, 1, 2) is replaced with (3, 4, 3). This change seems to be irrelevant because P1 and P3 could still, in theory, form a coalition to obtain (6, 3, 5). However, the CES algorithm would yield the outcome (5, 5, 3) in this modified game, deviating markedly from the outcome of the CES algorithm in the original game.

The difference is attributable to P2's newfound credible threat against the coalition of P1 and P3. Specifically, if P1 chooses action $R$, P2 could now credibly choose action $c$, with the expectation that the coalition $\{1,3\}$ would subsequently choose the best-response action $i$, leading to the outcome (3, 4, 3).\footnote{For the sake of simplicity, this example presupposes that payoff transfers are not feasible. However, a similar game could be constructed even if transfers were permitted.} Given that (5, 5, 3) is the outcome of the IR reference system in the parallel game $\Gamma_{\{1,2\}}$, as shown in Figure~\ref{fig:3pl-3} (G), the outcome (3, 4, 3) of the parallel game $\Gamma_{\{1,3\}}$ ceases to be IR for P1 because P1 strictly prefers (5, 5, 3) to (3, 4, 3). Consequently, the `on-path' CES for this modified game can be summarized as
\[ [\{L\},\{a,c\},\{e,g,j,k\};\{1,2\},3],\]
wherein P1 and P2 form a coalition, agent P1 chooses $L$, agent P2 chooses $a$ and $c$, and P3 chooses $e$, $g$, $j$, and $k$ at the respective decision nodes. The outcome of this CES is (5, 5, 3). 

This example clearly demonstrates how a credible threat by P2 effectively mitigates any potential for P3 to destabilize the $\{1,2\}$ coalition; that is, the parallel game coalition $\{1,3\}$ is not IR with respect to the parallel game coalition $\{1,2\}$.

\section{Applications}
\label{sec:applications}

\subsection{Oligopolistic competition with mergers}
\label{subsec:oligopoli}

Let $\Theta(\mathcal{P}, c, r)$ denote a coalitional oligopolistic competition with the option for firms to form coalitions such as mergers and cartels. Formally, $\Theta(\mathcal{P}, c, r)$ is defined as a cooperative strategic game $\Gamma=(P, T , I, u, \Sigma, H)$ with the following parameters and their corresponding interpretations. Each player $i\in P$ is called a firm. $\mathcal{P}$ denotes the inverse demand function, $c$ the cost function, and $r$ the sharing rule in case of a coalition among two or more firms. Note that $\Theta(\mathcal{P}, c, r)$ is a general model of (dynamic) oligopolistic competition that includes the standard $n$-player Cournot, Bertrand, and Stackelberg competitions. 

Under `standard' assumptions that are commonly invoked in non-cooperative models of oligopolistic competition, Theorem~\ref{thm:imperfect} implies that a CES exists in coalitional oligopolistic games.\footnote{The standard assumptions are the ones that guarantee the existence of Nash equilibrium in non-cooperative models of quantity or price competition; see, e.g., \citet{roberts1976}, \citet{novshek1985}, and \citet{shapiro1989}.} 

Note that there are many ways of modelling $n$-firm strategic interaction via cooperative strategic games based on how payoffs are defined. Next, I provide a simple example of Stackelberg competition in which there are a leader and $2$ followers who observe the quantity choice of the leader and then choose their quantities simultaneously. The purpose of the example is to illustrate a CES in a simple but non-trivial coalitional oligopolistic competition.

For ease of comparison with the Cournot competition and the standard Stackelberg competition, assume that the cost functions are symmetric. In this Stackelberg model, it is clear that the grand coalition (i.e., monopoly) payoff is greater than the total payoff of the firms when they act independently. Thus, there are incentives for the firms to form the grand coalition. 

Formally, consider a setting in which player 1 (or firm 1) is the leader, and each player $j \in \{2,3\}$ is a follower. Player $1$ chooses a quantity $q_1\geq 1$, and upon observing $q_1$, each follower $j$ simultaneously chooses a quantity $q_j\geq 0$. Let $Q=\sum^{3}_{i=1} q_i$ denote the total market quantity. The inverse demand function is given by $\mathcal{P}(Q)=a-bQ$, where $a> 0$ and $b> 0$. The cost function for each player $i$ is $c(q_i)=cq_i$, where $c>0$. 

The profit function for firm $i$ is given by $\Pi_i(Q)=\mathcal{P}(Q)q_i-c(q_i)$. If a coalition $C$ forms through a merger or collusion among two or more firms, the coalitional profit is $\Pi_C(Q)=\mathcal{P}(Q)q_C-c(q_C)$, where $q_C=\sum_{i\in C} q_i$. The individual profit for each `agent' firm $i$ in coalition $C$ is specified as defined as
\[
\Pi_{i|C}(Q)=\Pi^S_i(Q)+\frac{\Pi_C(Q)-\sum_{i\in C}\Pi^S_i(Q)} {|C|},
\] 
where $\Pi^S_i(Q)$ refers to the standard Stackelberg payoff for firm $i$ when all firms operate independently and non-cooperatively. In simpler terms, the firms within a merger distribute the net gains or losses from the coalition equally among themselves.\footnote{Although one could consider various vNM utility functions for the firms, the above-defined sharing rule is not implausible. Given that the firms are symmetric and that the leader enjoys higher profits than the followers in a non-cooperative Stackelberg setting, this rule ensures that the leader accrues a greater share of the profits if it engages in a coalition with followers.}

\begin{proposition}[Three-firm Stackelberg competition]
	\label{prop:stackelberg}
	In the Stackelberg model with a leader and two followers as described above, there is a unique CES outcome in which the leader receives a payoff of $\frac{(a-c)^2}{8b}$ and the followers each receive $\frac{(a-c)^2}{32b}$. The path of play for a CES in this game can be summarized as 
	\[ [(\frac{a-c}{2 b},\frac{a-c}{8 b},\frac{a-c}{8 b});1, \{2,3\}],\]
	where the leader does not collude with any other firm and produce $\frac{a-c}{2 b}$, while the two followers all collude and produce each $\frac{a-c}{8 b}$.	
\end{proposition}

For the proof, refer to section~\ref{subsec:proof4} in the Appendix. To offer an intuitive perspective on this result, Figure~\ref{fig:stackelberg} illustrates the payoff distributions for the three firms under both the CES and the non-cooperative Stackelberg equilibrium, given a specific set of parameters in this three-player game.

\begin{figure}[t]
\begin{center}
\begin{tikzpicture}
\begin{axis}[
    ymin=0, ymax=150,
    height=6.2cm, width=11.5cm,
	x tick label style={
		/pgf/number format/1000 sep=},
	ylabel= {The payoffs},
	enlargelimits=0,
	%legend style={at={(0.5,-0.1)},
	%anchor=north,legend columns=-1},
	ybar interval=0.6,
    %nodes near coords,
    %nodes near coords align={vertical},
    symbolic x coords={CES, Non-cooperative Stackelberg,c}
   ]
\addplot[style={bblue,fill=bblue,mark=none}]
	coordinates {(CES,125) (Non-cooperative Stackelberg,83)
		 (c,41) };
\addplot[style={rred,fill=rred,opacity=0.7,mark=none}]
	coordinates {(CES,31) (Non-cooperative Stackelberg,27)
		(c,39)  };
\addplot[style={ggreen,fill=ggreen,opacity=0.5,mark=none}]
	coordinates {(CES,31) (Non-cooperative Stackelberg,27)
		(c,39)  };
\addplot[red,line legend,sharp plot, update limits=false]
        coordinates {(CES,27) (c,27)};
\legend{Leader, Follower, Follower}
\end{axis}
\end{tikzpicture}
\caption{An illustration of CES and non-cooperative Stackelberg equilibrium payoffs in a three-firm competition, where $\frac{(a-c)^2}{b}=1000$.}
\label{fig:stackelberg}
\end{center}
\end{figure}

Surprisingly, Proposition~\ref{prop:stackelberg} shows that the unique CES outcome in the three-firm cooperative Stackelberg competition coincides with the outcome of the standard two-firm non-cooperative Stackelberg competition, even though the total profits are maximized when all firms collude. The leader, in strategically anticipating the potential collusion among the followers, derives greater benefits from this ensuing positive externality than it would by joining the followers' coalition to establish a monopoly. This apparent puzzle is not unique to the game described above; it has been studied for a long time (see, e.g., Salant et al. \citeyear{salant1983}).

\subsection{Logrolling}

The literature on logrolling has its roots in the influential works of \citet{buchanan1962} and \citet{riker1973}. The latter's vote trading paradox demonstrates that mutually beneficial vote trades can result in Pareto inferior outcomes for all parties involved, including the traders themselves.\footnote{For an in-depth discussion on this subject, the reader is referred to \citet{brams2003}.} More recently, \citet{casella2019} propose a vote trading model in which voters have separable preferences over proposals, and each proposal is decided by majority voting. Their finding is both general and striking. Specifically, they show that given any finite set of voters (assumed to be myopic) and proposals, and given any separable preferences and an initial vote profile, there exists a sequence of strictly payoff-improving trades that leads to a stable vote profile, meaning no further strictly payoff-improving trades can be made.

Let $L(K, v, r)$ denote a logrolling game, formally defined as a cooperative extensive form game with possibly imperfect information, $\Gamma=(P, T , I, u, \Sigma, H)$, supplemented by the following parameters and their respective interpretations. Each player $i\in P$ is called a voter. The set $K=\{1,2,...,\bar{k}\}$ specifies the range of proposals available for voting, and $r$ denotes the voting rule employed to accept or reject a proposal, such as majority rule. Each voter $i$ is allocated $v_i(k)$ votes to cast either in favor of or against proposal $k$.  Let $v=(v_1,...,v_n)$ be the initial vote profile. Given the voting rule, a pure strategy profile results in an outcome $o=(o_1, o_2,...,o_{\bar{k}})$, where each proposal $k$ is either accepted ($o_k=1$) or rejected ($o_k=0$). With a slight abuse of notation, $u_i(o)$ denotes $u_i(s')$, where $s'$ is the pure strategy profile that leads to outcome $o$. The concept of coalition formation in this context can be understood as an exchange of votes among the members of the coalition, either through a physical transfer as in \citet{casella2019} or via alternative agreements.

Next, I turn to the question of the existence of cooperative equilibrium systems in logrolling games. The following corollary follows directly from Theorem~\ref{thm:perfect} (perfect information games) and Theorem~\ref{thm:imperfect} (imperfect information games).
\begin{cor}
	\label{cor:logrolling}
	Every logrolling game with possibly imperfect information possesses a CES. If the game is of perfect information, then there is a CES that includes only pure strategies.
\end{cor}
In essence, Corollary~\ref{cor:logrolling} states that for any logrolling game---with any finite set of voters, any number of proposals, any initial vote profile, any type of voting rule, and any nature of preferences (either separable or not)---a CES is guaranteed to exist. This result substantially extends the existence result of \citet{casella2019} in four main dimensions: (i) the introduction of dynamically rational voters, (ii) the generality in the type of voting rule, (iii) the flexibility in the nature of voter preferences, and (iv) the inclusion of imperfect information games. Particularly noteworthy is the model's incorporation of dynamically rational voters. This extension is significant primarily because it precludes the possibility of voters engaging in myopically payoff-improving trades that could paradoxically reduce their overall utility.

In their working paper, \citet{casella2018} extend their analysis to include farsighted vote trading. They find that such trading may not necessarily lead to a stable outcome. The divergence between their results and those outlined in Corollary~\ref{cor:logrolling} arises from differences in both the frameworks and the underlying concepts. The notion of farsightedness they employ is consistent with the discussions found in the literature review in section~\ref{subsec:farsighted}. This differs from the application of dynamic rationality in the context of cooperative extensive form games.

\subsection{Monetary transfers and `corruption'}

In this subsection, I extend cooperative extensive form games to include strategic scenarios akin to `corruption' by incorporating wealth as an explicit parameter.  Let $\Gamma=(P, T , I, u, \Sigma, H, W)$ be a coalitonal extensive form game with wealth, where $W=(W_1, W_2,$ $..., W_n)$ denotes a wealth profile, with each $W_i \in \mathbb{R}_{+}$ representing the wealth of player $i$. Mutatis mutandis, the inclusion of wealth as an additional parameter does not undermine the validity of previous existence results. Specifically, Theorem~\ref{thm:perfect} and Theorem~\ref{thm:imperfect} continue to hold under this extended framework. 

Wealth operates as a mechanism by which non-individually rational reference systems can be rendered individually rational. Specifically, players may transfer wealth to incentivize other agents to cooperate at particular information sets, where they would otherwise have no rational motive to do so. This incorporation of wealth into the utility function creates a trade-off, balancing the act of transferring wealth against achieving a more favorable outcome. By incorporating wealth transfers, the model equips social scientists with a tool required to explore the strategic effects of situations commonly referred to as corruption. In essence, the model provides a framework for capturing transactions that effectively `but' an individual's cooperation across a range of strategic contexts.

\begin{figure}
	\centering
	\begin{tikzpicture}[font=\footnotesize,edge from parent/.style={draw,thick}]
	% Two node styles: solid and hollow
	\tikzstyle{solid node}=[circle,draw,inner sep=1.2,fill=black];
	\tikzstyle{hollow node}=[circle,draw,inner sep=1.2];
	% Specify spacing for each level of the tree
	\tikzstyle{level 1}=[level distance=15mm,sibling distance=50mm]
	\tikzstyle{level 2}=[level distance=15mm,sibling distance=25mm]
	\tikzstyle{level 3}=[level distance=15mm,sibling distance=15mm]
	% The Tree
	\node(0)[hollow node]{}
	child{node[solid node]{}
		child{node[solid node]{}
			child{node[below]{$\begin{pmatrix}0\\60\\40\\0\end{pmatrix}$} edge from parent node[left]{$F$}}
			child{node[below]{$\begin{pmatrix}10\\30\\60\\0\end{pmatrix}$} edge from parent node[right]{$A$}}
			edge from parent node[above left]{$F$}
		}
		child{node[below]{$\begin{pmatrix}30\\40\\30\\0\end{pmatrix}$}
			edge from parent node[above right]{$A$}
		}
		edge from parent node[above left]{Enter Market S}
	}
	child{node[solid node]{}
		child{node[below]{$\begin{pmatrix}20\\0\\0\\100\end{pmatrix}$}
			edge from parent node[above left]{$F$}
		}
		child{node[below]{$\begin{pmatrix}60\\0\\0\\70\end{pmatrix}$}
			edge from parent node[above right]{$A$}
		}
		edge from parent node[above right]{Enter Market L}
	};
	% information sets
	%\draw[loosely dotted,very thick](0-1-1)to[out=-15,in=195](0-2-1);
	%\draw[loosely dotted,very thick](0-1-2)to[out=-15,in=195](0-2-2);
	% movers
	\node[above,yshift=2]at(0){Firm 1};
	%\foreach \i in {1,2} \node[above,yshift=2]at(0-\i){Firm 2};
	\node[left,yshift=2]at(0-1){Firm 2};
	\node[right,yshift=2]at(0-2){Firm 4};
	\node[left,yshift=2]at(0-1-1){Firm 3};
	\end{tikzpicture}	
	\caption{International market entry game with an initial wealth distribution $W=(0,100,0,0)$}
	\label{fig:market_corruption}	
\end{figure}

To illustrate the incorporation of wealth transfers into this framework, consider Figure~\ref{fig:market_corruption}---a slight modification of the market entry game previously presented in Figure~\ref{fig:market}. Recall that in the CES of the game in Figure~\ref{fig:market_corruption}, Firms 1 and 4 collude, resulting in Firm 1 entering market L and Firm 4 choosing action A. The resulting outcome of this CES is $(60, 0, 0, 70)$  (for additional details, refer to subsection~\ref{subsec:credible_threats}).

Now, assume that the initial wealth profile is $W=(0,100,0,0)$ and, for the sake of simplicity, that the utility is linear in money. With these modifications, Firm 2 becomes willing to transfer 31 units of its wealth to Firm 1 to incentivize cooperation.\footnote{Here, I impose the simplifying constraint that only integer unit transfers are permitted.} This transfer clearly leaves Firm 1 better off, as it gains wealth, and it also benefits Firm 2, which would otherwise achieve a utility of zero in the original CES. Under these new conditions, the coalition $\{1, 2\}$ forms according to the CES algorithm, given the initial wealth structure. In this coalition, Firm 1 chooses action S, and Firm 2 chooses action A. The resulting CES outcome is now $(30, 40, 30, 0)$, and the final wealth profile adjusts to $W=(31, 69, 0, 0)$.

This example serves to highlight how the inclusion of wealth transfers can substantially alter strategic interactions and their resulting outcomes. The ability to incentivize cooperation through wealth transfers introduces a complex dynamic to the process of strategic coalition formation, one that can be effectively captured within this extended framework.

\section{Conclusions}

The present paper contributes to the existing literature by introducing a unified framework, called cooperative extensive form games. This framework accommodates not only the externalities and synergies arising from coalition formation but also addresses the dynamics of coalition formation and strategic behavior in a manner more comprehensive than prior models. Specifically, the framework allows for the formation of coalitions both among members of a sub-coalition and with external players. Within this framework, I introduce a novel solution concept called cooperative equilibrium system (CES). Unlike the Nash equilibrium, which is susceptible to multilateral profitable deviations, the CES is robust to both unilateral and multilateral credible deviations. This solution concept takes into account not only the profitability of deviations but also their credibility, which is defined with respect to an appropriate counterfactual that accounts for chain reactions triggered by the initial deviation.

The outcome of a CES outcome may be Pareto inefficient. This is evident since any non-cooperative extensive form game can be reformulated as a cooperative strategic game. An open question then arises: Is there a general class of coalitional utility functions, possibly derived from a well-defined cooperative game (e.g., in characteristic function form), that ensures Pareto efficiency of cooperative equilibrium systems?

I have identified multiple avenues for future research that can extend or modify the existing framework of cooperative strategic games. First, more nuanced coalition dynamics involving varying levels of cooperation---such as partial or mixed cooperation---warrant thorough analysis. Second, the integration of complex contracts into cooperative extensive form games adds additional layers of complexity that merit investigation, particularly when preferences exhibit cycling. Third, the role of disagreement points, whether they are exogenously defined or determined by specific rules, offers a broader context for understanding the dynamics of strategic coalition formation. Fourth, an examination of generalized cost structures, potentially including both monetary and psychological costs associated with coalition formation, provides avenues for expanding the current understanding of cooperative strategic games. Lastly, the refinement and generalization of cooperative equilibrium systems, coupled with the incorporation of wealth transfers, introduce complexities to the existing framework that necessitate further examination.

The framework and solution concept proposed in this paper have broad applicability across multiple disciplines. Political scientists could employ it to explore the dynamics of conflict and cooperation among nations. Likewise, the model is relevant for computer scientists working on cooperative behaviors in multi-agent systems. Biologists may find the framework useful for studying evolutionary patterns in species and genes. As previously elaborated in section~\ref{subsec:oligopoli}, the framework is also particularly relevant for economists investigating cooperation in strategic contexts, including airline alliances and oligopolistic cartels.\footnote{The applied researcher may find it useful to check Martin Antonov's implementation of the CES for three-player cooperative strategic games on the Game Theory Explorer \citep{savani2015}.}

\bibliographystyle{chicago}
\bibliography{me}

\section*{Appendix}
\section{Proofs}
\subsection{Proof of Proposition~\ref{prop:CES_SPNE}}
\label{subsec:proof3}

\begin{proof}
Let $\Gamma=(P, T , I, u, S, H)$ be a cooperative extensive form game with perfect information in which  the only feasible coalitions are singleton coalitions. This implies that $\Gamma$ is a non-cooperative extensive form game with perfect information. Notably, there does not exist any parallel game in $\Gamma$.
Hence, the non-cooperative reference system at a non-terminal node $x$ is always the credible reference system at $x$.

Under these conditions, Definition~\ref{def:CES} simplifies as follows: in a CES, a player $i=I(x)$, at each non-terminal node $x\in X$, chooses the action that maximizes $i$'s utility, given the CES of subgames starting at $x'\in f(x)$. Part (1) of Definition~\ref{def:CES} remains the same: a CES in a subgame with only one player prescribes playing the strategy that maximizes the player's utility in this subgame. These observations imply that CES strategies can be found via backward induction. Therefore, those strategies must coincide with subgame perfect equilibrium strategies in perfect information games.
\end{proof}

\subsection{Proof of Theorem~\ref{thm:perfect}}
\label{subsec:proof1}

The core of this proof rests on three key observations: (i) the proposed maximizations at each step of the CES algorithm are well-defined, (ii) the algorithm avoids indefinite cycles, and (iii) the algorithm produces a CES after a finite number of iterations.

\begin{enumerate}
	\item \textbf{Base case}: Consider $(n,\bar{x}) = (1,1)$ and $(\sigma^*,\pi^*)$ be a CES at $\Gamma(x)$ such that $\pi^*=i$ and $\sigma^*(x)\in \argmax_{s\in S} u_i(s)$. Note that the set $\argmax_{s\in S} u_i(s)$ is non-empty because there exists at least a pure action (out of finitely many pure actions) of player $i$ that maximizes player $i$'s utility at $x$. 
  
	\item \textbf{Induction step}: Assume that a CES exists for all subgames with parameters $(m,\bar{y})$ such that $1\leq m\leq n$, $1\leq\bar{y}\leq \bar{x}$, and $(n,\bar{x}) \neq (1,1)$ and $(m,\bar{y}) \neq (n,\bar{x})$. This is a sound assumption because all these parameters are finite.

	The subsequent steps demonstrate the existence of a CES for the subgame $\Gamma(x)$ with parameters $(n,\bar{x})$.
 	
	\begin{enumerate}
		\item Observe that given $[\sigma^*(x'),\pi^*(x')]$, $b^*_i(x)$  exists because
		\[
		\max_{b'_i(x)\in \Delta A_i(x)} u_{i}(b'_i(x)|\sigma^*(x'),\pi^*(x'))
		\]
		\[
		=\max_{b'_i(x)\in A_i(x)} u_{i}(b'_i(x)|\sigma^*(x'),\pi^*(x')),
		\]
		where $A_i(x)$ is a finite set; though, this does not preclude the existence of a (completely) mixed behavior strategy maximizer. Thus, the non-cooperative reference system $r_0(x)\coloneqq[\tau^{i}(x),\pi^{i}(x)]$ is well-defined.

		\item The reference systems, with the exception of non-cooperative reference systems, are cooperative equilibrium systems of relevant parallel games at $x$. Observe that each parallel game $\Gamma_{P_C}$ at $x$ has strictly fewer players than $\Gamma(x)$. Moreover, each subgame of the parallel game $\Gamma_{P_C}$ (excluding $\Gamma_{P_C}$ itself) possesses  strictly fewer nodes than $\Gamma_{P_C}$. Additionally, each parallel game of a subgame of $\Gamma_{P_C}$ has strictly fewer players than the original subgame. In simpler terms, for any given game, every parallel game possesses strictly fewer players, and each subgame possesses strictly fewer nodes than the original game. This iterative reduction in the magnitude of the parameters continues until the game under consideration converges to the one specified in the base case. Given the finite number of parallel games and subgames of a game, the CES for all such games including the subgame $\Gamma(x)$ is well-defined. It is noteworthy that the induction hypothesis assumes the existence of a CES for each $\Gamma_{P_C}$; the coherence of this assumption has now been substantiated.
        
        As a result, $(r_j(x))^{\bar{j}}_{j=1}$ where $r_j(x)\coloneqq[\tau^{C_j}(x),\pi^{C_j}(x)]$ 	is a finite, well-defined, and nondecreasing sequence from the perspective of $i$'s payoff.
			
		\item Assume the non-cooperative reference system $r^*_0(x)$ to be individually rational (IR). Given this assumption, inductively define a sequence of IR reference systems starting from $r^*_0(x)$. 

        Assume that for some $j \geq 0$, $r_j(x)$ is IR and denoted as $r^*_j(x)$. Let $j' > j$ be the smallest integer such that $r_{j'}(x)$ is IR with respect to $r^*_j(x)$. Let $j'>j$ be the smallest integer such that $r_{j'}(x)$ is IR with respect to  $r^*_j(x)$; that is, $u_{i'}(\tau^{C_{j'}}(x)|\pi^{C_{j'}}(x)) > u_{i'}(\tau^{C_{j}}(x)|\pi^{C_{j}}(x))$ for every agent $i'\in \bar{C}_{j'}$, where $C_{j'} \subseteq \bar{C}_{j'}$ and $\bar{C}_{j'}\in \pi^{C_{j'}}(x)$. In essence, each agent in the coalition $C_{j'}$ strictly prefers the CES of the parallel game $\Gamma_{P_{C_{j'}}}$ over that of $\Gamma_{P_{C_{j}}}$. In addition, $j'$ is chosen such that there is no IR reference system between $r^*_{j'}(x)$ and $r^*_j(x)$.
        
        Because $(r_j(x))^{\bar{j}}_{j=1}$ is a finite and nondecreasing sequence for $i$ and the sequence of IR reference systems is a strictly increasing subsequence of $(r_j(x))^{\bar{j}}_{j=0}$, there exists the credible reference system, denoted by $r^*_{\bar{l}}(x)=[\tau^{C_{\bar{l}}}(x),\pi^{C_{\bar{l}}}(x)]$, that maximizes $i$'s utility.
	\end{enumerate}
\end{enumerate}

Next, it is left to show that the CES algorithm does not enter any indefinite cycles, thereby ensuring its termination after a finite number of steps.  In the base case, the algorithm performs a standard utility maximization over a finite set of actions, precluding the existence of cycles. The induction step requires further elaboration concerning the potential for cycles. Importantly, this step operates recursively, assuming the existence of a CES for `smaller' games (i.e., games with fewer players or nodes). At each node, the algorithm compares the CES for each parallel game, beginning with the non-cooperative reference system (as outlined in Step 2.a). It then computes the credible IR reference system at the node through an inductive process (Steps 2.b and 2.c). This procedure does not include any cycles and will terminate after a finite number of iterations for two reasons: (i) each parallel game possesses finitely many players and actions, and any parallel game derived from a subgame possesses strictly fewer players due to coalition formation; and (ii) the credible IR reference system for each active player exists because there are finitely many such reference systems and the sequence of IR reference systems is a finite strictly increasing sequence as shown above. Thus, the induction step is also free from cycles.

Finally, I establish that the CES algorithm outputs a CES after executing a finite number of steps. Note that a CES of $\Gamma(x)$ is defined as system $[\sigma^*, \pi^*]$ where $[\sigma^*(x), \pi^*(x)]=r^*_{\bar{l}}(x)$, which extends $[\sigma^*(x'), \pi^*(x')]$ where $x'\in F(x)$. The existence of such a CES, $[\sigma^*, \pi^*]$, is guaranteed due to two reasons: (i) at each non-terminal node $x\in X$, including the root of $\Gamma$, the credible reference system $r^*_{\bar{l}}(x)$ at $x$ exists, as previously shown, and (ii) the game $\Gamma$ consists of a finite set of nodes. \qed

\subsection{The CES algorithm: Imperfect information games}
\label{subsec:CES_algo_imperfect}
Let $\Gamma$ be an extensive form game with imperfect information. I next define the CES algorithm which outputs a system on each subtree $T(h)$ starting at some information set $h$ by inducting on the number of players ($n$) and number of information sets ($\bar{h}$) in the successor information sets of $h$ including $h$ itself (and excluding the terminal nodes). For example, if $h$ is a penultimate information set, then $(n,\bar{h}) = (1,1)$. Let $i=I(h)$ be the active player at information set $h$.

\begin{enumerate}
	\item \textbf{Base case}: Let $(n,\bar{h}) = (1,1)$ and $g_{P'}$ be the  subgame at $root(h)$ where $P'$ is the set of players in the subgame.\footnote{Note that $g_{P'}$ is the only element in $G(h)$.} CES at $T(h)$ is defined by the pair $(\sigma^*(h),\pi^*(h))\coloneqq(\sigma^*,\pi^*)$ such that $\sigma^*(h)$ is a subgame perfect equilibrium in $g_{P'}$ and $\pi^*=P'$ in which no player forms a coalition.\footnote{Observe that $\sigma^*(h)$ is a strategy profile in $g_{P'}$.} 
	
	\item \textbf{Induction step}: 
	
	Assume that CES is defined for all subtrees with parameters $(m,\bar{y})$ satisfying  $1\leq m\leq n$, $1\leq\bar{y}\leq \bar{h}$ such that $(n,\bar{h}) \neq (1,1)$ and $(m,\bar{y}) \neq (n,\bar{h})$. By assumption, $[\sigma^*(h'),\pi^*(h')]$ is defined for all $h'\in F(h)$ where $h$ is the root of subtree $T(h)$. The solution is extended to subtree $T(h)$ with parameters $(n,\bar{h})$ as follows. I first define reference systems to iteratively compare the solutions of all parallel trees with the non-cooperative choice of $i$ at $h$.
	
	\begin{enumerate}
		\item The non-cooperative reference system at $h$, $r_0(h)$: Let $r_0(h)\coloneqq[\tau^{i}(h),\pi^{i}(h)]$ which is defined as the extension of $[\sigma^*(h'), \pi^*(h')]$ with $h'\in f'(h)$ to the  subgame starting at node $root(h)$ such that $[\tau^{i}(h),\pi^{i}(h)]$ is an SPNE in this subgame where each player in subgame $\Gamma(h)$ excluding those who act at an information set $h'\in F'(h)$ choose their strategies non-cooperatively.\footnote{It is worth noting that an SPNE is defined with respect to players, which may be coalitions, which are given by partition $\pi^*(h')$. In addition, the players/agents who act at an information set $h'\in F(h)$ choose their strategies according to the CES assumed by the induction hypothesis.}
		
		\item Reference systems: For any $(n,\bar{h})$ consider parallel tree $T_{P_C}(h)$ where $C\ni i$ is a feasible non-singleton coalition. Because $T_{P_C}(h)$ has strictly fewer players than $T(h)$ and has $\bar{h}$ information sets, its CES, $[\tau^{C}(h),\pi^{C}(h)]$, is defined by assumption.
		
		Let $(r_j(h))^{\bar{j}}_{j=1}$ be a sequence for $i$ where $r_j(h)\coloneqq[\tau^{C_j}(h),\pi^{C_j}(h)]$ satisfying (i) for every two indices $j$ and $j'$ with $j<j'$, we have $u_{i}(\tau^{C_j}(h)|\pi^{C_{j}}(h))\leq u_{i}(\tau^{C_{j'}}(h)|\pi^{C_{j'}}(h))$ (i.e., nondecreasing sequence for $i$), and 
		
		\noindent (ii) if $u_{i}(\tau^{C_j}(h)|\pi^{C_{j}}(h)) $ $= u_{i}(\tau^{C_{j'}}(h)|\pi^{C_{j'}}(h))$ and $C_j\subsetneq C_{j'}$, then $r_j$ precedes $r_{j'}$ for parallel tree coalitions $C_j$ and $C_{j'}$ containing $i$.\footnote{Ties are broken arbitrarily.}
		
		\item IR reference systems: The non-cooperative reference system $r_0(h)$ is IR by definition. The following IR reference systems are defined inductively as follows.
		
		Assume that $r_j(h)$ is IR, denoted as $r^*_j(h)$, for some $j\geq 0$. Let $j'>j$ be the smallest number such that $r_{j'}(h)$ is IR with respect to  $r^*_j(h)$---i.e., $u_{i'}(\tau^{C_{j'}}(h)|\pi^{C_{j'}}(h)) > u_{i'}(\tau^{C_{j}}(h)|\pi^{C_{j}}(h))$ for every $i'\in \bar{C}_{j'}$ where $C_{j'} \subseteq \bar{C}_{j'}$ and $\bar{C}_{j'}\in \pi^{C_{j'}}(h)$.
		
		Let $r^*_{\bar{l}}(h)$ be the credible reference system, which maximizes $i$'s utility. Note that $r^*_{\bar{l}}(h)=[\tau^{C_{\bar{l}}}(h),\pi^{C_{\bar{l}}}(h)]$. It may be that $r^*_0(h)$ where $\bar{l}=0$ is the only IR reference system.
	\end{enumerate}	
	CES of $T(h)$ is defined as $[\sigma^*, \pi^*]$ where $[\sigma^*(h), \pi^*(h)]=r^*_{\bar{l}}(h)$, which extends $[\sigma^*(h'), \pi^*(h')]$ where $h'\in F(h)$. When $h$ is the root of the game tree of $\Gamma$, $[\sigma^*, \pi^*]$ is said to be CES of $\Gamma$. $\blacksquare$
\end{enumerate}

\subsection{Proof of Theorem~\ref{thm:imperfect}}
\label{subsec:proof2}

The proof strategy for the existence of a CES in imperfect information games largely mirrors that employed for perfect information games, as outlined in the proof of Theorem~\ref{thm:perfect}. Thus, this subsection focuses exclusively on aspects that differentiate the two proofs. To be precise, the essence of this proof is built upon three main observations: (i) each maximization problem in the CES algorithm is well-defined, (ii) the algorithm precludes the presence of indefinite cycles, and (iii) the algorithm yields a CES within a finite number of iterations.

\begin{enumerate}
	\item \textbf{Base case}: Observe that in this case $g_{P'}$ denotes the  subgame at $root(h)$, where $P'$ represents the set of players in the subgame. A CES $(\sigma^*,\pi^*)$ at subtree $T(h)$ exists for two reasons: firstly, no player in $P'$ joins a coalition, and secondly the number of players in $g_{P'}$ is finite. Thus, game $g_{P'}$ possesses an SPNE, $\sigma^*(h)$.
	
	\item \textbf{Induction step}: Under the induction hypothesis, it is assumed that a CES is defined for all subtrees characterized by parameters $(m,\bar{y})$ satisfying  $1\leq m\leq n$ and $1\leq\bar{y}\leq \bar{h}$, subject to the conditions that  $(n,\bar{h}) \neq (1,1)$ and $(m,\bar{y}) \neq (n,\bar{h})$. 
 
	\begin{enumerate}
		\item The non-cooperative reference system at $h$, denoted $r_0(h)\coloneqq[\tau^{i}(h),\pi^{i}(h)]$: As in the perfect information case, the non-cooperative reference system includes a non-cooperative best-response of $i$ at $h$, which exists because the subgame starting at node $root(h)$ has an SPNE.
		
		\item Reference systems: Analogous to the perfect information case, the induction hypothesis ensures that a CES of every parallel tree $T_{P_C}(h)$ is well-defined. Thus, the nondecreasing sequence  $(r_j(h))^{\bar{j}}_{j=0}$ for player $i$ is also well-defined. 
		
		\item IR reference systems are defined inductively, and the sequence of IR  reference systems is strictly increasing with respect to player $i$'s payoff, as in the perfect information case. Given that there are finitely many IR reference systems, the credible reference system $r^*_{\bar{l}}(h)$ exists.
	\end{enumerate}	
\end{enumerate}

The reason for the absence of indefinite cycles in the CES algorithm is analogous to that in the perfect information setting. In summary, no cycles arise in either the base case or any inductive step for two primary reasons: (i) each parallel tree in the algorithm contains a finite number of players and actions, and a parallel tree derived from a subtree possesses strictly fewer players than the original subtree, due to the formation of some coalitions in the parallel tree; (ii) the sequence of IR reference systems for player $i$ is a strictly increasing and finite sequence, thereby ensuring the existence of the credible IR reference system for player $i$.

Finally, it is left to show that CES algorithm outputs a CES within a finite number of iterations. Analogous to the argument presented in section~\ref{subsec:proof1}, the existence of a CES $[\sigma^*, \pi^*]$ is guaranteed for two reasons: (i) at every information set $h$, including the root of $\Gamma$, the credible reference system $r^*_{\bar{l}}(h)$ exists, and (ii) the total number of information sets in $\Gamma$ is finite. \qed

\subsection{Proof of Proposition~\ref{prop:stackelberg}}
\label{subsec:proof4}
\begin{proof}
First, note that under full collusion the payoff of the monopoly would be $\frac{(a-c)^2}{4b}$. In the two-firm non-cooperative Stackelberg model the payoffs of the leader and the follower are $\frac{(a-c)^2}{8b}$ and $\frac{(a-c)^2}{16b}$, respectively. 

Second, in the three-player Stackelberg model described above the non-cooperative subgame perfect equilibrium payoffs of the leader and each of the followers are $\frac{(a-c)^2}{12b}$ and $\frac{(a-c)^2}{36b}$, respectively. To see this, notice that the reaction function of player 2 and player 3 are given by $\frac{a+c-b (q_1+q_3)}{2 b}$ and $\frac{a-b (q_1+q_2)+c}{2 b}$, respectively. Solving them simultaneously for any given $q_1$, we obtain $\frac{a-c-b q_1}{3 b} $, which is the Stackelberg best-response function for each follower. As a result, the leader's (non-cooperative) Stackelberg equilibrium production level is $\frac{a-c}{2 b}$ and the followers each produce $\frac{a-c}{6 b}$.

Notice that full collusion maximizes total profits (e.g., $\frac{(a-c)^2}{12b}+\frac{2(a-c)^2}{36b}<\frac{(a-c)^2}{4b}$). It is also clear that the followers would collude given the production of the leader. This is because each follower receives $\frac{(a-c)^2}{36b}$ in the three-player non-cooperative Stackelberg model whereas if the followers collude then they would each receive $\frac{1}{2}\frac{(a-c)^2}{16b}=\frac{(a-c)^2}{32b}>\frac{(a-c)^2}{36b}$. The main question left to answer is whether the leader can be better off by not colluding with the followers, anticipating that the followers would collude anyway. If all three firms collude, then the leader would receive $\frac{13(a-c)^2}{108b}$ since we have
\begin{equation}
\frac{(a-c)^2}{12b} + \frac{1}{3}\left (\frac{(a-c)^2}{4b}-\frac{(a-c)^2}{12b}-\frac{2(a-c)^2}{36b}\right) = \frac{13(a-c)^2}{108b}.\footnote{The leader would receive even less payoff in the case in which the leader colludes with only with one of the followers.}
\end{equation}

Note that $\frac{13(a-c)^2}{108b}<\frac{(a-c)^2}{8b}$, which is the leader's payoff in the case in which only the followers collude. This is because the case in which the followers collude is equivalent to the standard two-firm Stackelberg competition in which the leader receives $\frac{(a-c)^2}{8b}$ and the follower receives $\frac{(a-c)^2}{16b}$. Thus, the non-cooperative reference system of the leader, in which only the followers collude, is the credible reference system. As a result, in the three-player Stackelberg model there is a unique CES outcome in which the leader receives $\frac{13(a-c)^2}{108b}$ and the followers each receive $\frac{(a-c)^2}{32b}$. Note that when the followers collude there are many ways to produce $q_2+q_3$, so there are multiple cooperative equilibrium systems; though, in all of them the followers collude against the leader, and they all receive the same payoffs.
\end{proof}

\section{Illustration of the notation}
\begin{figure}[h!]
	\centering
		\begin{tikzpicture}[font=\footnotesize,edge from parent/.style={draw,thick}]
		% Two node styles: solid and hollow
		\tikzstyle{solid node}=[circle,draw,inner sep=1.2,fill=black];
		\tikzstyle{hollow node}=[circle,draw,inner sep=1.2];
		% Specify spacing for each level of the tree
		\tikzstyle{level 1}=[level distance=15mm,sibling distance=50mm]
		\tikzstyle{level 2}=[level distance=15mm,sibling distance=25mm]
		\tikzstyle{level 3}=[level distance=15mm,sibling distance=15mm]
		% The Tree
		\node(0)[hollow node]{}
		child{node[solid node]{}
			child{node[solid node]{}
				child{node[below]{} edge from parent node[left]{$e$}}
				child{node[below]{} edge from parent node[right]{$f$}}
				edge from parent node[above left]{$a$}
			}
			child{node[solid node]{}
				child{node[below]{} edge from parent node(s)[left]{$g$}}
				child{node[below]{} edge from parent node(t)[right]{$h$}}
				edge from parent node[above right]{$b$}
			}
			edge from parent node[above left]{$L$}
		}
		child{node[solid node]{}
			child{node[solid node]{}
				child{node[below]{} edge from parent node(m)[left]{$i$}}
				child{node[below]{} edge from parent node(n)[right]{$j$}}
				edge from parent node[above left]{$c$}
			}
			child{node[solid node]{}
				child{node[below]{} edge from parent node[left]{$k$}}
				child{node[below]{} edge from parent node[right]{$l$}}
				edge from parent node[above right]{$d$}
			}
			edge from parent node[above right]{$R$}
		};
		% information sets
		%\draw[loosely dotted,very thick](0-1-1)to[out=-15,in=195](0-2-1);
		%\draw[loosely dotted,very thick](0-1-2)to[out=-15,in=195](0-2-2);
		% movers
		\node[above,yshift=2]at(0){1};
		\node[left,yshift=2]at(0){$x_0$};
		\foreach \i in {1,2} \node[above,yshift=2]at(0-\i){2};
		\node[left,yshift=2]at(0-1){$x_1$};
		\node[right,yshift=2]at(0-2){$x_2$};
		\node[above,yshift=2]at(0-1-1){3};
		\node[left,yshift=2]at(0-1-1){$x_3$};
		\node[below]at(0-1-1-1){$z_1$};
		\node[below]at(0-1-1-2){$z_2$};
		
		\node[above,yshift=2]at(0-1-2){3};
		\node[right,yshift=2]at(0-1-2){$x_4$};
		\node[below]at(0-1-2-1){$z_3$};
		\node[below]at(0-1-2-2){$z_4$};
		
		\node[above,yshift=2]at(0-2-1){3};
		\node[left,yshift=2]at(0-2-1){$x_5$};
		\node[below]at(0-2-1-1){$z_5$};
		\node[below]at(0-2-1-2){$z_6$};
		
		\node[above,yshift=2]at(0-2-2){3};
		\node[right,yshift=2]at(0-2-2){$x_6$};
		\node[below]at(0-2-2-1){$z_7$};
		\node[below]at(0-2-2-2){$z_8$};
		
		%\foreach \i in {1,2,3,4} \node[above,yshift=2]at(0-1-\i){3};
		%	\node at($.5*(s)+.5*(t)$){1};
		%	\node at($.5*(m)+.5*(n)$){1};
		\end{tikzpicture}

	\caption{A three-player cooperative strategic game, where $x_0$ is the root of the game tree, $x_0$--$x_6$ are non-terminal nodes, and $z_1$--$x_8$ are terminal nodes.}
	\label{fig:3pl_standard_game}
\end{figure}

\begin{figure} [h!]
	\centering
		\begin{tikzpicture}[font=\footnotesize,edge from parent/.style={draw,thick}]
		% Two node styles: solid and hollow
		\tikzstyle{solid node}=[circle,draw,inner sep=1.2,fill=black];
		\tikzstyle{hollow node}=[circle,draw,inner sep=1.2];
		% Specify spacing for each level of the tree
		\tikzstyle{level 1}=[level distance=15mm,sibling distance=50mm]
		\tikzstyle{level 2}=[level distance=15mm,sibling distance=25mm]
		\tikzstyle{level 3}=[level distance=15mm,sibling distance=15mm]
		% The Tree
		\node(0)[hollow node]{}
		child{node[solid node]{}
			child{node[solid node]{}
				child{node[below]{} edge from parent node[left]{$e$}}
				child{node[below]{} edge from parent node[right]{$f$}}
				edge from parent node[above left]{$a$}
			}
			child{node[solid node]{}
				child{node[below]{} edge from parent node(s)[left]{$g$}}
				child{node[below]{} edge from parent node(t)[right]{$h$}}
				edge from parent node[above right]{$b$}
			}
			edge from parent node[above left]{$L$}
		}
		child{node[solid node]{}
			child{node[solid node]{}
				child{node[below]{} edge from parent node(m)[left]{$i$}}
				child{node[below]{} edge from parent node(n)[right]{$j$}}
				edge from parent node[above left]{$c$}
			}
			child{node[solid node]{}
				child{node[below]{} edge from parent node[left]{$k$}}
				child{node[below]{} edge from parent node[right]{$l$}}
				edge from parent node[above right]{$d$}
			}
			edge from parent node[above right]{$R$}
		};
		% information sets
		%\draw[loosely dotted,very thick](0-1-1)to[out=-15,in=195](0-2-1);
		%\draw[loosely dotted,very thick](0-1-2)to[out=-15,in=195](0-2-2);
		% movers
		\node[above,yshift=2]at(0){1,2};
		\foreach \i in {1,2} \node[above,yshift=2]at(0-\i){1,2};
		\node[above,yshift=2]at(0-1-1){3};
		\node[above,yshift=2]at(0-1-2){3};
		\node[above,yshift=2]at(0-2-1){3};
		\node[above,yshift=2]at(0-2-2){3};

		\node[below]at(0-1-1-1){$z_1$};
		\node[below]at(0-1-1-2){$z_2$};
		\node[below]at(0-1-2-1){$z_3$};
		\node[below]at(0-1-2-2){$z_4$};
		\node[below]at(0-2-1-1){$z_5$};
		\node[below]at(0-2-1-2){$z_6$};
		\node[below]at(0-2-2-1){$z_7$};
		\node[below]at(0-2-2-2){$z_8$};
		
		%\foreach \i in {1,2,3,4} \node[above,yshift=2]at(0-1-\i){3};
		%	\node at($.5*(s)+.5*(t)$){1};
		%	\node at($.5*(m)+.5*(n)$){1};
		\end{tikzpicture}

	\caption{A coalition between player 1 and player 2.}
	\label{fig:coalition-forming}
\end{figure}

\begin{figure}
	\centering
	\resizebox{1.02\textwidth}{!}{

		\begin{tikzpicture}[font=\footnotesize,edge from parent/.style={draw,thick}]
		% Two node styles: solid and hollow
		\tikzstyle{solid node}=[circle,draw,inner sep=1.2,fill=black];
		\tikzstyle{hollow node}=[circle,draw,inner sep=1.2];
		% Specify spacing for each level of the tree
		\tikzstyle{level 1}=[level distance=15mm,sibling distance=50mm]
		\tikzstyle{level 2}=[level distance=15mm,sibling distance=25mm]
		\tikzstyle{level 3}=[level distance=15mm,sibling distance=15mm]
		\node[draw] at (-4.5,0) {$\Gamma$};
		% The Tree
		\node(0)[hollow node]{}
		child{node[solid node]{}
			child{node[solid node]{}
				child{node[below]{} edge from parent node[left]{$e$}}
				child{node[below]{} edge from parent node[right]{$f$}}
				edge from parent node[above left]{$a$}
			}
			child{node[solid node]{}
				child{node[below]{} edge from parent node(s)[left]{$g$}}
				child{node[below]{} edge from parent node(t)[right]{$h$}}
				edge from parent node[above right]{$b$}
			}
			edge from parent node[above left]{$L$}
		}
		child{node[solid node]{}
			child{node[solid node]{}
				child{node[below]{} edge from parent node(m)[left]{$i$}}
				child{node[below]{} edge from parent node(n)[right]{$j$}}
				edge from parent node[above left]{$c$}
			}
			child{node[solid node]{}
				child{node[below]{} edge from parent node[left]{$k$}}
				child{node[below]{} edge from parent node[right]{$l$}}
				edge from parent node[above right]{$d$}
			}
			edge from parent node[above right]{$R$}
		};
		% information sets
		%\draw[loosely dotted,very thick](0-1-1)to[out=-15,in=195](0-2-1);
		%\draw[loosely dotted,very thick](0-1-2)to[out=-15,in=195](0-2-2);
		% movers
		\node[above,yshift=2]at(0){1};
		 \node[above,yshift=2]at(0-1){2};
		\node[above,yshift=2]at(0-2){2};
		\node[above,yshift=2]at(0-1-1){3};
		\node[above,yshift=2]at(0-1-2){3};
		\node[above,yshift=2]at(0-2-1){3};
		\node[above,yshift=2]at(0-2-2){3};

		\node[below]at(0-1-1-1){$z_1$};
		\node[below]at(0-1-1-2){$z_2$};
		\node[below]at(0-1-2-1){$z_3$};
		\node[below]at(0-1-2-2){$z_4$};
		\node[below]at(0-2-1-1){$z_5$};
		\node[below]at(0-2-1-2){$z_6$};
		\node[below]at(0-2-2-1){$z_7$};
		\node[below]at(0-2-2-2){$z_8$};
		
		%\foreach \i in {1,2,3,4} \node[above,yshift=2]at(0-1-\i){3};
		%	\node at($.5*(s)+.5*(t)$){1};
		%	\node at($.5*(m)+.5*(n)$){1};
		\end{tikzpicture}
	
	\qquad
	
		\begin{tikzpicture}[font=\footnotesize,edge from parent/.style={draw,thick}]
		% Two node styles: solid and hollow
		\tikzstyle{solid node}=[circle,draw,inner sep=1.2,fill=black];
		\tikzstyle{hollow node}=[circle,draw,inner sep=1.2];
		% Specify spacing for each level of the tree
		\tikzstyle{level 1}=[level distance=15mm,sibling distance=50mm]
		\tikzstyle{level 2}=[level distance=15mm,sibling distance=25mm]
		\tikzstyle{level 3}=[level distance=15mm,sibling distance=15mm]
		% The Tree
		\node[draw] at (-4.5,0) {$\Gamma_{P_{\{1,2\}}}$};
		\node(0)[hollow node]{}
		child{node[solid node]{}
			child{node[solid node]{}
				child{node[below]{} edge from parent node[left]{$e$}}
				child{node[below]{} edge from parent node[right]{$f$}}
				edge from parent node[above left]{$a$}
			}
			child{node[solid node]{}
				child{node[below]{} edge from parent node(s)[left]{$g$}}
				child{node[below]{} edge from parent node(t)[right]{$h$}}
				edge from parent node[above right]{$b$}
			}
			edge from parent node[above left]{$L$}
		}
		child{node[solid node]{}
			child{node[solid node]{}
				child{node[below]{} edge from parent node(m)[left]{$i$}}
				child{node[below]{} edge from parent node(n)[right]{$j$}}
				edge from parent node[above left]{$c$}
			}
			child{node[solid node]{}
				child{node[below]{} edge from parent node[left]{$k$}}
				child{node[below]{} edge from parent node[right]{$l$}}
				edge from parent node[above right]{$d$}
			}
			edge from parent node[above right]{$R$}
		};
		% information sets
		%\draw[loosely dotted,very thick](0-1-1)to[out=-15,in=195](0-2-1);
		%\draw[loosely dotted,very thick](0-1-2)to[out=-15,in=195](0-2-2);
		% movers
		\node[above,yshift=2]at(0){1,2};
		\foreach \i in {1,2} \node[above,yshift=2]at(0-\i){1,2};
		\node[above,yshift=2]at(0-1-1){3};
		\node[above,yshift=2]at(0-1-2){3};
		\node[above,yshift=2]at(0-2-1){3};
		\node[above,yshift=2]at(0-2-2){3};

		\node[below]at(0-1-1-1){$z_1$};
		\node[below]at(0-1-1-2){$z_2$};
		\node[below]at(0-1-2-1){$z_3$};
		\node[below]at(0-1-2-2){$z_4$};
		\node[below]at(0-2-1-1){$z_5$};
		\node[below]at(0-2-1-2){$z_6$};
		\node[below]at(0-2-2-1){$z_7$};
		\node[below]at(0-2-2-2){$z_8$};
		
		%\foreach \i in {1,2,3,4} \node[above,yshift=2]at(0-1-\i){3};
		%	\node at($.5*(s)+.5*(t)$){1};
		%	\node at($.5*(m)+.5*(n)$){1};
		\end{tikzpicture}

	}
	\\
	\resizebox{1.02\textwidth}{!}{

		\begin{tikzpicture}[font=\footnotesize,edge from parent/.style={draw,thick}]
		% Two node styles: solid and hollow
		\tikzstyle{solid node}=[circle,draw,inner sep=1.2,fill=black];
		\tikzstyle{hollow node}=[circle,draw,inner sep=1.2];
		% Specify spacing for each level of the tree
		\tikzstyle{level 1}=[level distance=15mm,sibling distance=50mm]
		\tikzstyle{level 2}=[level distance=15mm,sibling distance=25mm]
		\tikzstyle{level 3}=[level distance=15mm,sibling distance=15mm]
		\node[draw] at (-4.5,0) {$\Gamma_{P_{\{1,3\}}}$};
		% The Tree
		\node(0)[hollow node]{}
		child{node[solid node]{}
			child{node[solid node]{}
				child{node[below]{} edge from parent node[left]{$e$}}
				child{node[below]{} edge from parent node[right]{$f$}}
				edge from parent node[above left]{$a$}
			}
			child{node[solid node]{}
				child{node[below]{} edge from parent node(s)[left]{$g$}}
				child{node[below]{} edge from parent node(t)[right]{$h$}}
				edge from parent node[above right]{$b$}
			}
			edge from parent node[above left]{$L$}
		}
		child{node[solid node]{}
			child{node[solid node]{}
				child{node[below]{} edge from parent node(m)[left]{$i$}}
				child{node[below]{} edge from parent node(n)[right]{$j$}}
				edge from parent node[above left]{$c$}
			}
			child{node[solid node]{}
				child{node[below]{} edge from parent node[left]{$k$}}
				child{node[below]{} edge from parent node[right]{$l$}}
				edge from parent node[above right]{$d$}
			}
			edge from parent node[above right]{$R$}
		};
		% information sets
		%\draw[loosely dotted,very thick](0-1-1)to[out=-15,in=195](0-2-1);
		%\draw[loosely dotted,very thick](0-1-2)to[out=-15,in=195](0-2-2);
		% movers
		\node[above,yshift=2]at(0){1,3};
		\foreach \i in {1,2} \node[above,yshift=2]at(0-\i){2};
		\node[above,yshift=2]at(0-1-1){1,3};
		\node[above,yshift=2]at(0-1-2){1,3};
		\node[above,yshift=2]at(0-2-1){1,3};
		\node[above,yshift=2]at(0-2-2){1,3};

		\node[below]at(0-1-1-1){$z_1$};
		\node[below]at(0-1-1-2){$z_2$};
		\node[below]at(0-1-2-1){$z_3$};
		\node[below]at(0-1-2-2){$z_4$};
		\node[below]at(0-2-1-1){$z_5$};
		\node[below]at(0-2-1-2){$z_6$};
		\node[below]at(0-2-2-1){$z_7$};
		\node[below]at(0-2-2-2){$z_8$};
		
		%\foreach \i in {1,2,3,4} \node[above,yshift=2]at(0-1-\i){3};
		%	\node at($.5*(s)+.5*(t)$){1};
		%	\node at($.5*(m)+.5*(n)$){1};
		\end{tikzpicture}

		\qquad

		\begin{tikzpicture}[font=\footnotesize,edge from parent/.style={draw,thick}]
		% Two node styles: solid and hollow
		\tikzstyle{solid node}=[circle,draw,inner sep=1.2,fill=black];
		\tikzstyle{hollow node}=[circle,draw,inner sep=1.2];
		% Specify spacing for each level of the tree
		\tikzstyle{level 1}=[level distance=15mm,sibling distance=50mm]
		\tikzstyle{level 2}=[level distance=15mm,sibling distance=25mm]
		\tikzstyle{level 3}=[level distance=15mm,sibling distance=15mm]
		\node[draw] at (-4.5,0) {$\Gamma_{P_{\{1,2,3\}}}$};
		% The Tree
		\node(0)[hollow node]{}
		child{node[solid node]{}
			child{node[solid node]{}
				child{node[below]{} edge from parent node[left]{$e$}}
				child{node[below]{} edge from parent node[right]{$f$}}
				edge from parent node[above left]{$a$}
			}
			child{node[solid node]{}
				child{node[below]{} edge from parent node(s)[left]{$g$}}
				child{node[below]{} edge from parent node(t)[right]{$h$}}
				edge from parent node[above right]{$b$}
			}
			edge from parent node[above left]{$L$}
		}
		child{node[solid node]{}
			child{node[solid node]{}
				child{node[below]{} edge from parent node(m)[left]{$i$}}
				child{node[below]{} edge from parent node(n)[right]{$j$}}
				edge from parent node[above left]{$c$}
			}
			child{node[solid node]{}
				child{node[below]{} edge from parent node[left]{$k$}}
				child{node[below]{} edge from parent node[right]{$l$}}
				edge from parent node[above right]{$d$}
			}
			edge from parent node[above right]{$R$}
		};
		% information sets
		%\draw[loosely dotted,very thick](0-1-1)to[out=-15,in=195](0-2-1);
		%\draw[loosely dotted,very thick](0-1-2)to[out=-15,in=195](0-2-2);
		% movers
		\node[above,yshift=2]at(0){1,2,3};
		\foreach \i in {1,2} \node[above,yshift=2]at(0-\i){1,2,3};
		\node[above left]at(0-1-1){1,2,3};
		\node[above right]at(0-1-2){1,2,3};
		\node[above left]at(0-2-1){1,2,3};
		\node[above right]at(0-2-2){1,2,3};

		\node[below]at(0-1-1-1){$z_1$};
		\node[below]at(0-1-1-2){$z_2$};
		\node[below]at(0-1-2-1){$z_3$};
		\node[below]at(0-1-2-2){$z_4$};
		\node[below]at(0-2-1-1){$z_5$};
		\node[below]at(0-2-1-2){$z_6$};
		\node[below]at(0-2-2-1){$z_7$};
		\node[below]at(0-2-2-2){$z_8$};
		
		%\foreach \i in {1,2,3,4} \node[above,yshift=2]at(0-1-\i){3};
		%	\node at($.5*(s)+.5*(t)$){1};
		%	\node at($.5*(m)+.5*(n)$){1};
		\end{tikzpicture}

}
	\caption{An illustration of an original game and its three parallel games at the root $x_0$.}
	\label{fig:parallelgames}	
\end{figure}

\clearpage

\begin{figure}[h]
	\centering
	%\resizebox{1.02\textwidth}{!}{

		\begin{tikzpicture}[font=\footnotesize,edge from parent/.style={draw,thick}]
		% Two node styles: solid and hollow
		\tikzstyle{solid node}=[circle,draw,inner sep=1.2,fill=black];
		\tikzstyle{hollow node}=[circle,draw,inner sep=1.2];
		% Specify spacing for each level of the tree
		\tikzstyle{level 1}=[level distance=15mm,sibling distance=50mm]
		\tikzstyle{level 2}=[level distance=15mm,sibling distance=25mm]
		\tikzstyle{level 3}=[level distance=15mm,sibling distance=15mm]
		% The Tree
		\node[draw] at (-4.5,0.5) {$[\sigma(x_0, g_{x_0}),\pi(x_0, g_{x_0})]$};
		\node(0)[hollow node]{}
		child{node[solid node]{}
			child{node[solid node]{}
				child{node[below]{} edge from parent[->,solid, thick] node[left]{$e$}}
				child{node[below]{} edge from parent[black, solid, thick] node[right]{$f$}}
				edge from parent[solid] node[above left]{$a$}
			}
			child{node[solid node]{}
				child{node[below]{} edge from parent[->,solid, thick] node(s)[left]{$g$}}
				child{node[below]{} edge from parent[solid] node(t)[right]{$h$}}
				edge from parent[->,solid, thick] node[above right]{$b$}
			}
			edge from parent[->,solid, thick] node[above left]{$L$}
		}
		child{node[solid node]{}
			child{node[solid node]{}
				child{node[below]{} edge from parent[solid] node(m)[left]{$i$}}
				child{node[below]{} edge from parent[->,solid, thick] node(n)[right]{$j$}}
				edge from parent[solid] node[above left]{$c$}
			}
			child{node[solid node]{}
				child{node[below]{} edge from parent[solid] node[left]{$k$}}
				child{node[below]{} edge from parent[->,solid, thick] node[right]{$l$}}
				edge from parent[->, solid, thick] node[above right]{$d$}
			}
			edge from parent[solid] node[above right]{$R$}
		};
		% information sets
		%\draw[loosely dotted,very thick](0-1-1)to[out=-15,in=195](0-2-1);
		%\draw[loosely dotted,very thick](0-1-2)to[out=-15,in=195](0-2-2);
		% movers
		\node[above,yshift=2]at(0){1,2,3};
		\foreach \i in {1,2} \node[above,yshift=2]at(0-\i){1,2,3};
		\node[above left]at(0-1-1){1,2,3};
		\node[above right]at(0-1-2){1,2,3};
		\node[above left]at(0-2-1){1,2,3};
		\node[above right]at(0-2-2){1,2,3};
		
				\node[below]at(0-1-1-1){$z_1$};
				\node[below]at(0-1-1-2){$z_2$};
				\node[below]at(0-1-2-1){$z_3$};
				\node[below]at(0-1-2-2){$z_4$};
				\node[below]at(0-2-1-1){$z_5$};
				\node[below]at(0-2-1-2){$z_6$};
				\node[below]at(0-2-2-1){$z_7$};
				\node[below]at(0-2-2-2){$z_8$};
		%\foreach \i in {1,2,3,4} \node[above,yshift=2]at(0-1-\i){3};
		%	\node at($.5*(s)+.5*(t)$){1};
		%	\node at($.5*(m)+.5*(n)$){1};
		\end{tikzpicture}

			\begin{tikzpicture}[font=\footnotesize,edge from parent/.style={draw,thick}]
			% Two node styles: solid and hollow
			\tikzstyle{solid node}=[circle,draw,inner sep=1.2,fill=black];
			\tikzstyle{hollow node}=[circle,draw,inner sep=1.2];
			% Specify spacing for each level of the tree
			\tikzstyle{level 1}=[level distance=15mm,sibling distance=50mm]
			\tikzstyle{level 2}=[level distance=15mm,sibling distance=25mm]
			\tikzstyle{level 3}=[level distance=15mm,sibling distance=15mm]
			% The Tree
			\node[draw] at (-4.5,-0.5) {$[\sigma(x_0, g_{x_1}),\pi(x_0, g_{x_1})]$};
			\node[draw] at (4.5,-0.5) {$[\sigma(x_0, g_{x_2}),\pi(x_0, g_{x_2})]$};
			\node(0)[gray, hollow node]{}
			child{node[solid node]{}
				child{node[black, solid node]{}
					child{node[below]{} edge from parent[ black, solid, thick] node[left]{$e$}}
					child{node[below]{} edge from parent[->,black, solid, thick] node[right]{$f$}}
					edge from parent[->, solid, black, thick] node[above left]{$a$}
				}
				child{node[black, solid node]{}
					child{node[below]{} edge from parent[->,solid, black, thick] node(s)[left]{$g$}}
					child{node[below]{} edge from parent[black, solid, thick] node(t)[right]{$h$}}
					edge from parent[black, solid, thick] node[above right]{$b$}
				}
				edge from parent[solid, gray] node[above left]{\textcolor{gray}{$L$}}
			}
			child{node[solid node]{}
				child{node[black, solid node]{}
					child{node[below]{} edge from parent[black, solid, thick] node(m)[left]{$i$}}
					child{node[black, below]{} edge from parent[->,solid, black, thick] node(n)[right]{$j$}}
					edge from parent[->,solid, black, thick] node[above left]{$c$}
				}
				child{node[black, solid node]{}
					child{node[black, below]{} edge from parent[->,black, solid, thick] node[left]{$k$}}
					child{node[below]{} edge from parent[black, solid, thick] node[right]{$l$}}
					edge from parent[black, solid, thick] node[above right]{$d$}
				}
				edge from parent[solid, gray] node[above right]{\textcolor{gray}{$R$}}
			};
			% information sets
			%\draw[loosely dotted,very thick](0-1-1)to[out=-15,in=195](0-2-1);
			%\draw[loosely dotted,very thick](0-1-2)to[out=-15,in=195](0-2-2);
			% movers
			\node[above,yshift=2]at(0){\textcolor{gray}{1}};
			\foreach \i in {1} \node[above,yshift=2]at(0-\i){2,3};
			\foreach \i in {2} \node[above,yshift=2]at(0-\i){2,3};
			\node[above,yshift=2]at(0-1-1){2,3};
			\node[above,yshift=2]at(0-1-2){2,3};
			\node[above,yshift=2]at(0-2-1){2,3};
			\node[above,yshift=2]at(0-2-2){2,3};
			
				\node[below]at(0-1-1-1){$z_1$};
				\node[below]at(0-1-1-2){$z_2$};
				\node[below]at(0-1-2-1){$z_3$};
				\node[below]at(0-1-2-2){$z_4$};
				\node[below]at(0-2-1-1){$z_5$};
				\node[below]at(0-2-1-2){$z_6$};
				\node[below]at(0-2-2-1){$z_7$};
				\node[below]at(0-2-2-2){$z_8$};	
			%\foreach \i in {1,2,3,4} \node[above,yshift=2]at(0-1-\i){3};
			%	\node at($.5*(s)+.5*(t)$){1};
			%	\node at($.5*(m)+.5*(n)$){1};
			\end{tikzpicture}		
			
			\begin{tikzpicture}[font=\footnotesize,edge from parent/.style={draw,thick}]
			% Two node styles: solid and hollow
			\tikzstyle{solid node}=[circle,draw,inner sep=1.2,fill=black];
			\tikzstyle{hollow node}=[circle,draw,inner sep=1.2];
			% Specify spacing for each level of the tree
			\tikzstyle{level 1}=[level distance=15mm,sibling distance=50mm]
			\tikzstyle{level 2}=[level distance=15mm,sibling distance=25mm]
			\tikzstyle{level 3}=[level distance=15mm,sibling distance=15mm]
			% The Tree
			\node[draw] at (-6.2,-2.3) {$[\sigma(x_0, g_{x_3}),\pi(x_0, g_{x_3})]$};
			\node(0)[gray, hollow node]{}
			child{node[gray, solid node, color=gray]{}
				child{node[black, solid node]{}
					child{node[below]{} edge from parent[->, black, solid, thick] node[left]{$e$}}
					child{node[below]{} edge from parent[black, solid, thick] node[right]{$f$}}
					edge from parent[] node[above left]{\textcolor{gray}{$a$}}
				}
				child{node[black, solid node]{}
					child{node[below]{} edge from parent[solid, black, thick] node(s)[left]{$g$}}
					child{node[below]{} edge from parent[->, black, solid, thick] node(t)[right]{$h$}}
					edge from parent[] node[above right]{\textcolor{gray}{$b$}}
				}
				edge from parent[solid, gray] node[above left]{\textcolor{gray}{$L$}}
			}
			child{node[gray, solid node, color=gray]{}
				child{node[black, solid node]{}
					child{node[below]{} edge from parent[->, black, solid, thick] node(m)[left]{$i$}}
					child{node[black, below]{} edge from parent[solid, black, thick] node(n)[right]{$j$}}
					edge from parent[] node[above left]{\textcolor{gray}{$c$}}
				}
				child{node[black, solid node]{}
					child{node[black, below]{} edge from parent[black, solid, thick] node[left]{$k$}}
					child{node[below]{} edge from parent[->,black, solid, thick] node[right]{$l$}}
					edge from parent[] node[above right]{\textcolor{gray}{$d$}}
				}
				edge from parent[solid, gray] node[above right]{\textcolor{gray}{$R$}}
			};
			% information sets
			%\draw[loosely dotted,very thick](0-1-1)to[out=-15,in=195](0-2-1);
			%\draw[loosely dotted,very thick](0-1-2)to[out=-15,in=195](0-2-2);
			% movers
			\node[above,yshift=2]at(0){\textcolor{gray}{1}};
			\foreach \i in {1,2} \node[above,yshift=2]at(0-\i){\textcolor{gray}{2}};
			\node[above,yshift=2]at(0-1-1){3};
			\node[above,yshift=2]at(0-1-2){3};
			\node[above,yshift=2]at(0-2-1){3};
			\node[above,yshift=2]at(0-2-2){3};
			
				\node[below]at(0-1-1-1){$z_1$};
				\node[below]at(0-1-1-2){$z_2$};
				\node[below]at(0-1-2-1){$z_3$};
				\node[below]at(0-1-2-2){$z_4$};
				\node[below]at(0-2-1-1){$z_5$};
				\node[below]at(0-2-1-2){$z_6$};
				\node[below]at(0-2-2-1){$z_7$};
				\node[below]at(0-2-2-2){$z_8$};	
			%\foreach \i in {1,2,3,4} \node[above,yshift=2]at(0-1-\i){3};
			%	\node at($.5*(s)+.5*(t)$){1};
			%	\node at($.5*(m)+.5*(n)$){1};
			\end{tikzpicture}	
					
%}
	\caption{An illustration of a system $[\sigma(x_0,\,\cdot\,),\pi(x_0,\,\cdot\,)]$. }
	\label{fig:systems}	
 \floatfoot{Note: For each $j\in \{1,2,...,6\}$, $g_{x_j}$ is in $G(x_0)$, which is the set of all subgames of the entire game. For the labels of the nodes, refer to Figure~\ref{fig:3pl_standard_game}. Also note that $[\sigma(x_0, g_{x_1}),\pi(x_0, g_{x_1})]$ is a system for the subgame starting at $x_1$ and $[\sigma(x_0, g_{x_2}),\pi(x_0, g_{x_2})]$ is a system for the subgame starting at $x_2$ etc.}
\end{figure}

\clearpage

\begin{figure}[h]
	\centering
	\resizebox{1.02\textwidth}{!}{
	\begin{tikzpicture}[font=\footnotesize,edge from parent/.style={draw,thick}]
	% Two node styles: solid and hollow
	\tikzstyle{solid node}=[circle,draw,inner sep=1.2,fill=black];
	\tikzstyle{hollow node}=[circle,draw,inner sep=1.2];
	% Specify spacing for each level of the tree
	\tikzstyle{level 1}=[level distance=15mm,sibling distance=50mm]
	\tikzstyle{level 2}=[level distance=15mm,sibling distance=25mm]
	\tikzstyle{level 3}=[level distance=15mm,sibling distance=15mm]
	% The Tree
	\node[draw] at (-4.5,0) {$r_0(x_0)$};
	\node(0)[hollow node]{}
	child{node[solid node]{}
		child{node[solid node]{}
			child{node[below]{} edge from parent[->, black, solid, very thick] node[left]{$e$}}
			child{node[below]{} edge from parent[black, solid, thick] node[right]{$f$}}
			edge from parent[thick] node[above left]{$a$}
		}
		child{node[solid node]{}
			child{node[below]{} edge from parent[solid, thick] node(s)[left]{$g$}}
			child{node[below]{} edge from parent[->, black, solid, very thick] node(t)[right]{$h$}}
			edge from parent[->, black, solid, very thick] node[above right]{$b$}
		}
		edge from parent[solid] node[above left]{$L$}
	}
	child{node[solid node]{}
		child{node[solid node]{}
			child{node[below]{} edge from parent[black, solid, thick] node(m)[left]{$i$}}
			child{node[below]{} edge from parent[->,solid, thick] node(n)[right]{$j$}}
			edge from parent[->,thick] node[above left]{$c$}
		}
		child{node[solid node]{}
			child{node[below]{} edge from parent[->,black, solid, thick] node[left]{$k$}}
			child{node[below]{} edge from parent[black, solid, thick] node[right]{$l$}}
			edge from parent[black, solid, thick] node[above right]{$d$}
		}
		edge from parent[->, solid, thick] node[above right]{$R$}
	};
	% information sets
	%\draw[loosely dotted,very thick](0-1-1)to[out=-15,in=195](0-2-1);
	%\draw[loosely dotted,very thick](0-1-2)to[out=-15,in=195](0-2-2);
	% movers
	\node[above,yshift=2]at(0){1};
	\foreach \i in {1} \node[above,yshift=2]at(0-\i){2,3};
	\foreach \i in {2} \node[above,yshift=2]at(0-\i){2};
	\node[above,yshift=2]at(0-1-1){2,3};
	\node[above,yshift=2]at(0-1-2){2,3};
	\node[above,yshift=2]at(0-2-1){3};
	\node[above,yshift=2]at(0-2-2){3};
	
		\node[below]at(0-1-1-1){$z_1$};
		\node[below]at(0-1-1-2){$z_2$};
		\node[below]at(0-1-2-1){$z_3$};
		\node[below]at(0-1-2-2){$z_4$};
		\node[below]at(0-2-1-1){$z_5$};
		\node[below]at(0-2-1-2){$z_6$};
		\node[below]at(0-2-2-1){$z_7$};
		\node[below]at(0-2-2-2){$z_8$};	
	%\foreach \i in {1,2,3,4} \node[above,yshift=2]at(0-1-\i){3};
	%	\node at($.5*(s)+.5*(t)$){1};
	%	\node at($.5*(m)+.5*(n)$){1};
	\end{tikzpicture}		
		\qquad
	\begin{tikzpicture}[font=\footnotesize,edge from parent/.style={draw,thick}]
% Two node styles: solid and hollow
\tikzstyle{solid node}=[circle,draw,inner sep=1.2,fill=black];
\tikzstyle{hollow node}=[circle,draw,inner sep=1.2];
% Specify spacing for each level of the tree
\tikzstyle{level 1}=[level distance=15mm,sibling distance=50mm]
\tikzstyle{level 2}=[level distance=15mm,sibling distance=25mm]
\tikzstyle{level 3}=[level distance=15mm,sibling distance=15mm]
% The Tree
\node[draw] at (-4.5,0) {$r_1(x_0)$};
\node(0)[hollow node]{}
child{node[solid node]{}
	child{node[solid node]{}
		child{node[below]{} edge from parent[->,solid, thick] node[left]{$e$}}
		child{node[below]{} edge from parent[black, solid, thick] node[right]{$f$}}
		edge from parent[->,solid, very thick] node[above left]{$a$}
	}
	child{node[solid node]{}
		child{node[below]{} edge from parent[->,solid, thick] node(s)[left]{$g$}}
		child{node[below]{} edge from parent[solid] node(t)[right]{$h$}}
		edge from parent[solid] node[above right]{$b$}
	}
	edge from parent[->,solid,very thick] node[above left]{$L$}
}
child{node[solid node]{}
	child{node[solid node]{}
		child{node[below]{} edge from parent[black, solid, thick] node(m)[left]{$i$}}
		child{node[below]{} edge from parent[->,solid, thick] node(n)[right]{$j$}}
		edge from parent[->, solid, very thick] node[above left]{$c$}
	}
	child{node[solid node]{}
		child{node[below]{} edge from parent[->,black, solid, thick] node[left]{$k$}}
		child{node[below]{} edge from parent[black, solid, thick] node[right]{$l$}}
		edge from parent[black, solid, thick] node[above right]{$d$}
	}
	edge from parent[solid, thick] node[above right]{$R$}
};
% information sets
%\draw[loosely dotted,very thick](0-1-1)to[out=-15,in=195](0-2-1);
%\draw[loosely dotted,very thick](0-1-2)to[out=-15,in=195](0-2-2);
% movers
\node[above,yshift=2]at(0){1,2};
\foreach \i in {1,2} \node[above,yshift=2]at(0-\i){1,2};
\node[above,yshift=2]at(0-1-1){3};
\node[above,yshift=2]at(0-1-2){3};
\node[above,yshift=2]at(0-2-1){3};
\node[above,yshift=2]at(0-2-2){3};

		\node[below]at(0-1-1-1){$z_1$};
		\node[below]at(0-1-1-2){$z_2$};
		\node[below]at(0-1-2-1){$z_3$};
		\node[below]at(0-1-2-2){$z_4$};
		\node[below]at(0-2-1-1){$z_5$};
		\node[below]at(0-2-1-2){$z_6$};
		\node[below]at(0-2-2-1){$z_7$};
		\node[below]at(0-2-2-2){$z_8$};
		
%\foreach \i in {1,2,3,4} \node[above,yshift=2]at(0-1-\i){3};
%	\node at($.5*(s)+.5*(t)$){1};
%	\node at($.5*(m)+.5*(n)$){1};
\end{tikzpicture}	
	}
	\\
	\resizebox{1.02\textwidth}{!}{
				\begin{tikzpicture}[font=\footnotesize,edge from parent/.style={draw,thick}]
		% Two node styles: solid and hollow
		\tikzstyle{solid node}=[circle,draw,inner sep=1.2,fill=black];
		\tikzstyle{hollow node}=[circle,draw,inner sep=1.2];
		% Specify spacing for each level of the tree
		\tikzstyle{level 1}=[level distance=15mm,sibling distance=50mm]
		\tikzstyle{level 2}=[level distance=15mm,sibling distance=25mm]
		\tikzstyle{level 3}=[level distance=15mm,sibling distance=15mm]
		% The Tree
		\node[draw] at (-4.5,0) {$r_2(x_0)$};
		\node(0)[hollow node]{}
		child{node[solid node]{}
			child{node[solid node]{}
				child{node[below]{} edge from parent[->,solid,very thick] node[left]{$e$}}
				child{node[below]{} edge from parent[black, solid, thick] node[right]{$f$}}
				edge from parent[solid] node[above left]{$a$}
			}
			child{node[solid node]{}
				child{node[below]{} edge from parent[->,solid,very thick] node(s)[left]{$g$}}
				child{node[below]{} edge from parent[solid] node(t)[right]{$h$}}
				edge from parent[->,solid,very thick] node[above right]{$b$}
			}
			edge from parent[->,solid,very thick] node[above left]{$L$}
		}
		child{node[solid node]{}
			child{node[solid node]{}
				child{node[below]{} edge from parent[solid] node(m)[left]{$i$}}
				child{node[below]{} edge from parent[->,solid,very thick] node(n)[right]{$j$}}
				edge from parent[solid] node[above left]{$c$}
			}
			child{node[solid node]{}
				child{node[below]{} edge from parent[solid] node[left]{$k$}}
				child{node[below]{} edge from parent[->,solid,very thick] node[right]{$l$}}
				edge from parent[->, solid, very thick] node[above right]{$d$}
			}
			edge from parent[solid] node[above right]{$R$}
		};
		% information sets
		%\draw[loosely dotted,very thick](0-1-1)to[out=-15,in=195](0-2-1);
		%\draw[loosely dotted,very thick](0-1-2)to[out=-15,in=195](0-2-2);
		% movers
		\node[above,yshift=2]at(0){1,2,3};
		\foreach \i in {1,2} \node[above,yshift=2]at(0-\i){1,2,3};
		\node[above left]at(0-1-1){1,2,3};
		\node[above right]at(0-1-2){1,2,3};
		\node[above left]at(0-2-1){1,2,3};
		\node[above right]at(0-2-2){1,2,3};
		
				\node[below]at(0-1-1-1){$z_1$};
				\node[below]at(0-1-1-2){$z_2$};
				\node[below]at(0-1-2-1){$z_3$};
				\node[below]at(0-1-2-2){$z_4$};
				\node[below]at(0-2-1-1){$z_5$};
				\node[below]at(0-2-1-2){$z_6$};
				\node[below]at(0-2-2-1){$z_7$};
				\node[below]at(0-2-2-2){$z_8$};
		%\foreach \i in {1,2,3,4} \node[above,yshift=2]at(0-1-\i){3};
		%	\node at($.5*(s)+.5*(t)$){1};
		%	\node at($.5*(m)+.5*(n)$){1};
		\end{tikzpicture}
		
		\qquad
		
		\begin{tikzpicture}[font=\footnotesize,edge from parent/.style={draw,thick}]
% Two node styles: solid and hollow
\tikzstyle{solid node}=[circle,draw,inner sep=1.2,fill=black];
\tikzstyle{hollow node}=[circle,draw,inner sep=1.2];
% Specify spacing for each level of the tree
\tikzstyle{level 1}=[level distance=15mm,sibling distance=50mm]
\tikzstyle{level 2}=[level distance=15mm,sibling distance=25mm]
\tikzstyle{level 3}=[level distance=15mm,sibling distance=15mm]
% The Tree
\node[draw] at (-4.5,0) {$r_3(x_0)$};
\node(0)[hollow node]{}
child{node[solid node]{}
	child{node[solid node]{}
		child{node[below]{} edge from parent[->,solid,very thick] node[left]{$e$}}
		child{node[below]{} edge from parent[black, solid, thick] node[right]{$f$}}
		edge from parent[->,solid,very thick] node[above left]{$a$}
	}
	child{node[solid node]{}
		child{node[below]{} edge from parent[->,solid,very thick] node(s)[left]{$g$}}
		child{node[below]{} edge from parent[solid] node(t)[right]{$h$}}
		edge from parent[solid] node[above right]{$b$}
	}
	edge from parent[solid] node[above left]{$L$}
}
child{node[solid node]{}
	child{node[solid node]{}
		child{node[below]{} edge from parent[->,solid,very thick] node(m)[left]{$i$}}
		child{node[below]{} edge from parent[solid] node(n)[right]{$j$}}
		edge from parent[solid] node[above left]{$c$}
	}
	child{node[solid node]{}
		child{node[below]{} edge from parent[solid] node[left]{$k$}}
		child{node[below]{} edge from parent[->,solid,very thick] node[right]{$l$}}
		edge from parent[->, solid, very thick] node[above right]{$d$}
	}
	edge from parent[->,solid,very thick] node[above right]{$R$}
};
% information sets
%\draw[loosely dotted,very thick](0-1-1)to[out=-15,in=195](0-2-1);
%\draw[loosely dotted,very thick](0-1-2)to[out=-15,in=195](0-2-2);
% movers
\node[above,yshift=2]at(0){1,3};
\foreach \i in {1,2} \node[above,yshift=2]at(0-\i){2};
\node[above,yshift=2]at(0-1-1){1,3};
\node[above,yshift=2]at(0-1-2){1,3};
\node[above,yshift=2]at(0-2-1){1,3};
\node[above,yshift=2]at(0-2-2){1,3};

		\node[below]at(0-1-1-1){$z_1$};
		\node[below]at(0-1-1-2){$z_2$};
		\node[below]at(0-1-2-1){$z_3$};
		\node[below]at(0-1-2-2){$z_4$};
		\node[below]at(0-2-1-1){$z_5$};
		\node[below]at(0-2-1-2){$z_6$};
		\node[below]at(0-2-2-1){$z_7$};
		\node[below]at(0-2-2-2){$z_8$};
%\foreach \i in {1,2,3,4} \node[above,yshift=2]at(0-1-\i){3};
%	\node at($.5*(s)+.5*(t)$){1};
%	\node at($.5*(m)+.5*(n)$){1};
\end{tikzpicture}	
		
}
	\caption{A sequence of reference systems at $x_0$ with respect to a given system, $(\sigma,\pi)$, where player 1's preferences are ordered as follows: $u_1(z_6) \leq u_1(z_1) \leq u_1(z_3) \leq u_1(z_8)$.}
	\label{fig:reference_points}	
\end{figure}
\end{document}